\documentstyle[11pt]{article}

\def\tabaddress#1{{\small\it\begin{tabular}[t]{c}#1
\\[1.2ex]\end{tabular}}}
\def\UPCMAT{\it Departamento de Matem\'atica Aplicada y Telem\'atica\\
    Edificio C-3, Campus Norte UPC\\
   C/ Jordi Girona 1\\
   E-08034 BARCELONA\\
   SPAIN}
\font\fr=eufm10 scaled \magstep 1 %(caracteres goticos)
\font\es=msbm10                   %(caracteres ``doble barra´´)
\newtheorem{teor}{Theorem}
\newtheorem{prop}{Proposition}
\newtheorem{corol}{Corollary}
\newtheorem{lem}{Lemma}
\newtheorem{definition}{Definition}

\newtheorem{require}{Requirement}
\newtheorem{post}{Postulate}

\def\beq{\begin{equation}}
\def\eeq{\end{equation}}
\def\bea{\begin{eqnarray}}
\def\eea{\end{eqnarray}}
\def\beann{\begin{eqnarray*}}
\def\eeann{\end{eqnarray*}}
\def\beasn{\begin{sneqnarray}}
\def\eeasn{\end{sneqnarray}}
\def\ben{\begin{enumerate}}
\def\een{\end{enumerate}}
\def\bit{\begin{itemize}}
\def\eit{\end{itemize}}
\def\dst{\(\displaystyle}
\def\proof{( {\sl Proof} )\quad}

\def\derpar#1#2{\frac{\partial{#1}}{\partial{#2}}}

\def\map#1{\mathrel{\mathop{\to}\limits^{#1}}}
\def\mapping#1{\mathrel{\mathop{\longrightarrow}\limits^{#1}}}

\def\forta#1{\mathop{\mathpalette\@vereq\sim}\limits_{#1}}

\def\qed{\ifvmode\removelastskip\fi
{\unskip\nobreak\hfil\penalty50\hbox{}\nobreak\hfil
\hbox{\vrule height1.2ex width1.2ex}\parfillskip=0pt
\finalhyphendemerits=0 \par\smallskip}}
\def\buit{\hbox{\rm\O}}

\def\vf{\mbox{\fr X}}

\def\d{{\rm d}}
\def\C{{\cal C}}
\def\P{{\cal P}}
\def\D{{\cal D}}
\def\E{{\cal E}}
\def\H{{\cal H}}
\def\Hr{\tilde{\cal H}}
\def\He{{\cal H}^{ext}}
\def\Hc{{\cal H}^C}
\def\GL{{\bf GL}(n,\Complex )}
\def\SL{{\bf SL}(n,\Complex )}
\def\ML{{\bf ML}(n,\Complex )}

\def\sta{|\psi \rangle }
\def\Op{{\rm O}}
\def\h{{\rm h}}
\def\curv{{\bf \Omega}}
\def\Zahl{\mbox{\es Z}}
\def\Real{\mbox{\es R}}

\def\Complex{\mbox{\es C}}
\def\inn{\mathop{i}\nolimits}

\def\Tan{{\rm T}}

\def\Lie{\mathop{\rm L}\nolimits}
\def\Cinfty{{\rm C}^\infty}

\catcode`@=12
\def\LF{{\mit\Lambda}_{\Omega}}
\def\rps{(\tilde C,\tilde\Omega )}
\title{MATHEMATICAL FOUNDATIONS OF GEOMETRIC QUANTIZATION}
\author{\sc Arturo Echeverr\'ia-Enr\'iquez,
   \\
   {\sc Miguel C. Mu\~noz-Lecanda\thanks{MATMCML@MAT.UPC.ES}},
   \\
   {\sc Narciso Rom\'an-Roy\thanks{MATNRR@MAT.UPC.ES}},
   \\
   {\sc Carles Victoria-Monge\thanks{MATCVM@MAT.UPC.ES}}
   \\
   \tabaddress{\UPCMAT}}
\date{ }

\begin{document}
\maketitle

\begin{abstract}
In this review
the foundations of Geometric Quantization are explained and discussed.
In particular, we want to clarify the mathematical aspects related to
the
geometrical structures involved in this theory: complex line bundles,
hermitian connections, real and complex polarizations, metalinear
bundles and
bundles of densities and half-forms.
In addition, we justify all the steps followed in the geometric
quantization programme,
 from the standpoint definition to the structures which are successively

introduced.
\end{abstract}

\vfill \hfill
\vbox{\raggedleft AMS s.\,c.\,(1991): 58F06, 55R10, 53C15}\null

\thispagestyle{empty}
\setcounter{page}{0}

\clearpage
\tableofcontents

\section{Introduction}

There are two kinds of theories for describing
the dynamical behaviour of a physical system:
{\it Classical} and {\it Quantum} theories.
The quantum description is obtained from the classical one
following an appropriate procedure, which is called {\sl quantization}
of the system.

In Theoretical Physics, there are different ways to quantize a classical

theory;
such as,  {\it canonical quantization}, {\it Feynmann-path integral
quantization}
{\it Weyl-Wigner quantization}, {\it Moyal quantization},
and other methods derived from these ones.
In many cases, the first of them is a direct and easy way of
quantization.

Canonical quantization is based in the so-called {\it Dirac's rules for
quantization}.
It is applied to ``simple'' systems:
finite number of degrees of freedom and
``flat'' classical phase spaces (an open set of $\Real^{2n}$).
The lines of the method are mainly the following
\cite{AM-78}, \cite{CDL-77}, \cite{Di-pqm}, \cite{GP-78}:
\begin{itemize}
\item
{\sl Classical description} (starting data).
The system is described by the {\it Hamiltonian} or {\it canonical
formalism}:
its classical phase space is locally coordinated
by a set of {\it canonical coordinates} $(q^j,p_j)$,
the {\it position} and {\it momentum} coordinates.
Classical observables are real functions $f(q^j,p_j)$.
Eventually, a Lie group $G$ of symmetries acts on the system.
\item
{\sl Quantum description}.
The quantum phase space is a complex Hilbert space $\H$.
Quantum observables are self-adjoint operators acting on $\H$, ${\cal
O}(\H )$
\footnote{
The Hilbert space is complex in order to take into account the
interference phenomena of wave functions representing the quantum
states.
The operators are self-adjoint in order to assure their eigenvalues are
real.}.
The symmetries of the system are realized by a group of unitary
operators $U_G(\H )$.
\item
{\sl Quantization method}.
As a Hilbert space we take the space of square integrable complex
functions
of the configuration space; that is, functions depending only on the
position coordinates,
$\psi(q^j)$. The quantum operator associated with $f(q^j,p_j)$
is obtained by replacing $p_j$ by $-i\hbar \derpar{}{q^j}$,
and hence we have the correspondence
$f(q^j,p_j) \mapsto \Op_f(q^j,-i\hbar \derpar{}{q^j})$.
In this way, the classical commutation rules
between the canonical coordinates are assured to have
a quantum counterpart: the commutation rules
between the quantum operators of position and momentum
(which are related to the ``uncertainty principle'' of Quantum
Mechanics).
\end{itemize}
Nevertheless, canonical quantization involves several problems.
The principal ones are the following:
\begin{itemize}
\item
As we have said, it can be applied to finite dimensional systems
with ``flat'' classical phase spaces. Some difficulties arise when this
is not the case.
\item
The method exhibits a strong coordinate dependence:
it needs the existence of global canonical coordinates and depends on
their choice,
that is, it is not invariant under canonical transformations.
In addition, the result of quantization depends on the
order on which $q^j$ and $p_j$ appear in the expression of the classical

observables.
\item
The procedure is easy for simple systems, but serious difficulties arise

when we deal with
constrained systems or systems with internal degrees of freedom.
\item
There are several ways to obtain the quantization of a system: the
so-called
{\it Schr\"odinger representation}, {\it Bargmann-Fock representation},
etc. Canonical quantization does not provide a unified frame for all of
them.
\end{itemize}

In order to solve these and other questions,
a new theory, called {\sl geometric quantization}, was developed in the
70's.
Its main goal is to set a relation between
classical and quantum mechanics from a geometrical point of view,
taking as a model the canonical quantization method.
In this sense, it is a theory which removes
some ambiguities involved in the canonical quantization procedure. Thus,

for instance:
\begin{itemize}
\item
It gives a unified frame for the various kinds of representations.
\item
It generalizes the quantization procedure for classical
phase spaces which are not necessarily ``flat''
(and, even, without being a cotangent bundle).
\item
Since it is a geometrical theory, it is a coordinate-free quantization
procedure.
\item
It clarifies the analogies between the mathematical
structures involved in classical and quantum theories.
\end{itemize}

On the other hand, a relevant feature
of geometric quantization is its close relationship
with the theory of irreducible unitary representations of Lie groups
\cite{AK-71}, \cite{Ki-62}, \cite{Ko-70}.
This relation can be understood in the following way:
in the geometrical description of a regular system
the classical phase space is, in a lot of cases, a symplectic manifold
$(M,\Omega )$. The classical observables are the real smooth
functions ${\mit\Omega}^0(M)$. Suppose $G$ is a Lie group of symmetries
with a strongly symplectic action on $(M,\Omega )$.
Geometric quantization tries to establish a correspondence between the
categories
$(M,{\mit\Omega}^0(M))$ and $(\H ,{\cal O}(\H ))$
and such that the group of symmetries $G$
is realized as a group of unitary operators $U_G(\H )$ \cite{Go-80}.
The situation is summarized in the following diagram:
$$
\begin{array}{ccccc}
& G & \longrightarrow & (M,\Omega ) &
\\
{\rm irr. \ unit. \ rep. \ } & \bigg\downarrow & &
\bigg\downarrow & {\rm geom. \ quant.\ }
\\
& U_G(\H ) & \longrightarrow & \H &
\end{array}
$$
Then, the way of constructing irreducible unitary representations
of $G$ (by the orbit method) is related to the way of constructing the
Hilbert space $\H$
(made of quantum states) from $(M,\Omega )$.

The first works on geometric quantization
are due to J.M. Souriau \cite{So-69}, B. Kostant \cite{Ko-70} and I.E.
Segal \cite{Se-60},
although many of their ideas were based on
previous works by A.A. Kirillov \cite{Ki-62}, \cite{Ki-76}.
Nowadays, their results constitute what is known as {\sl prequantization

procedure}.

Nevertheless, the so-obtained quantum theory is unsatisfactory
 from the point of view of the irreducibility of the
quantum phase space, so that another basic structure
in the geometric quantization programme has to be introduced:
the so-called {\sl polarization},
(this concept was due to B. Kostant and J.M. Souriau in the real case,
and to L. Auslander and B. Kostant in the complex one).
Concerning to this question, the relation between  quantizations of the
same
system arising from different choices of polarizations
was also studied: it is performed by means of the so-called
{\it Blattner-Kostant-Sternberg kernels}
(see \cite{Bl-73}, \cite{Bl-75}, \cite{Bl-77}, \cite{GS-77}).

Once the polarization condition is imposed,
in a lot of cases, other structures have to be added,
because  the inner product between {\it polarized quantum states}
is not well-defined in general.
The key is to introduce the {\sl bundles of densities and half-forms}
\cite{Wn-77}
in order to define the inner product between quantum states.
Finally, in many cases, the so-called {\sl metaplectic correction} must
be done
for obtaining the correct energy levels of the quantum theory
\cite{Bl-73}, \cite{GS-77}, \cite{Ko-74}.

Although the geometric quantization programme was initially developed
for quantizing regular systems; that is, symplectic manifolds,
it was applied soon to {\sl presymplectic manifolds},
as an attempt for giving a geometric framework for the canonical
quantization
rules which P.A.M. Dirac applied to {\sl singular systems}
(see, for instance,  \cite{AS-86}, \cite{Blau-88a}, \cite{Blau-88b},
\cite{Go-86},
\cite{GS-81},\cite{Lo-90}, \cite{Ml-89}, \cite{Sn-83}, \cite{Tu-91}).
Nevertheless, the method shows important limitations;
mainly, the noncommutativity of the procedures of constraining
and quantizing. In order to overcome these problems
new geometrical structures were introduced, which led to the so-called
{\sl BRST quantization} (from {\it Becchi-Rouet-Stora-Tuytin})
\cite{ALN-90}, \cite{ALN-91}, \cite{DEGST-91}, \cite{Ib-90},
\cite{Ko-77}, \cite{Lo-92},
\cite{Tu-92a}, \cite{Tu-92b}.

In addition, geometric quantization has been extended in order to be
applied to
{\sl Poisson manifolds} \cite{Va-91}, \cite{Va-97}.
The origin of this question is that, in the most general cases,
the phase space of classical dynamical systems are not symplectic
manifolds merely,
but Poisson manifolds. As a generalization of these ideas, quantization
of
{\sl Jacobi manifolds}  has been also considered recently \cite{CLM-96},

\cite{LMP-97}.
The interest of this topic is that, from the mathematical point of view,

Jacobi manifolds are the natural generalization of Poisson manifolds
(in particular, of symplectic, cosymplectic and Lie-Poisson manifolds);
and their physical interest lies on their relation with the
{\sl Batalin-Vilkovisky algebras}.

As a final remark, it is interesting to point out that
geometric quantization is a theory developed essentially
for the quantization of finite dimensional systems.
Few things are known about the geometric quantization of
field theories, which is a topic under research.

The aim of this paper is to give a mathematical detailed description of
geometric quantization. In particular, our study concerns
just to the ``standard theory''; that is, we only consider geometric
quantization
of symplectic manifolds (up to the metaplectic correction).

We pay special attention to several questions, namely:
\begin{itemize}
\item
The analysis of the mathematical aspects related
to the structures involved in the geometric
quantization theory, such as complex line bundles, hermitian
connections,
real and complex polarizations, metalinear bundles and
bundles of densities and half-forms.
\item
The justification of all the steps followed in the geometric
quantization programme,
 from the standpoint definition to the structures which are successively

introduced.
\end{itemize}

Next, we give some indications on the organization of the paper.

In section 2, we begin with a discussion on the ideas and postulates on
which the canonical
quantization is based and which justify the steps of the geometric
quantization programme.
Section 3 contains a careful and detailed exposition
of the mathematical concepts needed in the first stage of
prequantization.
Next we begin the geometric quantization programme, properly said.
So, section 4 is devoted to explain its first steps
which lead to the so-called prequantization procedure.
Once the problems arising above have been discussed,
a new structure for geometric quantization is introduced and justified
in section 5: the concept of {\it polarization}, its
properties as well as their application to quantization.
Section 6 is devoted to introduce new mathematical structures:
the {\it metalinear structure} and the {\it bundle of densities} and
{\it half-forms}.
These structures are then used in order to complete the quantization
programme.
The final section is devoted to discuss some problems concerning to
the geometric quantization of constrained systems.

All manifolds are assumed to be finite dimensional,
paracompact, connected and $C^\infty$. All maps are $C^\infty$.
Sum over crossed repeated indices is understood.
As far as possible, we follow the notation of references
\cite{AM-78} and \cite{AMR-mtaa}.
In particular, if $(M,\Omega )$ is a symplectic manifold ($\dim M =
2n$),
the Hamiltonian vector fields are defined as $\inn(X_f)\Omega = \d f$.
Then, we have
$$
\{ f,g \} = \Omega (X_f,X_g) =-\inn(X_f)\inn(X_g)\Omega =-X_f(g) =
X_g(f)
$$
and $X_{\{f,g\}} =[X_g,X_f]$.
In a chart of canonical coordinates the expression of the symplectic
form is
$\Omega = \d q^j \wedge \d p_j$, and
$$
\{ f,g \} =
\derpar{f}{q^j}\derpar{g}{p_j}-\derpar{f}{p_j}\derpar{g}{q^j}
$$

\section{Preliminary statements}

Since our goal is to construct a geometrical theory
of quantization based on the canonical quantization programme,
we will take as the standpoint model
the geometrical framework which describes the {\it canonical formalism}
of the classical physical systems  \cite{AM-78}, \cite{Ar-mmmc}.
Then, a set of postulates is stated in order to construct
the corresponding quantum description.

There are several ways of choosing a set of axioms
or postulates for Quantum Mechanics (see, for example,
\cite{CDL-77}, \cite{Di-pqm}, \cite{GP-78}, \cite{Ja-fqm},
\cite{Mk-mfqm},
\cite{Mo-qmdo}, \cite{Se-pgqm}, \cite{Vn-mfqm}).
In general, these postulates can be arranged into three groups:
those which we can call the {\sl ``kinematical'' postulates},
the {\sl``dynamical'' postulate} and the {\sl ``statistical''
postulates}.
Nevertheless, in this paper we are only interested in the
``kinematical'' and,
eventually, in the ``dynamical'' aspects of quantum theory
and, therefore, we omit (if possible) any reference to ``statistical''
considerations.

Thus, this section is devoted to state and comment
those postulates of Quantum Mechanics in which the standpoint
definition of the geometric quantization programme is based.
Many details of this presentation can be found also in
\cite{Ki-gq} and \cite{Tu-96}.

\subsection{The postulates of Quantum Mechanics}

\subsubsection{On the space of states}

In the canonical formalism of Classical Mechanics,
the phase space of a physical system is assumed to be a manifold ${\cal
M}$
which is endowed with a symplectic structure $\Omega$
or, more generically, with a Poisson structure $\{\ ,\ \}$.
Every point of this manifold represents a {\it (classical) physical
state} of the system.
Since, as we will see in section \ref{prequa},
the geometric quantization program deals with symplectic manifolds,
if ${\cal M}$ is Poissonian,
remembering that a Poisson manifold is the union of symplectic
manifolds,
(its {\it symplectic leaves} \cite{LM-87}), with this programme
we quantize every symplectic leaf, which is only a partial
representation of the phase space
\footnote{
See the comments in the Introduction about
geometric quantization of Poisson manifolds.
}.
This is an essential fact because, given a symplectic manifold
$(M,\Omega )$,
there is a natural way to define a Hilbert space associated with $M$
and a set of self-adjoint operators acting on it and satisfying
the suitable conditions which we will discuss afterwards.

In contrast to this fact,
in Quantum Mechanics the framework for the description of a physical
system
is a separable complex Hilbert space.
In general, there are different ways of working, mainly: the {\it
Hilbert space formulation}
and the {\it projective Hilbert space formulation}
\footnote{
There is also a third possibility:
the so-called {\it unit sphere formulation}.
Nevertheless we do not consider it here,
since it is not relevant for our presentation
of geometric quantization.
You can see a detailed exposition of it,
as well as its use in an alternative
presentation of geometric quantization, in
\cite{Tu-85} and \cite{Tu-96}.
}.
Next we are going to compare both of them.
\begin{enumerate}
\item
{\bf The Hilbert space formulation}:

The initial framework will be a Hilbert space $\H$.
At first, it seems reasonable to identify each element
$\sta \in \H$ as a quantum state, but as it is well known,
the dynamical equations in Quantum Mechanics are linear
in the sense that the set of solutions is a linear subspace.
Hence  we cannot identify each element $\sta \in \H$ as a quantum state
since,
for every $\lambda \in \Complex$, $\lambda \not= 0$,
there is no way to choose between the solutions
$\sta$ and $\lambda\sta$, so they must represent the same quantum state
\footnote{
When $\mid \lambda \mid = 1$,
$\lambda$ will be called a {\it phase factor}.
}.
Then, the true quantum states are really rays in that Hilbert space.
Hence, in this formulation there is a redundancy which has to be
taken into account when we define the quantum states.
\item
{\bf The projective Hilbert space formulation}:

If we want to eliminate this redundancy in the definition of quantum
states,
the only way is passing to the {\it projective Hilbert space}
${\bf P}\H$ which is made of the complex lines in $\H$:
$$
\sta_{\Complex} := \{ \lambda \sta \ \mid \ \lambda \in \Complex
\ , \ \sta \in \H \}
$$
We have the following projection
$$
\begin{array}{ccc}
\H -\{ 0 \} & \mapping{\pi} & {\bf P}\H
\\
\sta & \mapsto & \sta_{\Complex}
\end{array}
$$
Hence, in this picture, a quantum state is given by an unique element of

${\bf P}\H$.
It must be remarked that, like in the classical situation,
${\bf P}\H$ is a differentiable manifold but, in contrast, it is
infinite-dimensional.
\end{enumerate}

We can summarize this discussion in the following postulate:

\begin{post}
In the framework of Quantum Mechanics, a physical system is described by

a
{\sl separable (complex) Hilbert space} $\H$.
Every state of this system at time $t$ is represented by a
{\sl ray} $|\psi (t) \rangle_{\Complex}$ belonging to the Hilbert space.

Any element $|\psi (t) \rangle $ (different from zero) of this ray is
called a {\rm vector state}.
\label{pos1}
\end{post}

{\bf Comment}:
\begin{itemize}
\item
There is a one-to-one correspondence between
the states in this postulate and the so-called
{\it pure states} in several axiomatic formulations of Quantum Mechanics

\cite{Mk-mfqm}, \cite{Mo-qmdo}, \cite{Se-pgqm}, \cite{Vn-mfqm},
\cite{Wl-tgqm}.
This correspondence is established in the following way:
the pure states are the projection operators over the one-dimensional
subspaces of $\H$,
meanwhile the so-called {\it mixed states} are convex combinations of
projection operators
(which are not necessarily  projection operators).
\end{itemize}

\subsubsection{On the observables}

In the classical picture, a {\it physical observable}
(that is, a measurable quantity) is a real smooth function
$f \in {\mit\Omega}^0(M)$ and the result of
a measure of a classical observable is the value taken by the
representative
function on a point (state) of the classical phase space.
In contrast:

\begin{post}
In the framework of Quantum Mechanics,
every observable of a physical system is represented by a
{\sl self-adjoint linear operator} which acts on its associated Hilbert
space
\footnote{
Technical difficulties concerning the domains of unbounded self-adjoint
operators
can be ignored hereafter since they are not important
for the purposes of geometric quantization.
}.

The result of a measure of a quantum observable
is an eigenvalue of the corresponding operator.
\label{pos2}
\end{post}

\subsubsection{On the dynamics}

The above two postulates establish the ``kinematical'' framework
for the description of Quantum Mechanics.
The following step lies in stating the dynamical equations.
In the canonical formulation of the classical theory,
the usual way is to give a function $H \in {\mit\Omega}^0(M)$
containing the dynamical information of the system
(the {\it Hamiltonian function}) and hence to take the
{\it Hamilton equations} as the equations of motion.
Then, the dynamical evolution of an observable $f$ is given by
\beq
\frac{\d f}{\d t} = X_H(f) = \{ f,H \}
\label{clasev}
\eeq
In Quantum Mechanics we have two possible options:

\begin{post}
In the framework of Quantum Mechanics, the dynamics of the system is
defined
by a quantum observable $\Op_H$ called the
{\sl Hamiltonian operator} of the system. Then:
\begin{enumerate}
\item
{\rm Heisenberg picture \/}:
The dynamical evolution of the system is carried out by the quantum
observables and,
in the interval of time between two consecutive measures,
the evolution of every observable $\Op_f(t)$  is given by the
{\rm Heisenberg equation}
\footnote{
Observe the analogy with the classical equation
(\ref{clasev})}
$$
i\hbar\frac{\d}{\d t}{\Op_f(t)} = [\Op_f(t),\Op_H(t)]
$$
In this picture the states of the system are constant in time.
\item
{\rm Schr\"odinger picture \/}:
The dynamical evolution of the system is carried out by the states and,
in the interval of time between two consecutive measures, on each ray
$|\psi (t) \rangle_{\Complex}$, there is some representative vector
state
$|\psi (t) \rangle $ such that the evolution of the system is given by
the
{\rm Schr\"{o}dinger equation}
$$
i\hbar\frac{\d}{\d t}|\psi (t) \rangle = \Op_H|\psi (t) \rangle
$$
In this picture the observables of the system are  represented
by operators which are constant in time.
\end{enumerate}
\label{pos5}
\end{post}

Really, geometric quantization concerns only to the ``kinematical''
aspects of the quantum theory. Nevertheless, several
attempts have been made, trying to set up this dynamical postulate in
geometrical terms,
although we do not treat this subject on our exposition
(see, for instance \cite{ACP-82}, \cite{Il-98a}, \cite{Il-98b},
\cite{Tu-85}
for more information on this topic).

There are other postulates which are related with
the {\it probability interpretation} of Quantum Mechanics.
As we have said earlier, we do not consider them in this study
(for further information, see the references given at the beginning).

\subsection{The geometric quantization programme}

Given a symplectic manifold $(M,\Omega )$ (the phase space of a
classical system),
the aim of the quantization programme is to construct a Hilbert space
$\H$
(the space of states of a quantum system),
and associate a self-adjoint operator $\Op_f$
to every smooth function $f$ in a Poisson subalgebra of
${\mit\Omega}^0(M)$
\footnote{
The apparent modesty of this purpose is due to
the fact that it is impossible to represent the full Poisson algebra
${\mit\Omega}^0(M)$ in the conditions that we will specify soon
\cite{Vh-51}.
}.
In addition, as the set of self-adjoint operators
${\cal O}(\H )$ is also a Lie algebra with the bracket operation,
it seems reasonable to demand this correspondence
between classical and quantum observables to be a Lie algebra
homomorphism.

Next, we are going to justify another condition to be satisfied by this
representation.

\subsubsection{On the irreducibility of the space of states}

First of all we need to set the concept of irreducibility,
both in the classical and the quantum picture.

\begin{definition}
Let $(M,\Omega )$ be a symplectic manifold.
A set of smooth functions $\{ f_j \} \subset {\mit\Omega}^0(M)$
is said to be a {\rm complete set of classical observables}
iff every other function $g \in {\mit\Omega}^0(M)$
such that $\{ f_j,g \} = 0$, for all $f_j$, is constant.
\end{definition}

Observe that this imply that
the functions $\{ f_j \}$ separate points in $M$.
Moreover, for every $m \in M$, there exists an open set $U$
and a subset $\{ f'_i \}$ of $\{ f_j \}$ which is a local system of
coordinates on $U$.
Then, let $X_{f'_i}$ be the Hamiltonian vector fields
associated with these functions. This set is a local basis for the
vector fields in $M$
(notice that, in general, the subset $\{ f'_i \}$ is not a global
system of coordinates on $M$, since this would imply that $M$
is parallelizable. As a consequence, this notion is used only locally by

means
of canonical systems of coordinates).
Therefore, if $\{ {\varphi_i}_t \}$ are local one-parameter
groups of $\{ X_{f'_i} \}$ defined in an open set $V \subset M$,
and $S$ is a submanifold of $M$ with ${\rm dim}\, S<{\rm dim}\, M$,
then $S \cap V$ is not invariant by the action of
$\{ {\varphi_i}_t \}$, that is, $M$ is irreducible under the action of
this
set of local groups of diffeomorphisms.

The quantum analogy of this concept can be established as follows:

\begin{definition}
Let $\H$ be a Hilbert space.
A set of self-adjoint operators $\{ \Op_j \}$ (acting in $\H$)
is said to be a {\rm complete set of operators}
iff every other operator $\Op$ which commutes with
all of them is a multiple of the identity.

If $\H$ is considered as the quantum representation of
a physical system, then this set is called a
{\rm complete set of quantum observables}.
\end{definition}

Notice that the operators $\{ \Op_j \}$ and $\Op$
involved in this statement do not come necessarily
 from any set of classical observables.

As above, this concept can be  related to the irreducibility of $\H$
under
the action of this set as follows:

\begin{prop}
If a set of self-adjoint operators
$\{ \Op_j \}$ on $\H$ is a complete set of operators
then $\H$ is irreducible under the action of $\{ \Op_j \}$
(that is, every closed subspace of $\H$ which is invariant
under the action of this set is either equal to $\{ 0 \}$ or $\H$).
\end{prop}
{\sl (Proof)} \quad
Let $\{ \Op_j \}$ be a complete set and $F \subset \H$
a closed subspace invariant under the action
of this set. Let $\Pi_F$ be the projection operator
over $F$. Then $\Pi_F$ commutes with all the elements of
the complete set. In fact, if $\Op$ is a self-adjoint operator on $\H$
which leaves $F$ invariant (and therefore $F^\perp$ is also invariant),
and $\psi \in \H$, $\psi = \psi_1+\psi_2$,
with $\psi_1 \in F$ and $\psi_2 \in F^\perp$, we have:
$$
[\Op ,\Pi_F]\psi =
\Op (\Pi_F (\psi_1))+\Op (\Pi_F (\psi_2))-
\Pi_F(\Op (\psi_1))-\Pi_F(\Op (\psi_2)) =
\Op (\psi_1) -\Op (\psi_1) = 0
$$
since $\Pi_F (\psi_1) = \psi_1$ and $\Pi_F (\psi_2) = 0$.
Then $\Pi_F=\lambda Id_{\H}$, so $F=\{ 0\}$ if $\lambda =0$,
or $F=\H$ if $\lambda\not= 0$.
\qed

It is interesting to point out that if the operators
$\Op_j$ and $\Op$ are continuous, then the converse also holds
(see, for instance, \cite{Dd-ea} and \cite{La-ra}).

Taking into account the above discussion, we will demand:

\noindent
{\bf Irreducibility postulate (first version)}
{\it If $\{ f_j \}$ is a complete set of classical observables
of a physical system then, in the framework of Quantum Mechanics,
their associated quantum operators make up a complete set of quantum
observables
(which implies that the Hilbert space $\H$ is irreducible
under the action of the set $\{ \Op_{f_j} \}$).}

{\bf Comment}:
\begin{itemize}
\item
Let $M=\Real^{2n}$ be the classical phase space,
and $(q^j,p_j)$ the position and momentum canonical coordinates.
Then the uniparametric groups associated with the
Hamiltonian fields $X_{q^j}$ and $X_{p_j}$,
are the groups of translations in position and momentum
which act irreducibly in the classical phase space,
that is, there are no proper submanifolds of $M$ invariant by these
actions.
In the old canonical quantization scheme, $\H = {\cal L}^2(\Real^n)$
(the space of $q^j$-dependent square integrable functions),
and the self-adjoint operators corresponding to the complete set
$(q^j,p_j)$ are
$$
\Op_{q^j} = q^j \quad , \quad \Op_{p_j} = -i\hbar \derpar{}{q^j}
$$
This set of operators is a complete set of quantum observables,
and this is the translation to the quantum picture
of the irreducibility of the phase space.
The above operators satisfy the following commutation rules:
$$
[\Op_{q^k},\Op_{q^j}] = 0
\ , \
[\Op_{p_k},\Op_{p_j}] = 0
\ , \
[\Op_{q^k},\Op_{p_j}] = i\hbar \delta_{k j} {\rm Id}
\ \footnote
{This last equality is related to the known {\it uncertainty principle}
of Quantum Mechanics.}
$$
\end{itemize}

\subsubsection{Classical and quantum symmetries}

A relevant concept in Physics is the notion of {\it symmetry} of a
system.
Next we are going to discuss this subject,
both from the classical and the quantum point of view.
As we will see at the end, we can relate this discussion to the
irreducibility postulate.

Let $(M,\Omega )$ be a symplectic manifold.
A {\it symmetry} of the system described by $(M,\Omega )$
is an element $g$ of a Lie group $G$ which acts symplectically on
$(M,\Omega )$.
So, every symmetry is represented by a symplectomorphism
$\phi_g \colon M \to M$ (that is, such that $\phi_g^*\Omega =\Omega$).
Let ${\bf Sp}(M,\Omega )$ be the group of these symplectomorphisms.
A group of symmetries of $(M,\Omega )$ is then represented
by a subgroup of ${\bf Sp}(M,\Omega )$.

Remember that, if $f \in {\mit\Omega}^0(M)$,
then every local uniparametric group $\{ \varphi_t \}$
associated with the Hamiltonian vector field $X_f$
is a group of local symmetries of $(M,\Omega )$.

The quantum counterpart of this concept is the following:

\begin{definition}
{\rm (Wigner):}
A {\rm symmetry} of the quantum description of a physical system is a
map
${\bf P}\H \to {\bf P}\H$ such that:
\begin{description}
\item[{\rm (i)}] \
It is  bijective.
\item[{\rm (ii)}]
It preserves the map
$$
\begin{array}{ccc}
{\bf P}\H \times {\bf P}\H & \longrightarrow & \Real^+ \cup \{ 0 \}
\\
(\pi \sta , \pi |\psi '\rangle ) & \mapsto &
\frac{\mid \langle \psi | \psi '\rangle \mid^2}
{\| \psi \|^2 \| \psi '\|^2}
\end{array}
$$
which means that the``transition probabilities'' are conserved.
\end{description}
\label{qsym}
\end{definition}

{\bf Comments}:
\begin{itemize}
\item
Incidentally, we can point out that
this definition means that a symmetry of the quantum description
is an isometry in ${\bf P}\H$ for the {\it Fubini-Studdi metric}
\cite{Wl-85}.
\item
The space ${\bf P}\H$ is naturally endowed with a strongly symplectic
form.
It can be proved that a bijection $g \colon {\bf P}\H \to {\bf P}\H$
conserving this symplectic structure and the natural complex
structure of ${\bf P}\H$ is a quantum symmetry
(if $\H$ is finite-dimensional, then the converse is also true:
every quantum symmetry preserves the natural symplectic form
and the complex structure of ${\bf P}\H$)
(see \cite{Tu-87} and the references quoted therein).
\end{itemize}

These comments reveal the analogy between the classical and
quantum concepts of symmetry.

A way of realizing quantum symmetries
is to projectivize unitary or antiunitary operators on $\H$.
The following proposition proves that this is actually the only possible

way.

\begin{prop}
{\rm (Wigner):}
Let $g$ be a symmetry of the quantum description of a physical system.
Then:
\begin{description}
\item[{\rm (i)}] \
There exists either a unitary or (alternatively) an antiunitary
operator $U_g$ on $\H$ such that $U_g$ induces $g$, that is, for
every $\sta \in \H -\{ 0 \}$, we have $\pi(U_g\sta ) = g(\pi \sta
)$.
\item[{\rm (ii)}]
If $U_g$ and $U'_g$ are unitary or antiunitary operators
on $\H$ which induce $g$, then $U_g$ and $U'_g$ differ on a phase
factor.
\end{description}
\label{sg}
\end{prop}

\begin{corol}
If $g_1$, $g_2$ are quantum symmetries and
$U_{g_1}$, $U_{g_2}$ are unitary operators on $\H$
inducing those symmetries, then
$U_{g_1g_2}=\alpha (g_1,g_2)U_{g_1}U_{g_2}$;
where $\alpha (g_1,g_2) \in {\bf U}(1)$.
\end{corol}

Let $G$ be a connected group of quantum symmetries.
As a consequence of the previous results, if ${\bf U}(\H )$ denotes
the set of unitary operators on $\H$, there exists a subgroup
$G'\subset {\bf U}(\H )$ such that the following sequence is exact
$$
1 \to {\bf U}(1) \to G' \to G \to 1
$$
that is, $G'$ is a {\it central extension} of $G$ by ${\bf U}(1)$.

In other words, if ${\bf U}(\H )$ induces the quantum symmetries
and ${\bf PU}(\H ):={\bf U}(\H )/{\bf U}(1)$ is the group of
projectivized
unitary operators on $\H$, then, it is isomorphic to the group of
quantum symmetries, and every subgroup of quantum symmetries
is isomorphic to a quotient of a subgroup of ${\bf U}(\H )$ by ${\bf
U}(1)$.

Summarizing, the situation is the following:
Let $G$ be a Lie group, then it is a group of symmetries of the
physical system, both for the classical and the quantum descriptions,
if we have the following representations of $G$:
\begin{itemize}
\item
As classical symmetries, by a representation as symplectomorphisms
acting on $(M,\Omega )$ (the phase space of the system in the classical
picture).
\item
As quantum symmetries, by a representation as unitary transformations
acting on $\H$
(the space supporting the quantum states of the system in the quantum
picture).
\end{itemize}
Then, we can state the following version of the
irreducibility postulate:

\noindent
{\bf Irreducibility postulate (second version)}
{\it Suppose $G$ is a group of symmetries of a physical system
both for the classical and the quantum descriptions.
If $G$ acts transitively on $(M, \Omega )$
(by means of the corresponding group of symplectomorphisms),
then the Hilbert space $\H$ is an irreducible representation space
for a ${\bf U}(1)$-central extension of the corresponding
group of unitary transformations.
}

Suppose $G$ is a Lie group which acts on $(M,\Omega )$
and the action is {\it strongly symplectic},
that is the fundamental vector fields associated
to the Lie algebra of $G$ by this action are
global Hamiltonian vector fields $\{ X_{f_i} \} \subset {\cal X}_h(M)$.
In this case, the connection between the first and second version
of the irreducibility postulate can be established as follows:
if the action is transitive, then the Hamiltonian functions $\{ f_i \}$
make up a
complete set of classical observables for $(M,\Omega )$
\footnote{
If the action is transitive and symplectic but not strongly symplectic
(or Hamiltonian),
then the fundamental vector fields are locally Hamiltonian
and the corresponding locally Hamiltonian functions
make up a local complete set of classical observables.
}.
Conversely, if $\{ f_i \}$ is a complete set of classical observables,
then these functions can be thought as the generators of a
group $G$ of infinitesimal symplectomorphisms
whose action on $(M,\Omega )$ is strongly symplectic and
transitive \cite{Ki-gq}.

Now, a remaining question is the following:
{\sl can every classical symmetry of a physical system
be translated into a quantum symmetry?}.
This question can be reformulated and generalized in a more precise way.

In fact, since every classical symmetry must be
a symplectomorphism of the classical phase space $(M,\Omega )$,
the maximal set of classical symmetries is
the group of all symplectomorphisms ${\bf Sp}(M,\Omega )$.
In an analogous way, since every quantum symmetry must be
a projective unitary transformation of ${\bf P}\H$,
the maximal set of them is the group of projective unitary
transformations ${\bf PU}(\H )$.
Then, the question can be generalized in the following terms:
{\sl if $(M,\Omega )$ represents the classical phase space of a physical

system
and $\H$ is the Hilbert space for the quantum description,
are the groups ${\bf Sp}(M,\Omega )$ and ${\bf PU}(\H )$ isomorphic?}.

As we will remark at the end of the following section, in general the
answer is negative
\cite{AM-78}, \cite{Go-80}, \cite{Gr-46}, \cite{Vh-51}:
on the one hand, there is no way to associate
an element of ${\bf PU}(\H )$ to every element of ${\bf Sp}(M,\Omega )$.

Physically this means that, in some cases,
if $G$ is a group of symmetries of a classical system,
in the quantization procedure some symmetries are preserved
but other ones are broken. These situations are called
{\sl quantum anomalies} in the physical literature
(see, for instance, \cite{GS-dggt} for a more detailed explanation on
this topic).
On the other hand, neither every quantum symmetry
comes necessarily from a classical one.
In physical terms this is related to the fact that it may have
unitary operators which have not classical counterpart.

Finally, a relevant fact is that, as it is clear from the discussion
made in this section,
geometric quantization of a classical system is closely related to the
study of irreducible
representations of a Lie group
\footnote{
And, really, first developments of geometric quantization
arose from works on the second problem.}.
Pioneering works on these topics are \cite{Ki-62} (for nilpotent
groups),
\cite{AK-71}, \cite{Be-rglr} (for solvable Lie groups)
and \cite{Bl-73}, \cite{Bl-77}, \cite{RW-rsg} (for semisimple groups).
(For a summarized guideline of some of these methods see,
for instance, \cite{Ki-gq} and \cite{Wn-77}).

\subsection{The standpoint definition of the
            geometric quantization programme}

Taking into account the previous postulates and the discussion
made in the above paragraphs, we can say that
the objective of the geometric quantization programme
is to try of finding a correspondence between the set of pairs
{\sl (Symplectic manifolds $(M,\Omega )$,  smooth real functions
$\Cinfty (M)$)}
and {\sl (Complex Hilbert spaces $\H$, self-adjoint operators ${\cal
O}(\H )$)};
or, in a more general way, a functor between the categories
{\sl (Symplectic manifolds $(M,\Omega )$, symplectomorphisms ${\bf
Sp}(M,\Omega )$)}
and {\sl (Complex Hilbert spaces $\H$, unitary operators ${\bf U}(\H
)$)}.
This functorial relation must satisfy certain properties.

Hence, we can establish the following standpoint definition:

\begin{definition}
A {\rm full quantization} of the classical system $(M,\Omega )$
is a pair $(\H_Q,O)$ where:
\begin{description}
\item[{\rm (a)}]
$\H_Q$ is a separable complex Hilbert space.
The elements $\sta \in \H_Q$ are the {\rm quantum wave functions}
and the elements $\sta_{\Complex} \in {\bf P}\H_Q$
are the {\rm quantum states} of the system. $\H_Q$ is called the
{\rm intrinsic Hilbert space} and ${\bf P}\H_Q$ is the
{\rm space of quantum states} of the system.
\item[{\rm (b)}]
$O$ is a one to one map, taking classical observables
(i.e., real functions $f \in {\mit\Omega}^0(M)$) to self adjoint
operators
$\Op_f$ on $\H_Q$, such that
\begin{description}
\item[{\rm (i)}] \ \
$\Op_{f+g} =\Op_f+\Op_g$.
\item[{\rm (ii)}] \
$\Op_{\lambda f} = \lambda \Op_f \ , \
\forall \lambda \in \Complex$ .
\item[{\rm (iii)}]
$\Op_1 = Id_{\H_Q}$
\item[{\rm (iv)}] \
$[\Op_f,\Op_g] = i \hbar \Op_{\{ f,g \}}$
\item[{\rm (v)}] \ \
If $\{ f_j \}$ is a complete set of classical observables
of $(M,\Omega )$, then $\H_Q$ has to be irreducible
under the action of the set $\{ \Op_{f_j} \}$.
Alternatively, suppose $G$ is a group of symmetries of a physical system

both for the classical and the quantum descriptions.
If $G$ acts transitively on $(M, \Omega )$,
then $\H_Q$ provides an irreducible representation space
for a ${\bf U}(1)$-central extension of the corresponding
group of unitary transformations.
\end{description}
The set of these operators is denoted ${\cal O}(\H_Q)$
and its elements are called {\rm quantum observables} or {\rm quantum
operators}.
\end{description}
\label{fquan}
\end{definition}

The justification of this definition lies in the discussion held in the
previous sections.
Thus, part (a) of definition arises as a consequence of postulate
\ref{pos1}.
On the other hand, part (b) is the translation of postulate
\ref{pos2} and, in relation to conditions listed there, we point out
that:
\begin{itemize}
\item
Conditions (i) and (ii) establish the {\it linearity} of the map $O$
which,
although it has not, in general, a physical interpretation,
it is a desirable property from the mathematical point of view.
\item
Condition (iii) gives account of the fact that,
if the result of a measurement has to be equal to $1$
in every state of the classical description of the system,
then we want the same result in the quantum description;
that is the only expected value has to be $1$, so the corresponding
operator must be the identity.
\item
Condition (iv) imposes that, moreover,
the map $O$ is a Lie algebra morphism (up to a factor).
\item
Finally, condition (v) is the irreducibility Postulate.
\end{itemize}

Hence, the quantization programme consists of constructing a Hilbert
space $\H_Q$
on which the Lie algebra of classical observables could be represented
irreducibly by self-adjoint operators on $\H_Q$
satisfying conditions in part (b) of definition \ref{fquan}.

It is important to point out that, as it is proved in
\cite{AM-78}, \cite{Go-80}, \cite{Gr-46} and \cite{Vh-51},
it is not possible to find a full quantization for every classical
system,
even in the case of $M=\Real^{2n}$. (In this last case, it is not
possible to quantize
all the classical observables of the system.
Then, the usual way is to quantize only a subset of all the classical
observables
which is called a {\it Hilbert subalgebra}. We will treat this feature
afterwards).

\section{Hermitian line bundles}

Before starting the explanation of the geometric quantization programme,

several geometric tools (which are basic in this task)
have to be known. This section deals with
the study and development of all these concepts.
General references are  \cite{Bl-gq}, \cite{Ga-83}, \cite{Ko-70} and
\cite{Wo-80}, .

\subsection{Complex line bundles}

\begin{definition}
Let $\pi :L \to M$ be a projection of manifolds.
$(L,\pi ,M)$ is said to be a {\rm complex line bundle} iff:
\begin{description}
\item[{\rm (i)}] \
For every $m \in M$, the fiber $L_m = \pi^{-1}(m)$
is a one dimensional complex vector space.
\item[{\rm (ii)}]
There exists an open covering $\{ U_l \}$ of $M$
and sections $s_l:U_l \to L$ such that the maps
\beann
\eta_l : & \Complex \times U_l & \to  \pi^{-1}(U_l)
\\
& (z,m) & \mapsto zs_l(m)
\eeann
are diffeomorphisms.
\end{description}
\label{clb}
\end{definition}

Observe that the family $\{ (U_l,\eta_l) \}$
is a bundle trivialization and that $s_l(m) \not= 0, \ \forall m \in
U_l$.

Taking into account that
$$
\Complex \times U_{l j} \mapping{\eta_j}
\pi^{-1}(U_{l j}) \mapping{\eta_l^{-1}} \Complex \times U_{l j}
$$
(where $U_{l j}=U_l \cap U_j$), we have that the {\it transition
functions}
$\Psi_{l j}$ are given by $\Psi_{l j} = \eta_l^{-1} \circ \eta_j$, that
is,
$$
\Psi_{l j}(z,m) = \eta_l^{-1}(\eta_j(z,m)) =
\eta_l^{-1}(zs_j(m)) =
\left(\frac {zs_j(m)}{s_l(m)},m\right)
$$
Observe that $\frac{s_j(m)}{s_l(m)}$ is well defined.
We can also write $\Psi_{l j}(z,m) =(zc_{l j}(m),m)$ where
\beann
c_{l j} : & U_{l j} & \hookrightarrow \Complex^*
\\
& m & \mapsto \frac{s_j(m)}{s_l(m)} = (\pi_1 \circ \Psi_{l j})(1,m)
\eeann
satisfying that $c_{l j}c_{jk} = c_{l k}$, in $U_{l jk}$,
$c_{ii} = 1$, and $c_{l j} = c_{ji}^{-1}$.
These functions $c_{l j}$ are also called ``transition functions''
and the above conditions {\it cocycle conditions}.

\begin{definition}
Two complex line bundles $(L,\pi ,M)$ and $(L',\pi ',M)$
are said to be {\rm equivalent} if there exists a fiber diffeomorphism
$\phi : L \to L'$ such that the restrictions to the fibers
$\phi_m : L_m \to L_m'$ are $\Complex$-linear and the induced map on $M$

is the identity.

In this case, we can construct trivializations with the same
transition functions, since they can be carried from one to the other
by means of the diffeomorphism.
We denote by ${\bf L}(M)$ the set of equivalence classes of complex line

bundles.
\label{clbe}
\end{definition}

In the set ${\bf L}(M)$
we consider the tensor product  (over $\Complex$, that is,
$(L \otimes L')_x := L_x \otimes_{\Complex}L'_x$);
then we have a group structure in which
the unit is the trivial bundle and the inverse $L^{-1}$ is the dual
$L^*$.
The transition functions of $L \otimes L'$ are the product of those of
$L$ and $L'$.
This is called the {\it Picard's group} of $M$.

Let $\sigma : M \to L$ be a section. We are going to analyze the
relation
between the restrictions of $\sigma$ to two open sets of a
trivialization.
Let $\{ U_l,\eta_l \}$ be a trivialization with transition functions
$c_{l j}$ and let $\sigma_l$ be the restriction of $\sigma$ to the open
set $U_l$.
Considering the following diagram
\beann
\Complex \times U_l \mapping{\eta_l} &\pi^{-1}(U_l)&
\\
& \Big\uparrow  \sigma_l &
\\
& U_l &
\eeann
the section $\sigma$ defines local functions (with values in $\Complex$)

as
$f_l := \pi_1 \circ \eta_l^{-1} \circ \sigma_l$;
and in the same way for $U_j$. Now, in $U_{l j}$ we have:
\beann
f_l(m)
&=&
(\pi_1 \circ \eta_l^{-1})(\sigma (m)) =
\pi_1((\eta_l^{-1} \circ \eta_j \circ \eta_j^{-1})(\sigma (m)))=
\pi_1(\Psi_{l j}(\eta_j^{-1}(\sigma (m))))
\\ &=&
\pi_1(\Psi_{l j}(f_j(m),m))=
\pi_1(c_{l j}(m)f_j(m),m) =
c_{l j}(m)f_j(m)
\eeann
that is, $f_l = c_{l j}f_j$.
Therefore, a section of $L$ induces, on each open set $U_l$,
a function (with values on $\Complex$) and the relation between the
functions defined
in two intersecting trivializing open sets
is obtained taking the product by the transition functions.

\subsection{Chern classes}

Let $(L,\pi ,M)$ be a complex line bundle and $\{ U_l,s_l \}$ a
trivialization
with transition functions $c_{l j}$. Let ${\cal F}$ be
the sheaf of germs of complex smooth functions
in the manifold $M$, and ${\cal F}^*$ the set of the nowhere-vanishing
ones,
considered as a sheaf of groups with the product.
The transition functions define a \v{C}ech $1$-cocycle in $M$,
$c:U_{l j} \mapsto c_{l j}$, with values in the sheaf ${\cal F}^*$,
and then it determines an element in the cohomology group
$\check{H}^1(M,{\cal F}^*)$.

\begin{prop}
The above \v{C}ech cohomology class
does not depend on the trivialization used
but only on the class of $L$ in ${\bf L}(M)$. Moreover, the assignment
${\bf L}(M) \longrightarrow \check{H}^1 (M,{\cal F}^*)$
is a group isomorphism.
\label{indep}
\end{prop}
{\sl (Proof)} \quad
The independence of the trivialization
is a consequence of the fact that
the union of two trivializations is
a trivialization whose associated covering is
a refinement of those of the initial trivializations.

The assignment is a group morphism because
the transition functions of the tensor product
are the ordinary product of the transition functions
of the factors.
\qed

Consider now the exact sequence of sheaves
$$
0 \to \Zahl \map{\epsilon} {\cal F} \map{e} {\cal F}^* \to 0
$$
where $\epsilon$ is the natural injection and
$e(f):=e^{2\pi if}$.
The corresponding cohomology sequence is
$$
\check{H}^1(M,\Zahl ) \to
\check{H}^1(M,{\cal F}) \to
\check{H}^1(M,{\cal F}^*) \to
\check{H}^2(M,\Zahl) \to
\check{H}^2(M,{\cal F})
$$
but ${\cal F}$ is a fine sheaf
(with partitions of the unity),
therefore the cohomology groups with values in
${\cal F}$ and degree greater than zero are null and then
$\check{H}^1(M,{\cal F}^*) \cong  \check{H}^2(M,\Zahl)$.

\begin{teor}
The group ${\bf L}(M)$ is canonically isomorphic to $
\check{H}^2(M,\Zahl )$.
\label{cais}
\end{teor}
{\sl (Proof)} \quad
It is a consequence of the last proposition and the above arguments.
\qed

\begin{definition}
If $[L]$ is an element of  ${\bf L}(M)$,
then the element $K([L])$ of $\check{H}^2(M,\Zahl)$
corresponding to $[L]$, will be called the
{\rm first Chern class} of $L$.
\label{chern}
\end{definition}

\subsection{Connections and curvature}

Let $\Gamma (L)$ be the ${\cal F}$-module of sections
of $(L,\pi ,M)$ and ${\cal X}^{\Complex}(M)$
the module of complex vector fields on $M$
(${\cal X}^{\Complex}(M) = {\cal X}(M) \otimes \Complex$).

\begin{definition}
A {\rm connection} in the complex line bundle
$(L,\pi ,M)$ is a $\Complex$-linear map
\beann
\nabla : & {\cal X}^{\Complex}(M) & \to
Hom_{\Complex}(\Gamma (L),\Gamma (L)) =
\Gamma (L) \otimes_{\Complex} \Gamma (L)^*
\\
& X & \mapsto \quad \quad \nabla_X
\eeann
satisfying that
\begin{description}
\item[{\rm (1)}] \
$\nabla_{fX} = f\nabla_X$
\item[{\rm (2)}]
$\nabla_X(fs) = (Xf)s + f\nabla_Xs$
\end{description}
that is, denoting by ${\cal F}^p$ the module of complex $p$-forms on
$M$,
an element of
$({\cal F}^1 \otimes_{{\cal F}^0} \Gamma (L))\otimes_{\Complex}\Gamma
(L)^*$
satisfying condition $(2)$.

Equivalently, a connection is a map
\beann
\nabla : & \Gamma(L) & \to
{\cal F}^1 \otimes_{{\cal F}^0} \Gamma (L)
\\
& s & \mapsto \quad \nabla s
\eeann
such that
\begin{description}
\item[{\rm (i)}] \
It is $\Complex$-linear
\item[{\rm (ii)}]
$\nabla (fs) = \d f \otimes s + f\nabla s$
\end{description}
(That is,
$\nabla \in ({\cal F}^1 \otimes_{{\cal F}^0} \Gamma (L))
\otimes_{\Complex} \Gamma (L)^*$
and, moreover, it satisfies condition (ii),
from which, taking
$\nabla_X s := \inn(X) \nabla s$,
the equivalence between both definitions
is immediate).

$\nabla_Xs$ is called the
{\rm covariant derivative}
of the section $s$ with respect to $X$
and the connection $\nabla$.
\label{nabla}
\end{definition}

\begin{prop}
$(\nabla_Xs)(m)$ depends only on $X_m$ and on the germ of $s$ in $m$.
\label{depen}
\end{prop}
{\sl (Proof)} \quad If $s$ has null germ in $m$, then it vanishes
in a neighborhood $U$ of $m$. Let $f \in {\cal F}^0$ be null in an
open set $V \subset U$ and taking the value equal to $1$ in the
complementary of an open set $W$ contained in $U$, such that $V
\subset W \subset U$. It is clear that $s = sf$ and we have that
$$ (\nabla s)(m) = (\nabla (fs))(m) = (\d f)(m) \otimes s(m) +
f(m)(\nabla s) = 0 $$ therefore two sections with the same germ in
$m$ have the same covariant derivative in $m$. On the other hand,
let $U$ and $V$ be open neighborhoods of $m$ in $M$ with $U
\subset V$ and $f \in {\cal F}^0$ such that $f \vert_U = 1$, $f
\vert_{M-V} = 0$. If $s$ is a section it is clear that $s$ and
$fs$ have the same germ in $m$ and then we have $$ (\nabla_X s)(m)
= (\nabla_X (fs))(m) = X_m(f)s(m) + f(m)\inn(X_m)(\nabla s)
=\inn(X_m)(\nabla s) $$ and then it only depends on $X_m$. \qed

In order to find the local expression of a connection,
let $\{ U_l,s_l \}$ be a trivialization of the bundle.
Every section in $U_l$ has the form $s=fs_l$;
hence, according to the property (ii),
in order to calculate $\nabla s$,
it suffices to calculate $\nabla s_l$.
Writing
$\nabla s_l = 2\pi i \omega^l  \otimes s_l$,
(where $\omega^l $ is an element of
${\cal F}^1$ in $U_l$)
\footnote{
Be careful: the summation convention does not apply in this expression
and
in those ones related with it.
},
in $U_{l j}$ we have that
$s_j = c_{l j}s_l$,
therefore
$$
\nabla s_j =
\nabla (c_{l j}s_l) =
\d c_{l j} \otimes s_l + c_{l j} \nabla s_l
$$
that is
$$
2\pi i \omega^j \otimes s_j =
2\pi i c_{l j} \omega^j \otimes s_l =
\d c_{l j} \otimes s_l + c_{l j} 2\pi i \omega^l  \otimes s_l
$$
therefore
$$
2\pi i c_{l j} \omega^j =
\d c_{l j} + c_{l j} 2\pi i \omega^l
$$
and hence, in $U_{l j}$,
$$
\omega^j =
\frac{1}{2\pi i} \frac{\d c_{l j}}{c_{l j}} + \omega^l
$$

Observe now that we can write the following relation
for the covariant derivative of a section:
\beq
\nabla_X s =
\nabla_X (fs_l) =
(X(f) + 2\pi i \langle X,\omega^l  \rangle f) s_l
\label{expcon}
\eeq

The family
$\{ U_l,\omega^l  \}$
is called
{\it connection $1$-form},
but this is not a global form in $M$.
Nevertheless, in $U_{l j}$
the equality
$\d \omega^j \vert_{U_{l j}} = \d \omega^l  \vert_{U_{l j}}$
holds,
hence there exists a global complex $2$-form
$\curv $ on $M$ such that
$\curv  \vert_{U_l} = \d \omega^l $.

\begin{definition}
The $2$-form $\curv $
so defined is called the
{\rm curvature form} of the connection $\nabla$.
\label{cur}
\end{definition}

The relation with the ``classical'' curvature
is as follows:

\begin{prop}
If $X_1, X_2 \in {\cal X}^{\Complex}(M)$
and $s \in \Gamma (L)$, then
$$
2\pi i \curv  (X_1,X_2)s =
(\nabla_{X_1}\nabla_{X_2} - \nabla_{X_2}\nabla_{X_1} -
\nabla_{[X_1,X_2]})s
$$
\label{clacur}
\end{prop}
{\sl (Proof)} \quad
Taking a trivialization
$\{ U_l,s_l \}$,
in the open set $U_l$ we have
$$
\curv  (X_1,X_2) =
\d \omega^l (X_1,X_2) =
X_1 \omega^l (X_2) - X_2 \omega^l (X_1) - \omega^l  ([X_1,X_2])
$$
but
\beann
\nabla_{X_1} \nabla_{X_2}s_l
&=&
\nabla_{X_1}(2\pi i \omega^l (X_2)s_l)=
2\pi i X_1(\omega^l (X_2))s_l + (2\pi i)^2 \omega^l  (X_2) \omega^l
(X_1)s_l
\\
\nabla_{X_2} \nabla_{X_1}s_l
&=&
\nabla_{X_2}(2\pi i \omega^l (X_1)s_l)=
2\pi i X_2(\omega^l (X_1))s_l + (2\pi i)^2 \omega^l  (X_1) \omega^l
(X_2)s_l
\\
\nabla_{[X_1,X_2]}s_l
&=&
2\pi i \omega^l ([X_1,X_2])s_l
\eeann
and thus the result follows immediately.
\qed

Another way of defining the curvature is the following:
we can extend the action of $\nabla$
to the $p$-forms on $M$  with values in $\Gamma (L)$
in the following way
$$
{\cal F}^p \otimes_{{\cal F}^0} \Gamma (L) \mapping{\nabla^p}
{\cal F}^{p+1} \otimes_{{\cal F}^0} \Gamma (L)
$$
by means of the expression
$$
\nabla^p(\alpha \otimes s) =
\d \alpha \otimes s + (-1)^p \alpha \wedge \nabla s
$$
Denoting
$\varpi := \nabla^1 \circ \nabla$,
we have that
\beann
\varpi (fs)&=&(\nabla^1 \circ \nabla )(fs) =
\nabla^1 (\d f \otimes s + f\nabla s)
\\
&=& \d (\d f)\otimes s - \d f \wedge \nabla s
+ \d f \wedge \nabla s + f(\nabla^1 \circ \nabla )s
=f\varpi (s)
\eeann
therefore $\varpi$ is ${\cal F}^0$-linear,
that is,
$$
\varpi \in ({\cal F}^2 \otimes_{{\cal F}^0} \Gamma (L))
\otimes_{{\cal F}^0} \Gamma (L)^*
\cong {\cal F}^2 \otimes_{{\cal F}^0} (\Gamma (L)
\otimes_{{\cal F}^0} \Gamma (L)^*)
$$
Let $\curv $ be the $2$-form
obtained from $\varpi$ by means of the
natural contraction of the last two factors of $\varpi$.
The so-obtained form is a complex $2$-form on $M$.
In order to see that it coincides
with the curvature form,
it suffices to calculate it in an open set $U_l$
of a trivialization $\{ U_l,s_l \}$:
\beann
(\nabla^1 \circ \nabla )s_l
&=&
\nabla^1 (\nabla s_l) =
\nabla^1 (\omega^l  \otimes s_l) =
\d \omega^l  \otimes s_l + \omega^l  \wedge \nabla s_l
\\
&=&
\d \omega^l  \otimes s_l + \omega^l  \wedge \omega^l \otimes s_l =
\d \omega^l  \otimes s_l
\eeann
therefore
$\varpi \vert_{U_l} = \d \omega^l  \otimes s_l \otimes s_l^*$,
where $s_l^*$ is the dual of $s_l$.
Hence, $\curv  = \d \omega^l $,
as we wanted to prove.
(We have not considered the factor
$2\pi i$ which depends on the definition of the connection form).

\subsection{Hermitian structures}

\begin{definition}
Let $(L,\pi ,M)$ be a complex line bundle
endowed with a connection $\nabla$.
A {\rm hermitian structure} on $(L,\pi ,M)$
is a correspondence such that assigns
a {\sl hermitian metric} $\h_m$ on $L_m$, for every $m \in M$,
in a differentiable way; that is, if $s, s'$ are differentiable
sections,
then the function $\h (s,s')$ is differentiable.
This is equivalent to say that $\h$ is an element of
$\Gamma(L^* \otimes \bar L^*)$,
(where $\bar L$ is the conjugate bundle of $L$),
that is, $\h$ satisfies condition $\h(s,s') = \overline{\h(s',s)}$.

Given a hermitian structure $\h$ on $(L,\pi ,M)$,
$\nabla$ is said to be a {\rm hermitian connection} with respect to $\h$

iff
$$
X(\h (s,s')) = \h (\nabla_X s,s') + \h (s,\nabla_X s')
\ , \
\forall X \in {\cal X}(M)
\ , \
\forall s,s' \in \Gamma (L)
$$
This is equivalent to say that $\nabla \h = 0$,
when the connection is extended to $\Gamma(L^* \otimes \bar L^*)$
in the usual way.
\label{hs}
\end{definition}

\begin{prop}
If $\nabla$ is a hermitian connection with respect to $\h$,
then the curvature $\curv $ of $\nabla$ is a real form
\footnote
{\quad
A vector field $X$ is {\it real} if $X(f) \in {\mit\Omega}^0(M) \ ,
\ \forall f \in {\mit\Omega}^0(M)$. A $1$-form $\alpha$ is real iff
$\alpha (X)$ is real, for every real vector field. And so on.
}.
\label{cureal}
\end{prop}
{\sl (Proof)} \quad We are going to see it for trivializing
neighborhoods. So, let $\{ U_l,s_l \}$ be a trivialization, $\{
\omega^l  \}$ the connection forms of $\nabla$ on $\{ U_l \}$ and
$X$ a real vector field; we have \beann X(\h (s_l,s_l)) &=& \h
(\nabla_X s_l,s_l) + \h (s_l,\nabla_X s_l)
\\ &=&
2\pi i \h (\omega^l (X)s_l,s_l) - 2\pi i \h (s_l,\omega^l (X)s_l) =
2\pi i (\omega^l (X) - \overline{\omega^l (X)}) \h (s_l,s_l)
\eeann
and therefore
$$
\omega^l (X) - \overline{\omega^l (X)} =
\frac{1}{2\pi i} \frac{X(\h (s_l,s_l))}{\h (s_l,s_l)}
$$
That is
$$
\omega^l  - \overline{\omega^l } =
\frac{1}{2\pi i} \frac{\d \h (s_l,s_l)}{\h (s_l,s_l)}
$$
hence $\d\omega^l  - \d\overline{\omega^l } =0$,
that is, $\d\omega^l $ is real,
for every $l$ and thus the curvature $\curv$ is real.
\qed

{\bf Comments:}
Let $(L,\pi ,M)$ be a complex line bundle
with hermitian metric $\h$ and hermitian connection $\nabla$.
\begin{enumerate}
\item
Let $\{ U_j,s_j \}$ be a trivialization of $L$ with
$\h (s_j,s_j)(m) = 1$, for every $m \in U_j$.
Then if $\omega^j$ is the connection form of $\nabla$ in $U_j$,
from the above equality we have that
$\omega^j - \overline{\omega^j} = 0$,
so they are real forms.

These special trivializations can be obtained from any one
by dividing every $s_j$ by its module.
\item
If $\{ U_j,s_j \}$ is one of such trivializations,
then $|c_{l j}(m)|=1$, for every $m \in U_{l j}$, since
$$
1 = \h (s_l,s_l) = \h (s_j,s_j) = c_{l j} \bar c_{l j} \h (s_j,s_j)
$$
then $c_{l j} = e^{if_{l j}}$ with $f_{l j} \colon U_{l j} \to \Real$
a differentiable function. So we have that every hermitian line bundle
admits a trivialization with transition functions
taking values in $S^1$, the group of isometries of $\h$.
\end{enumerate}

\subsection{Existence of hermitian connections}

Let $M$ be a differentiable manifold and $\curv $ a real closed two form

on $M$.

\begin{teor}
The necessary and sufficient condition for $\curv $ to be the curvature
$2$-form
of a hermitian connection $\nabla$ on a complex line bundle $(L,\pi ,M)$

endowed with a hermitian metric is that the cohomology class
$[\curv ] \in H^2(M,\Real )$ is integer, that is, the \v{C}ech
cohomology class
canonically associated with $[\curv ]$ belongs to the image of the
morphism
$\varepsilon^2 \colon \check{H}^2(M,\Zahl ) \to \check{H}^2(M,\Real )$
induced by the inclusion $\varepsilon \colon \Zahl \hookrightarrow
\Real$.

Moreover, in this case, $\curv $ is a representative of the image by
$\varepsilon^2$
of the Chern class of the bundle $(L,\pi ,M)$.
\label{cce}
\end{teor}

Before starting the proof, we remind how to construct
the canonical isomorphism between the
de Rham cohomology and the \v{C}ech cohomology
of degree two; and what means that $[\curv ]$ is integer.

Let $\curv  \in {\mit\Omega}^2(M)$ be
a real $2$-form with $\d\curv  = 0$ and
$[\curv ] \in H^2(M,\Real )$ its cohomology class.
We can associate an element of $\check{H}^2(M,\Real )$ to $[\curv ]$.
Let $\{ U_l \}$ be a contractible covering of $M$
\footnote
{\quad
What we understand by ``contractible covering'' is that
all the open sets of the covering
and all their intersections are contractible.
In order to prove its existence, it suffices to endow $M$ with a
Riemannian metric and to take a
covering by geodesically convex open sets.}.
Since $U_l$ are contractible, we have that
$\curv  \vert_{U_l} = \d \omega^l $,
where $\omega^l  \in {\mit\Omega}^1(U_l)$.
But $U_{l j}$ is also contractible and
$\d\omega^l  \vert_{U_{l j}} = \d\omega^j \vert_{U_{l j}}$,
hence $(\omega^j - \omega^l )\vert_{U_{l j}} = \d f^{l j}$,
where $f^{l j} \in {\mit\Omega}^0(U_{l j})$. In $U_{l jk}$ we have that
$$
(\d f^{l j} + \d f^{jk} - \d f^{l k})\vert_{U_{l jk}} =0
$$
therefore $f^{l j} + f^{jk} - f^{l k} = \alpha^{l jk}$ is constant in
$U_{l jk}$.

Let $a$ be the map defined by
$(U_l,U_j,U_k) \mapsto \alpha^{l jk}$.
We have that $a$ is a \v{C}ech cochain associated with the
cohomology class $[\curv ]$ and then we can construct a map
$$
\begin{array}{ccc}
H^2(M,\Real ) & \longrightarrow & \check{H}^2(M,\Real )
\nopagebreak
\\ {}
[\curv ] & \mapsto & [a]
\end{array}
$$
since $\d a = 0$, as
\beann
\d a (U_i,U_j,U_k,U_l)
&=&
a(U_j,U_k,U_l) - a(U_i,U_k,U_l) + a(U_i,U_j,U_l) - a(U_i,U_j,U_k) = 0
\eeann
It is evident that $[a]$ does not depend
on $\curv $ but only on its cohomology class,
since, taking into account that, if $\Omega' = \curv  + \d \eta$,
then ${\omega'}^l = \omega^l  + \eta$ in $U_l$. Therefore
$({\omega'}^j - {\omega'}^l)\vert_{U_{l j}} =
(\omega^j - \omega^l )\vert_{U_{l j}} = \d f^{l j}$
and hence both $\curv $ and ${\bf \Omega'}$
have the same associated cochain.

On the other hand, the natural injection
$\varepsilon : \Zahl \hookrightarrow \Real$ induces a morphism
\beann
\varepsilon^2:& \check{H}^2(M,\Zahl )& \longrightarrow
\check{H}^2(M,\Real )
\\
& [a] & \mapsto \quad [j(a)]
\eeann
therefore we obtain that $[\curv ]$ is in the image of $\varepsilon^2$
if, and only if, there exists a contractible covering $\{ U_l \}$ of
$M$,
a family of $1$-forms $\{ \omega^l  \}$
and a family of functions $\{ f^{l j} \}$ such that
$$
\curv \vert_{U_l} = \d \omega^l
\quad ; \quad
(\omega^j - \omega^l )\vert_{U_{l j}} = \d f^{l j}
\quad ; \quad
(f^{l j} + f^{jk} - f^{l k})\vert_{U_{l jk}} \in \Zahl
$$

Now we prove the theorem.

\noindent {\sl (Proof)} \quad
$(\Longrightarrow )$    \quad
Let $\curv $ be the curvature form of a connection $\nabla$
on a complex line bundle $(L,\pi ,M)$ with hermitian metric $\h$.
Let $\{ U_l,s_l \}$ be a trivialization of the bundle with $\h
(s_l,s_l)=1$.
We know that the forms $\omega^l $ are real, and since
$$
(\omega^j - \omega^l )\vert_{U_{l j}} =
\frac{1}{2\pi i} \frac{\d c_{l j}}{c_{l j}} =
\d f^{l j}
$$
then the functions $f^{l j}$ can be chosen real also.
But we have $2\pi i f^{l j} = \log c_{l j}$;
and since $c_{l j} c_{jk} = c_{l k}$ we obtain that
$\log c_{l j} + \log c_{jk} - \log c_{l k}$
is an integer multiple of $2\pi i$, hence $f^{l j} + f^{jk} - f^{l k}$
is an integer, so, as we have seen above, $[\curv ]$ is integer.

\quad
$(\Longleftarrow)$ \quad
Suppose $\curv $ is a closed $2$-form in $M$ such that
$[\curv ]$ is an integer class.
Let $\{ U_l \}$ be a contractible covering of $M$
and  $\{ \omega^l  \}$ and $\{ f^{l j} \}$ as above.
Now we have that
$f^{l j} + f^{jk} - f^{l k} \in \Zahl$.
Put $c_{l j} = e^{2\pi i f^{l j}}$ in $U_{l j}$.
Taking into account that $c_{l j} c_{jk} = c_{l k}$,
we deduce that a complex line bundle
$(L,\pi ,M)$ with transition functions $c_{l j}$ can be constructed.
The sections $s_l$ can be taken equal to $1$ on each $U_l$.
Once the bundle is constructed, we can take the connection $\nabla$
which is determined by the connection forms
$\omega^l$, and whose curvature form is obviously $\curv $.
The hermitian structure is given by
$$
\h_m(e,e') = z \bar z'
$$
where $e = (m,z)$, $e' = (m,z')$
in any open set of the trivialization containing $m$.
We are going to see that this is
a hermitian connection with respect to this metric. Consider
$X \in {\cal X}(M)$ and $s,s' \in \Gamma (L)$. If
$m \in M$ and $U_l$ is a trivializing open set with $m \in U_l$ and
$s \vert_{U_l} = f_ls_l$, $s' \vert_{U_l} = f_l 's_l$, then we have
\beann
(X(\h (s,s')))(m)
&=&
(X(\h (s,s')))\vert_{U_l}(m)=
(X(\h (f_l s_l,f_l 's_l)))\vert_{U_l}(m)
\\
&=&
(X(f_l \bar f_l '))\vert_{U_l}(m)=
(Xf_l )(m) \bar f_l '(m) + f_l (m)(X \bar f_l ')(m)
\eeann
On the other hand,
\beann
\h (\nabla_X s,s')\vert_{U_l}(m)
&=&
\h ((Xf_l )s_l + 2 \pi i f_l\omega^l  (X)
 s_l ,f_l 's_l )\vert_{U_l}(m)=
((Xf_l )\bar f_l ' + 2 \pi i f_l\omega^l  (X) \bar f_l ')(m)
\\
\h (s,\nabla_Xs')\vert_{U_l}(m)
&=&
(\h (f_l s_l ,(Xf_l ')s_l+ 2 \pi i f_l '\omega^l(X) s_l))(m)=
(f_l (X \bar{f_l '}) -  f_l 2 \pi i \bar{f_l '}\omega^l  (X))(m)
\eeann
because $\omega^l  (X)$ is real. Therefore
$$
X(\h (s,s')) =
\h (\nabla_X s,s') + \h (s,\nabla_Xs')
$$
\qed

\subsection{Classification of hermitian connections}

\subsubsection{Affine structure on the set of connections}

In the following
$(M,\Omega)$ will be a symplectic manifold and
$(L,\pi , M)$ a complex line bundle
endowed with an hermitian metric $\h$.

Denote by ${\rm Con}(L)$ the set of connections on the complex
line bundle $L$. If $\nabla , \nabla ' \in {\rm Con}(L)$, we have
that, for every $s \in \Gamma (L)$ and $f \in {\cal F}^0$, $$
(\nabla - \nabla ')(fs) = f(\nabla - \nabla ')s $$ Therefore $$
\nabla - \nabla '\in {\cal F}^1\otimes_{{\cal F}^0}\Gamma (L)
\otimes_{{\cal F}^0}\Gamma (L)^* \simeq \Gamma (L \times
L^*)\otimes_{{\cal F}^0}{\cal F}^1 \simeq \Gamma (\Complex \times
M)\otimes_{{\cal F}^0}{\cal F}^1 \simeq {\cal F}^1 $$ On the other
hand, if $\eta \in {\cal F}^1$ and $\nabla$ is a connection, so is
$\nabla + \eta$. So we have:

\begin{prop}
The ${\cal F}^0$-module of the one-forms
${\cal F}^1$
acts freely and transitively on the set
${\rm Con}(L)$.
\end{prop}

{\bf Comments:}
\begin{itemize}
\item
An equivalent way of stating the above proposition is the following:
${\rm Con}(L)$ has an affine structure modeled on the vector space
${\cal F}^1(M)$.
\item
Observe that if
$\{ U_j,s_j \}$
is a trivialization of the fiber bundle and
$\{ \omega^j \}$
are the connection forms of $\nabla$, then
$\{ \omega^j + \eta \}$
are the connection forms of $\nabla + \eta$.
\end{itemize}

Denote by ${\rm Con}(L,\h )$
the set of hermitian connections with respect to $\h$.
Then, if $\nabla \in {\rm Con}(L,\h )$
and $\eta \in {\cal F}^1$,
the connection $\nabla + \eta$
is hermitian with respect to $\h$
if, and only if, $\eta$ is real.
In order to see this, it suffices to remind that,
in a trivializing system $\{ U_j,s_j \}$
such that $\{ U_j \}$ is a contractible covering,
the connection forms are real for the hermitian connections.
Then we have proved that:

\begin{prop}
The ${\mit\Omega}^0$-module ${\mit\Omega}^1$
acts freely and transitively on the set
${\rm Con}(L,\h )$.
\end{prop}

Now, let ${\rm Con}(L,\h ,\curv )$
be the set of hermitian connections with curvature $\curv $.
If $\nabla$ and $\nabla ' = \nabla + \eta$
are two of these connections, taking into account the relation between
the corresponding connection forms and their curvature,
we deduce that $\d \eta = 0$. Thus:

\begin{prop}
The group $Z^1(M)$ of the differential 1-cocycles of $M$
acts freely and transitively on the set
${\rm Con}(L,\h ,\curv )$.
\end{prop}

\subsubsection{Equivalence of line bundles with connection}

Let $(L,\pi ,M)$ and $(L',\pi',M)$ be complex line bundles and
$\phi \colon L \to L'$ a complex line bundle diffeomorphism
inducing the identity on $M$.
If $s \colon M \to L$ is a section of $L$,
then $s' = \phi \circ s$ is a section of $L'$.
The map $\Gamma (\phi) \colon \Gamma (L) \to \Gamma (L')$
given by $\Gamma (\phi)(s) := \phi \circ s$
is an isomorphism of ${\cal F}^0$-modules,
because it is a diffeomorphism and is
$\phi$ is $\Complex$-linear on the fibers.

In order to calculate $\Gamma (\phi)$ locally,
let $\{ U_j,s_j \}$ and $\{ U_j,s'_j \}$ be
trivializing systems of $L$ and $L'$.
Then $\phi \circ s_j = \varphi_j s'_j$,
where $\varphi_j \colon U_j \to \Complex^*$
is a differentiable function with values in
$\Complex^*$, because
$\phi_m \colon L_m \to L_m'$
is a $\Complex$-isomorphism for every $m \in M$.
But $s_j = c_{kj}s_k$
and $s'_j = c'_{kj}s'_k$,
where $c_{kj}$ and $c'_{kj}$
are the transition functions of $L$ and $L'$ respectively.
Then, on $U_{jk}$ we have:
$$
\begin{array}{ccccc}
\phi \circ s_j &=& \varphi_j s'_j &=&
\varphi_j c'_{kj}s'_k
\\
\phi \circ s_j &=& \phi \circ (c_{kj}s_k) &=&
c_{kj}\phi \circ s_k = c_{kj}\varphi_k s'_k
\end{array}
$$
therefore $\varphi_j c'_{kj} = c_{kj} \varphi_k$,
that is $\varphi_j = c_{kj} \varphi_k (c'_{kj})^{-1}$.
Hence, $\Gamma (\phi)$ is represented by the family
$\{ U_j,\varphi_j \}$
with $\varphi_j \colon U_j \to \Complex^*$
satisfying
$\varphi_j = c_{kj} \varphi_k (c'_{kj})^{-1}$.

Let $\nabla$ be a connection in $L$.
We can construct a connection in $L'$ induced by $\nabla$ and $\phi$.

\begin{definition}
The {\rm connection induced} by $\nabla$ and $\phi$ in $L'$ is the
unique connection $\nabla '$ in $L'$
such that the following diagram commutes:
\beann
\Gamma (L) &\mapping{\Gamma (\phi)} & \Gamma (L)
\\
\nabla \Bigg\downarrow & & \Bigg\downarrow \nabla '
\\
{\cal F}^1 \otimes_{{\cal F}^0} \Gamma (L)
&\mapping{{\rm id} \otimes \Gamma (\phi)} &
{\cal F}^1 \otimes_{{\cal F}^0} \Gamma (L)
\eeann
\end{definition}

\begin{prop}
If $\nabla '$ is the connection induced by
$\nabla$ and the diffeomorphism $\phi$,
then both connections have the same curvature.
\end{prop}
{\sl (Proof)} \quad
Let $\{ U_j,s_j \}$ and $\{ U_j,s'_j \}$ be
trivializing systems of $L$ and $L'$.
We have
$$
\nabla s_j = 2\pi i \omega^j \otimes s_j
\ , \
\nabla ' s'_j = 2\pi i \omega^{,j} \otimes s'_j
$$
where $\{ \omega^j \}$ and $\{ \omega^{,j} \}$
are the connection forms of $\nabla$ and $\nabla '$
for the open covering $\{ U_j \}$.
According to the above notations,
$\phi \circ s_j = \varphi_j s'_j$,
then:
\beann
({\rm id}\otimes \Gamma (\phi))(\nabla s_j)
&=&
({\rm id}\otimes \Gamma (\phi))(2\pi i \omega^j \otimes s_j) =
2\pi i \omega^j \otimes \varphi_j s'_j
\\
\nabla '(\Gamma (\phi)s_j)
&=&
\nabla '(\varphi_j s_j) =
\d \varphi_j \otimes s'_j + 2\pi i \varphi_j \omega^{,j} \otimes s'_j
\eeann
 But $({\rm id}\otimes \Gamma (\phi)) \circ \nabla =
 \nabla '\circ \Gamma (\phi )$, then
$$
\omega^j = \frac{1}{2\pi i}\frac{\d \varphi_j}{\varphi_j} + \omega^{,j}
$$
hence $\d \omega^j = \d \omega^{,j}$; and the result follows.
\qed

{\bf Comments:}
\begin{enumerate}
\item
Observe that if $\nabla '$ is induced by $\nabla$ and $\phi$,
we have proved that their connection forms are related by
\beq
\omega^j = \frac{1}{2\pi i}\frac{\d \varphi_j}{\varphi_j} + \omega^{,j}
\label{numerado}
\eeq
where $\varphi_j \colon U_j \to \Complex^*$ is defined by
$\phi \circ s_j = \varphi_j s'_j$.
\item
Let $\{ U_j,s_j \}$ be a trivializing system of $L$.
Then $\{ U_j,s'_j=\phi \circ s_j \}$ is a trivializing system of $L'$,
because $\phi$ is a diffeomorphism.
In these trivializing systems the functions
$\varphi_j \colon U_j \to \Complex^*$
are identically equal to one, then
$\omega^j = \omega^{,j}$.
\item
Let $L$ and $L'$ be complex line bundles on $M$
and $\{ U_j,s_j \}$, $\{ U_j,s'_j \}$
trivializing systems.
Suppose
$\varphi_j~\colon U_j~\to~\Complex^*$
is a family of functions related by
$\varphi_j = c_{kj} \varphi_k (c'_{kj})^{-1}$
in $U_{jk}$.
Then there exists a unique diffeomorphism
$\phi \colon L \to L'$
which induces the family $\{ \varphi_j \}$.
In fact, if $l \in L$ and $\pi (l) \in U_j$, put
$$
\phi (l) = \frac{l}{s_j (\pi (l))}\varphi_j (\pi (l))s'_j (\pi (l))
$$
Observe that $\phi (l)$ is well defined:
since if $\pi (l) \in U_{jk}$, then the expression of $\phi (l)$
does not depend on the chosen index and
the calculus can be made using the trivializing systems.
Hence we have proved the following:

\begin{prop}
The necessary and sufficient condition for
a connection $\nabla '$ in $L'$ to be induced by
a connection $\nabla$ in $L$ and a diffeomorphism
from $L$ to $L'$, is that there exist trivializations
$\{ U_j,s_j \}$ and $\{ U_j,s'_j \}$
on $L$ and $L'$ and a family of functions
$\varphi_j \colon U_j \to \Complex^*$
satisfying $\varphi_j = c_{kj} \varphi_k (c'_{kj})^{-1}$
\end{prop}

\item
Now suppose that $L = L'$,
$\phi \colon L \to L$ is a diffeomorphism and
$\{ U_j,s_j \}$ a trivializing system of $L$.
In this case we have that
$\phi \circ s_j = \varphi_j s'_j$,
but $\varphi_j = c_{kj}\varphi_k(c_{kj})^{-1}=\varphi_k$ in $U_{jk}$.
Therefore there exists a global function
$\varphi \colon M \to \Complex^*$
such that $\phi \circ s = \varphi s$, for every $s \in \Gamma (L)$.

If $\nabla$ is a connection in $L$ and
$\nabla '$ is induced by $\nabla$ and $\phi$,
then the connection forms are related by (\ref{numerado}).
Now, repeating the arguments of the comment 3 above, we have that:
\begin{itemize}
\item
Given $\varphi \colon M \to \Complex^*$,
there exists a unique diffeomorphism
$\phi \colon L \to L$
such that $\Gamma (\phi )s = \varphi s$,
for $s \in \Gamma (L)$.
In this case, if $l \in L$,
we have that $\phi (l) = \varphi (\pi (l))l$.
\item
The necessary and sufficient condition for
the connection $\nabla '$ to be induced by
$\nabla$ in $L$ is that there exists one function
$\varphi \colon M \to \Complex^*$
such that, if $\{ U_j,s_j \}$ is a trivializing system of $L$,
their connection forms are related by (\ref{numerado}).
\end{itemize}
\end{enumerate}

Taking into account the above results and comments,
the equivalence we will use is the following:

\begin{definition}
Two complex line bundles with connection
$(L,\nabla )$ and $(L',\nabla ')$ on $M$
are said to be
{\rm equivalent}
iff there exists a diffeomorphism
$\phi \colon L \to L'$
such that $\nabla '$ is the connection induced
by $\nabla$ and $\phi$.
\label{lbce}
\end{definition}

{\bf Comment:}
Observe that, if
$(L,\nabla ) \simeq (L',\nabla ')$
by the diffeomorphism $\phi$,
taking trivializing systems of $L$ and $L'$,
$\{ U_j,s_j \}$ and $\{ U_j,s'_j=\phi \circ s_j \}$,
according to comment 2 above, we have
\beann
\nabla s_j
&=&
2\pi i \omega^j \otimes s_j
\\
\nabla ' (\phi \circ s_j)
&=&
\nabla' \varphi_j s'_j = 2\pi i \omega^j \otimes s'_j
\eeann
That is, the local connection forms associated with
$\nabla$ and $\nabla '$ are the same
with respect to the given trivializing systems.

\subsubsection{The case of hermitian line bundles}

In the same way as in definition \ref{lbce}, we state:

\begin{definition}
Two complex line bundles with hermitian connection
$(L,\h ,\nabla )$ and
$(L',\h ',\nabla ')$ on $M$
are said to be {\rm equivalent}
iff there exists a diffeomorphism
$\phi \colon L \to L'$ such that
$\h = \phi^*\h '$ and $\nabla '$
is the connection induced by $\nabla$ and $\phi$.
\label{clbhc}
\end{definition}

We are interested in studying the set of
equivalence classes of complex line bundles with hermitian connection.
Previously we need the following:

\begin{lem}
\begin{description}
\item[{\rm (a)}]
Let $\h_1$, $\h_2$ be
two hermitian metrics in $\Complex$.
There exists a $\Complex$-linear isomorphism
$\phi \colon \Complex \to \Complex$
such that, if $z_1,z_2 \in \Complex$,
we have that $\h_1(z_1,z_2) = \h_2(\phi (z_1),\phi (z_2))$;
that is $\h_1 = \phi^*\h_2$.
Moreover $\phi$ is real.
\item[{\rm (b)}]
Let $(L,\pi ,M)$ be a complex line bundle
and $\h_1$, $\h_2$
two hermitian metrics in $L$.
There exists a diffeomorphism
$\phi \colon L \to L$
such that, if $l_1,l_2 \in L$,
then $\h_1(l_1,l_2) = \h_2(\phi (l_1),\phi (l_2))$;
that is $\h_1 = \phi^*\h_2$.
In addition, $\phi$ is generated by a real function
$\varphi \colon M \to \Real^*$.
\item[{\rm (c)}]
Let $L$ and $L'$ be complex line bundles over $M$
with the same Chern class and $\h$ and $\h '$
hermitian metrics on $L$ and $L'$.
There exists a diffeomorphism $\phi \colon L \to L'$
such that, if $l_1,l_2 \in L$,
then $\h_1(l_1,l_2) = \h_2(\phi (l_1),\phi (l_2))$;
that is $\h_1 = \phi^*\h_2$.
\end{description}
\end{lem}
{\sl (Proof)} \quad
\begin{description}
\item[{\rm (a)}]
Let $z \in \Complex$ with $\h_1(z,z)=1$
and let $\lambda = \h_2(z,z)$.
Consider the isomorphism $\phi \colon \Complex \to \Complex$
given by $\phi (w) = \lambda^{1/2}w$.
Then $\phi$ satisfies the required conditions.
\item[{\rm (b)}]
If $m \in M$, then $\h_{1m}$ and $\h_{2m}$
are hermitian metrics on $L_m$.
Consider the function $\varphi \colon M \to \Real^+$ such that
$\h_{1m}(l_1,l_2) = \h_{2m}(\varphi (m)l_1,\varphi (m)l_2)$:
for every $l_1$ and $l_2$ in $L_m$. The function $\varphi$ exists
by the item (a), and it is differentiable because $h_1$ and $h_2$ are
differentiable.
The diffeomorphism $\phi \colon L \to L$ given by
$\phi (l) = \varphi (\pi (l))l$ satisfies the required condition.
\item[{\rm (c)}]
Let $\psi \colon L \to L'$ be a diffeomorphism.
Its existence is assured because $L$ and $L'$
have the same Chern class.
Let $\hat \h$ be the metric on $L'$ induced by $\psi$ and $\h$,
that is, $\h = \psi^* \hat \h$.
Let $\eta \colon L' \to L'$ be the diffeomorphism
mapping $\hat \h$ into $\h '$, that is,
$\hat \h = \eta ^* \h '$.
Then $\phi = \eta \circ \psi \colon L \to L'$
is a diffeomorphism and $\phi^* \h ' = \h$.
\end{description}
\qed

The equivalence of complex line bundles with
hermitian connection is given by the following:

\begin{prop}
Let $(L,\h )$ and $(L',\h ')$ be complex line bundles
with hermitian metric and having the same Chern class.
Let $\phi \colon L \to L'$ be the diffeomorphism
satisfying $\h = \phi^*\h '$.
If $\nabla$ is an hermitian connection in $L$ with
respect to $\h$, then $\nabla '$,
the connection induced by $\phi$ on $L'$,
is hermitian with respect to $\h '$.
\label{llprim}
\end{prop}
{\sl (Proof)} \quad
We have that
$\nabla ' \circ \Gamma (\phi ) =
({\rm id} \otimes \Gamma (\phi )) \circ \nabla$,
that is, if $X \in {\cal X}(M)$ and
$\sigma \in \Gamma (L)$, then
$\nabla '_X (\phi \circ \sigma ) = (\phi \circ \nabla_X)\sigma$.
Now, take $s,s' \in \Gamma (L')$,
then there exist $\sigma ,\sigma ' \in \Gamma (L)$
with $s = \phi \circ \sigma$,
$s' = \phi \circ \sigma '$,
and, if $X \in {\cal X}(M)$, we have
\beann
X(\h '(s,s'))
&=&
X(\h (\phi^{-1} \circ s,\phi^{-1} \circ s')) =
X(\h (\sigma ,\sigma '))=
\h (\nabla_X \sigma ,\sigma ') + \h (\sigma ,\nabla_X \sigma ')
\\
&=&
\h '(\phi \circ \nabla_X \sigma ,\phi \circ \sigma ') +
\h '(\phi \circ \sigma ,\phi \circ \nabla_X \sigma ')=
\h '(\nabla '_X s,s') + \h '(s,\nabla '_X s')
\eeann
\qed

{\bf Comments:}
\begin{enumerate}
\item
The last proposition proves that it is irrelevant
to take different hermitian metrics on a complex line bundle
or different complex line bundles with the same Chern class.
That is, if we fix $(L,\h )$, then in the quotient set
defined by the equivalence relation introduced in definition
\ref{clbhc},
every class has a representative of the form
$(L,\h ,\nabla )$.
\item
According to this, from now on we will take
a fixed complex line bundle $L$ with one fixed
hermitian metric $\h$ on $L$.
Then, all the results obtained in the following will refer to
complex line bundles having the same Chern class as~$L$.
\end{enumerate}

Now, the result we are interested in is:

\begin{prop}
Let $\nabla_1$ and $\nabla_2$ be equivalent connections on the
complex line bundle $(L,\pi ,M)$ with hermitian metric $\h$.
Suppose that $\nabla _1$ is hermitian. Then $\nabla_2$ is
hermitian if, and only if, the function $\varphi \colon M \to
\Complex^*$ relating the connection forms of $\nabla_1$ and
$\nabla_2$ has constant module. \label{ll}
\end{prop}
{\sl (Proof)} \quad
Let $\{ U_j,s_j \}$ be  a trivializing system
with $\h (s_j,s_j) = 1$.
We have that
\dst \omega^j_1 = \frac{1}{2\pi i}\frac{\d \varphi}{\varphi} +
\omega^j_2\) .
As every hermitian connection has real connection forms,
then \dst\frac{1}{2\pi i}\frac{\d \varphi}{\varphi}\)
must be a real form.
Then the statement is a consequence of the following lemma
whose proof is immediate.
\qed

\begin{lem}
Let $\varphi \colon M \to \Complex^*$ be a differentiable
function. The necessary and sufficient condition for \dst \frac{\d
\varphi}{\varphi}\) to be imaginary is that $\varphi$ has constant
module.
\end{lem}

Consider now the group
$B = \{ \varphi \colon M \to \Complex^* ; |\varphi | = const. \}$
with the product operation, and the morphism
$$
\begin{array}{cccc}
\eta \colon & B & \rightarrow & Z^1(M)
\\
& \varphi & \mapsto & \frac{1}{2\pi i}\frac{\d \varphi}{\varphi}
\end{array}
$$
$\ker \eta$ is made of the constant functions.
Let $C$ be the image of $\eta$.
According to the above discussion,
we have proved that:

\begin{prop}
The group \dst \frac{Z^1(M)}{C}\)
acts freely and transitively on the set
${\rm Con}(L,\h ,\curv )/ \sim$;
where $\sim$ is the equivalence of hermitian connections.
\end{prop}

\subsubsection{Calculation of $Z^1(M)/C$ and $C/B^1(M)$}

\begin{lem}
$B^1(M) \subset C \subset Z^1(M)$.
\end{lem}
{\sl (Proof)} \quad
Consider $f \in {\mit\Omega}^0(M)$. Then $\d f \in B^1(M)$.
Let $\varphi \colon M \to \Complex^*$ be the map
defined by $\varphi := e^{2\pi if}$.
So we have that
\dst\frac{1}{2\pi i} \frac{\d \varphi}{\varphi} = \d f\) ;
therefore the result follows.
(Observe that, in general,
$B^1(M) =C$ does not hold, as we will see later).
\qed

Therefore we have the following exact sequence of groups:
$$
0 \to \frac{C}{B^1(M)} \to \frac{Z^1(M)}{B^1(M)}
\to \frac{Z^1(M)}{C} \to 0
$$
and from here
$$
\frac{Z^1(M)}{C} \simeq \frac{Z^1(M)/B^1(M)}{C/B^1(M)}
= \frac{H^1(M,\Real )}{C /B^1(M)}
$$

Now, we are going to study the group $\frac{C}{B^1(M)}$
in order to characterize the last quotient.
Consider the natural injection $\varepsilon \colon \Zahl \to \Real$,
which is
understood as a morphism between constant sheaves over $M$. We have:

\begin{prop}
The morphism
$\varepsilon^1 \colon \check H^1(M,\Zahl ) \to \check H^1(M,\Real )$
(induced by $\varepsilon$) is injective.
\end{prop}
{\sl (Proof)} \quad
Consider the exact sequence of constant sheaves over $M$:
$$
0 \to \Zahl \map{\varepsilon} \Real \map{e} S^1 \to 0
$$
with $e(\alpha )= e^{2\pi i \alpha}$,
whose associated exact cohomology sequence is
$$
0 \to \check H^0(M,\Zahl ) \map{\varepsilon^0}
\check H^0(M,\Real ) \map{e^0}
\check H^0(M,S^1) \map{\partial^0}
\check H^1(M,\Zahl ) \map{\varepsilon^1}
\check H^1(M,\Real )
$$
that is
$$
0 \to \Zahl \map{\varepsilon} \Real \map{e} S^1
\map{\partial^0}
\check H^1(M,\Zahl ) \map{\varepsilon^1}
\check H^1(M,\Real ) \ldots
$$
hence $\ker \partial^0 = S^1$, and then
${\rm Im} \partial^0 =\ker \varepsilon^1 = 0$.
\qed

On the other hand,
we have a  canonical isomorphism between
$H^1(M,\Real )$ and $\check H^1(M,\Real )$
which is constructed in the following way:
consider $[\eta ] \in H^1(M,\Real )$
and $\eta \in [\eta ]$.
If $\{ U_j \}$ is a contractible covering of $M$, then there exist
$f_j \colon U_j \to \Real$ such that $\d f_j = \eta\mid_{U_j}$.
In $U_{jk}$ we have that $\d (f_j - f_k) = 0$,
then $f_j-f_k \mid_{U_{jk}} \in \Real$, that is, it is constant.
In this way we have the assignment
$[\eta] \mapsto \{ U_{jk} \mapsto f_j - f_k \}$, which is an
isomorphism.
Let $\alpha \colon \check H^1(M,\Real ) \to H^1(M,\Real )$ be
the inverse isomorphism.
Consider now the sequence of maps:
$$
\check H^1(M,\Zahl ) \map{\varepsilon^1}
\check H^1(M,\Real ) \map{\alpha}
H^1(M,\Real )
$$
Let $H^1(M,\Zahl )$ be the image of
$\check H^1(M,\Zahl )$ in $H^1(M,\Real )$
by $\alpha \circ \varepsilon^1$,
this subgroup is characterized in the following way:
$[\eta ] \in H^1(M,\Zahl )$ if, and only if, there exists a contractible

covering
$\{ U_j \}$ of $M$ and functions $f_j \colon U_j \to \Real$ such that
$\d f_j = \eta \mid_{U_j}$ and $f_j-f_k \mid_{U_{jk}} \in \Zahl$;
for every representative $\eta \in [\eta ]$.
Taking this into account we can characterize the group
\dst\frac{C}{B^1(M)}\) as follows:

\begin{prop}
\dst\frac{C}{B^1(M)}\) is isomorphic to $H^1(M,\Zahl )$.
\end{prop}
{\sl (Proof)} \quad
According to the definition of $C$, we have a natural injection
$$
\frac{C}{B^1(M)} \to \frac{Z^1(M)}{B^1(M)} = H^1(M,\Real )
$$
Consider $\varphi \colon M \to \Complex^*$ with $| \varphi | = r$.
Its associate element in $C$ is \dst\frac{1}{2\pi i}\frac{\d
\varphi}{\varphi}\) .
We denote by \dst\left[\frac{1}{2\pi i}\frac{\d
\varphi}{\varphi}\right]\)
its class in $\frac{C}{B^1(M)}$
and maintain this notation when it is considered in $H^1(M,\Real )$.
We are going to see that
\dst\left[\frac{1}{2\pi i}\frac{\d \varphi}{\varphi}\right] \in
H^1(M,\Zahl )\) .
The image of $\varphi$ is in the circumference
with radius equal to $r$ in the complex plane.
Consider the following open sets in this image
$$
U_1 = \{z \in \Complex \mid |z| = r , z \not= r \}
\quad , \quad
U_2 = \{z \in \Complex \mid |z| = r , z \not= -r \}
$$
Consider $U_j' =\varphi^{-1}(U_j)$
and
$\varphi_j = \varphi \mid_{U_j'}$ ($j=1,2$).
Taking determinations of the logarithm in $U_1$ and $U_2$,
we can construct differential functions $f_j \colon U_j \to \Real$
such that $\varphi _j = re^{2\pi i f_j}$.
We have that \dst\d f_j =\frac{1}{2\pi i}\frac{\d
\varphi_j}{\varphi_j}\) ,
and if $m \in U_1' \cap U_2'$, then
$re^{2\pi if_1(m)} = re^{2\pi if_2(m)}$, hence $f_1-f_2 \in \Zahl$.
Therefore \dst[\frac{1}{2\pi i}\frac{\d \varphi}{\varphi}] \in
H^1(M,\Zahl )\) ,
as we wanted.

On the other hand, if $[\eta ] \in H^1(M,\Zahl )$
and $\{ U_j \}$ is a contractible open covering of $M$,
let $f_j \colon U_j \to \Real$ be maps such that
$\eta \mid_{U_j} = \d f_j$ and $f_j-f_k \mid_{U_{jk}} \in \Zahl$.
Let $\varphi_j \colon U_j \to \Complex^*$
the map defined by $\varphi_j(m) := e^{2\pi if_j(m)}$.
If $m \in U_{jk}$ we have
$$
\frac{\varphi_j(m)}{\varphi_k(m)} = e^{2\pi i(f_j(m)-f_k(m))} = 1
$$
therefore $\varphi_j\mid_{U_{jk}} = \varphi_k\mid_{U_{jk}}$,
hence there exists $\varphi \colon M \to \Complex^*$
such that $\varphi \mid_{U_j} = \varphi_j$, for every $j$.
Moreover $|\varphi | = 1$, then
\dst\frac{1}{2\pi i}\frac{\d \varphi}{\varphi} \in C\) and
\dst\left[\frac{1}{2\pi i}\frac{\d \varphi}{\varphi}\right] \in
\frac{C}{B^1(M)}\) .
The image of
\dst\left[\frac{1}{2\pi i}\frac{\d \varphi}{\varphi}\right]\) in
$H^1(M,\Real )$
is $[\eta ]$, since
$$
\biggl( \eta - \frac{1}{2\pi i}\frac{\d \varphi}{\varphi} \biggr)
\mid_{U_j} = \d f_j - \d f_j = 0
$$
and the assertion holds.
\qed

In addition, taking into account the canonical isomorphisms
$\check H^1(M,\Real ) \simeq H^1(M,\Real )$,
$\check H^1(M,\Zahl ) \simeq H^1(M,\Zahl )$,
we have proved that:

\begin{teor}
The group
\dst\frac{\check H^1(M,\Real )}{\check H^1(M,\Zahl )}\)
acts freely and transitively on the set of hermitian connections in
$(L,\pi ,M)$ related to a given metric $\h$
and with curvature $\Omega$, module the equivalence of connections.
\end{teor}

And taking into account the influence of the hermitian metric
and the diffeomorphisms of $L$,
we have proved also the following:

\begin{teor}
Let $M$ be a differentiable manifold and
$\curv$ a real closed $2$-form in $M$. The group
\dst\frac{\check H^1(M,\Real )}{\check H^1(M,\Zahl )}\)
acts freely and transitively on the set of equivalence classes
of complex line bundles with hermitian connection
$(L,\nabla )$ in $M$
which have the same Chern class ${\rm c}(L)$
(that is, which are diffeomorphic to $L$)
and the same curvature $\curv$.
\end{teor}

\begin{corol}
If $M$ is simply connected,
then there exists only one equivalence class
of complex line bundles with hermitian connection $(L,\nabla )$
with the same Chern class and the same curvature.
\end{corol}

\subsubsection{Influence of the ker of the morphism
$\varepsilon^2 \colon \check H^2(M,\Zahl ) \to \check H^2(M,\Real )$}

Given a manifold $M$ and a real closed $2$-form $\curv$ on it,
if $L$ is a hermitian complex line bundle, we have studied
the equivalence classes of hermitian connections in $L$
with curvature $\curv$. In this case $\curv$
is a representative of the image of the Chern class
of $L$ by the morphism $\varepsilon^2$.
As ${\bf L}(M) \simeq \check H^2(M,\Zahl )$,
and $\varepsilon^2$ is not injective,
then the following problem arises:
there are non-diffeomorphic line bundles whose Chern classes
in $\check H^2(M,\Zahl )$ have the same image under $\varepsilon^2$.
Now we are going to study the effects of this problem
on the classification of complex line bundles
with connections $(L,\nabla )$ on $M$.

Consider the morphism
$\varepsilon^2 \colon \check H^2(M,\Zahl ) \to \check H^2(M,\Real )$
induced by $\varepsilon \colon \Zahl \to \Real$;
and denote $\bar G = \ker \varepsilon^2$.
Given $[\curv ] \in H^2(M,\Real ) \simeq \check H^2(M,\Real )$,
let $V$ be the antiimage of $[\curv ]$ by $\varepsilon^2$.
The group $\bar G$ acts freely an transitively in $V$ in the natural
way.
If $\sigma \in V$, let $P_\sigma$ be
the set of equivalence classes of complex line bundles
over $M$ with hermitian connection $(L,\nabla )$
whose Chern class is $\sigma$ and with curvature $\curv$.
We have that $\varepsilon^2(\sigma ) = [\curv ]$.
If $\sigma , \sigma' \in V$ with $\sigma \not= \sigma'$,
then $P_\sigma \cap P_{\sigma'} = \buit$.
Then we have a natural projection
$\pi \colon P = \bigcup_{\sigma \in V}P_\sigma \to V$.
Moreover, as we have seen above, the action of the group
\dst G =\frac{\check H^1(M,\Real )}{\check H^1(M,\Zahl )}\)
in $P$ preserves its fibers and is
free and transitive on these fibers.
Hence $(P,\pi ,V)$ is a $G$-principal bundle of sets.

The already known exact sequence of constant sheaves over $M$
$$
0 \to \Zahl \map{\varepsilon} \Real \map{e} S^1 \to 0
$$
gives the cohomology sequence
$$
0 \to \Zahl \map{\varepsilon} \Real \map{e^0} S^1
\map{\partial^0}
\check H^1(M,\Zahl ) \map{\varepsilon^1}
\check H^1(M,\Real ) \map{e^1}
\check H^1(M,S^1) \map{\partial_1} \ldots
$$
and taking into account that
${\rm Im}\partial^0 = \ker \varepsilon^1 = 0$
we have
$$
0 \to \check H^1(M,\Zahl ) \map{\varepsilon^1}
\check H^1(M,\Real ) \map{e^1}
\check H^1(M,S^1) \map{\partial^1}
\ker \varepsilon^2 \to 0
$$
therefore, as $\varepsilon^1$ is injective, we have
$$
0 \to \frac{\check H^1(M,\Real )}{\check H^1(M,\Zahl )}
\map{\bar e^1} \check H^1(M,S^1) \map{\partial^1} \bar G \to 0
$$
(where $\bar e^1$ is the natural morphism), that is
$$
0 \to G \map{\bar e^1} \check H^1(M,S^1) \map{\partial^1} \bar G \to 0
$$
Hence $\partial^1 \colon \check H^1(M,S^1) \to \bar G$
is a $G$-principal bundle of sets.

Next, we are going to construct, for each $\sigma \in V$, a bijection
$\phi$
between $P$ and $\check H^1(M,S^1)$.
Let $\eta \colon V \to P$, $\mu \colon \bar G \to \check H^1(M,S^1)$
be sections of $\pi$ and $\partial^1$ respectively.
If $L \in P$, we have that $\pi (L) \in V$.
There exists a unique $\bar g \in \bar G$
such that $\pi (L) = \sigma \bar g$.
On the other hand, there is a unique $g \in G$
such that $L = \eta (\pi (L))g$,
then we define $\phi (L)$ as
$\phi (L) := \mu(\bar g)g$. So,
$\phi$ is a map from $P$ to $\check H^1(M,S^1)$
such that the following diagram commutes:
$$
\begin{array}{ccccc}
& P & \mapping{\phi} & \check H^1(M,S^1) &
\\
\pi & \Bigg\downarrow & & \Bigg\downarrow  \partial^1 &
\\
& V & \mapping{\bar \phi} & \bar G &
\end{array}
$$
where, if $v \in V$,
$\bar \phi (v) = \bar h \in \bar G$
is the unique element such that
$v = \sigma \bar h$.
Observe that $\bar \phi$ is a bijection and $\phi$ is a bijection
on each fiber, hence $\phi$ is also a bijection.
In addition, $\phi$ is covariant with respect to
the actions of $G$ and $\bar G$,
that is, if $L \in P$, we have that
$\phi (Lg) = \phi (L)g$ and
$\bar \phi (v \bar g) = \bar \phi (v) \bar g$,
for $g \in G$ and $\bar g \in \bar G$.

Taking into account this bijection, we have proved the following:

\begin{teor}
Let $M$ be a differentiable manifold
and $\curv$ an integer real closed $2$-form.
The action of the group $\check H^1(M,S^1)$
on the set of equivalence classes of complex line bundles
with hermitian connection $(L,\nabla )$
with curvature $\curv$, is free and transitive.
\end{teor}

\begin{corol}
If $M$ is simply connected,
then this set of equivalence classes has a unique element
(since $\check H^1(M,S^1) = 0$,
because
$\check H^1(M,S^1) = {\rm Hom} (\pi^1(M),S^1) = 0$).
\end{corol}

\section{Prequantization}
\protect\label{prequa}

Now, we are ready to start the geometric quantization programme.

Our first goal in the geometric quantization programme
is to construct the intrinsic Hilbert space of the system.
With this aim, we follow a systematic procedure,
starting from the easiest possible model and modifying it in such a way
that
the situation is adapted to the rules of definition
\ref{fquan} (as it is done, for instance,
in \cite{Ki-76}, \cite{SW-76} and \cite{Wo-80}).

\subsection{First attempts to define quantum states and operators}

Let $(M,\Omega )$
be a $2n$-dimensional symplectic manifold.
The easiest way of constructing
a Hilbert space associated with it is to consider the algebra
of complex smooth functions with compact support in $M$
and the inner product defined on it by
\beq
\langle \varphi_1 |\varphi_2\rangle  := \int_M \varphi_1 \bar \varphi_2
\LF
\label{inprod}
\eeq
for a pair $\varphi_1, \varphi_2$ of such a functions,
where $\LF$ is the {\sl Liouville's volume form}
\dst \LF := (-1)^{\frac{1}{2}n(n-1)}\frac{1}{n!}\Omega^n\) .
With respect to this product, this algebra is a pre-Hilbert space.
Denote by $\C (M)$ its completion
\footnote{
Observe that $\C (M)$ coincides with the set of
square integrable smooth complex-valued functions
${\cal L}^2(M) \cap \Cinfty (M)$.}.

Our first attempt is to take $\C (M)$
as the intrinsic Hilbert space of the system.
Next, we want to define a set of self-adjoint operators ${\cal O}(\C
(M))$
such that
\begin{description}
\item[{\it (I)}] \
There is a one to one correspondence between the set of classical
observables
${\mit\Omega}^0(M)$ and ${\cal O}(\C (M))$.
\item[{\it (II)}]
The map $f \mapsto O_f$ satisfies conditions (i-v) of definition
\ref{fquan}.
\end{description}

In order to achieve (I), the simplest way is to construct ${\cal O}(\C
(M))$
from the set of {\it Hamiltonian vector fields} in $M$, ${\cal X}_H(M)$.

Then, for every $X_f \in {\cal X}_H(M)$, we construct an (unique)
operator
$\Op_f \in {\cal O}(\C (M))$ which is defined as follows:
$$
\Op_f := -i\hbar X_f
$$
where $X_f$ acts linearly on $\C (M)$ as a derivation of (real)
functions
 (i.e., taking into account that
 $\Cinfty (M) = {\mit\Omega}^0(M) \otimes\Complex)$.
Then, if $\psi \in \C (M)$, we have
$$
\Op_f\sta := -i\hbar X_f(\psi)
$$

Nevertheless, the map
$$
\begin{array}{cccccc}
O:&{\mit\Omega}^0 (M)&\to &{\cal X}_H(M)&\to &{\cal O}(\C (M))
\\
&f& \mapsto &X_f& \mapsto &\Op_f
\end{array}
$$
is not one to one since,
if $f$ is a constant function
then $\d f = 0$ and $X_f=0$,
hence functions differing in a constant
have the same associated operator.
In addition, even though properties (i) and (ii) of definition
\ref{fquan}
are satisfied, (iii) fails to be true.
In order to solve this,
the simplest correction consists in adding
an extra term in the above definition of
$\Op_f$, writing:
$$
\Op_f := -i\hbar X_f + f
$$
and so, for every $\psi \in \C (M)$,
$$
\Op_f\sta := -i\hbar X_f(\psi) +f\psi
$$

Now the properties (i-iii) hold but not (iv) because $$
[\Op_f,\Op_g]\sta = \hbar^2 [X_f,X_g](\psi) - 2i\hbar \{ g,f\}
(\psi )= i\hbar (-i\hbar X_{\{ f,g\}} + 2\{ f,g\} )(\psi) \not=
i\hbar \Op_{\{ f,g\}}\sta $$ Therefore, a new correction is
needed. Let $\theta \in {\mit\Omega}^1(U)$, ($U\subset M)$ be a
local {\it symplectic potential}, i.e., $\Omega = \d \theta$
(locally). Then we define \beq \Op_f := -i\hbar (X_f +
\frac{i}{\hbar}\langle X_f |\theta \rangle )+f \label{oper} \eeq
and so, for every $\psi \in \C (M)$, \beq \Op_f\sta := -i\hbar
(X_f + \frac{i}{\hbar}\langle X_f |\theta \rangle ) (\psi )+f\psi
\label{opersta} \eeq

Now, (i-iv) hold (see the proof of theorem \ref{prop})
and, at the moment, this is the final form of the quantum operator
associated with the classical observable $f$.
Observe that this is a local construction
and that the expression of this quantum operator
depends on the choice of the local symplectic potential.

\subsection{Space of states: definitions and justification}

The necessity of introducing a local symplectic potential
in order to have a correct definition of the
quantum operators leads to a new difficulty.
In fact, as we have said,
the construction of the operators is local.
This means that if
$U$, $U'$ are local charts of $M$
and $\theta$, $\theta '$
are the corresponding local symplectic potentials,
then in the intersection
they differ (locally) by an exact one-form,
that is,
$\theta' = \theta + \d \alpha$ ,
for some
$\alpha \in {\mit\Omega}^0(U \cap U')$
\footnote
{Physically, to change the local chart means that we are changing
the local reference system of the observer,
and we are allowing {\it gauge transformations}.
}.
Observe that $\Op_f\sta \not= \Op '_f\sta$.
Then, if we want the action of the quantum operator
on the vector states to be independent on the choice of $\theta$
(that is, to be independent on the local chart),
we have to impose that, if $\psi \in \C (M)$, then
$\psi$ and
$\psi '= e^{\frac{i \alpha}{\hbar}}\psi$
must represent the same vector state
of the intrinsic Hilbert space that we want to construct, because
$$
\Op '_f |\psi '\rangle =
(-i\hbar (X_f + \frac{i}{\hbar}\langle X_f,\theta '\rangle )
+ f)e^{\frac{i \alpha}{\hbar}}\psi =
e^{\frac{i \alpha}{\hbar}}
(-i\hbar (X_f + \frac{i}{\hbar}\langle X_f,\theta \rangle ) + f)\psi =
e^{\frac{i \alpha}{\hbar}} \Op_f\sta
$$
and then the relation between
$\Op '_f |\psi '\rangle$ and $\Op_f\sta$
is the same as between $\psi '$ and $\psi$.

The geometrical meaning of this property
of invariance is that the vector states $\sta$
of the intrinsic Hilbert space
cannot be just functions on $M$,
but, according to the results in section 3.4,
sections on a {\it complex line bundle}, $(L,\pi ,M)$,
with structural group $U(1)$.
It is relevant to mention that every complex line bundle
can be endowed with an hermitian metric $\h$ \cite{GHV-72},
which allows us to define an hermitian inner product
in the complex vector space of smooth sections $\Gamma (L)$.

We can summarize this discussion in the following statement
\cite{SW-76}:

\begin{require}
Let $(M,\Omega )$ be a symplectic manifold
(which represents, totally or partially,
the phase space of a physical system).
In the geometric quantization programme
the space of quantum states $\H_Q$ is constructed starting from the set
of
(smooth) sections, $\Gamma (L)$,
of a {\rm complex line bundle} over $M$, $(L,\pi ,M)$,
with $U(1)$ as structural group.

Then the complex line bundle
$(L,\pi ,M)$ is endowed with a smooth hermitian metric,
$\h \colon \Gamma (L) \times \Gamma (L) \to \Complex$.
The {\rm inner product} of sections
(with compact support) is defined by
\beq
\langle \psi_1 |\psi_2\rangle  :=
\biggl(\frac{1}{2\pi \hbar}\biggr)^n
\int_M \h (\psi_1,\psi_2) \LF
\label{inprod1}
\eeq
\label{pclb}
\end{require}

As a consequence,
a more subtle study concerning the
geometrical structures involved in the
quantization procedure is required.
In fact, if the quantum states are constructed from
sections of a line bundle $(L,\pi ,M)$,
and since the quantum operators
are defined from Hamiltonian vector fields in $M$,
it is necessary to clarify how
these vector fields ``act'' on sections of~$L$.

It is obvious that the natural way
is to introduce a connection $\nabla$
in $(L,\pi ,M)$
and use it to associate a
linear differential operator to each
vector field on $M$, acting on sections of $L$.
This operator is just the
{\it covariant derivative}
$\nabla_X$.
But, which kind of connections are suitable?.
Taking into account the expression
(\ref{opersta})
we can conclude that an immediate
solution to our problem consists in taking
the connection in such a way that,
if $\psi \in \Gamma (L)$,
\beq
\nabla_{X_f}\psi :=
(X_f + \frac{i}{\hbar}\langle X_f,\theta \rangle )\psi
\label{con}
\eeq
and so
$$
\Op_f\sta :=
(-i\hbar \nabla_{X_f} + f)\psi
$$
(observe that the product $f\psi$ is well defined).

Notice that, taking into account the expression (\ref{expcon}),
the equation (\ref{con}), as a condition on $\nabla$,
is equivalent to demand (locally) that
\dst\frac{\theta}{2\pi \hbar}\)
is the {\it connection 1-form} $\omega$ of $\nabla$.
But, since locally $\Omega=\d \theta$,
this is equivalent to demanding that
\beq
\frac{\Omega}{2\pi\hbar} = {\rm curv} \nabla \equiv \curv
\label{curv}
\eeq
We will use this condition later, in order to prove that the assignment
$f \to \Op_f$ is a morphism of Lie algebras (see section 4.3).

Hence, we can establish
\cite{SW-76}:

\begin{require}
In the geometric quantization programme,
the complex line bundle
$(L,\pi ,M)$
must be endowed with a
{\rm connection} $\nabla$
such that condition
(\ref{curv})
holds.

In addition, the hermitian metric
(introduced in requirement \ref{pclb})
and the connection have to be {\rm compatible}
in the following sense: $\h$ is $\nabla$-invariant, that is,
if $X \in {\cal X}(M)$ and $\psi_1,\psi_2 \in \Gamma (L)$, then
$$
X(\h (\psi_1,\psi_2)) =
\h (\nabla_X\psi_1,\psi_2) +
\h (\psi_1,\nabla_X\psi_2)
$$
(i.e., the connection is hermitian in relation to this metric).
\label{connec}
\end{require}

We can summarize this discussion in the following:

\begin{definition}
Let $(M,\Omega )$ be a symplectic manifold
(which represents, totally or partially,
the phase space of a physical system).
A {\rm Prequantization} of this system is a complex line bundle
$(L,\pi ,M)$ endowed with an hermitian metric $\h$
and a compatible connection $\nabla$ such that
\dst\frac{\Omega}{2\pi\hbar} = {\rm curv} \nabla\) .
If some prequantization exists for $(M,\Omega )$
we say that the system is {\rm prequantizable}
and the quintuple $(L,\pi ,M;\h ,\nabla )$
is said to be a {\rm prequantum line bundle} of the system.
\label{prequan}
\end{definition}

At this point, the question is
to study whether a classical system $(M,\Omega )$ is prequantizable or
not.
The answer to this question is given
by the following theorems
\cite{AM-78}, \cite{Ko-70}, \cite{SW-76}, \cite{Sn-80}:

\begin{teor}
{\rm (Weil's theorem) \cite{We-58}}.
Let $(M,\Omega )$ be a symplectic manifold.
Then, the system is prequantizable
if, and only if, \dst[\frac{\Omega}{2\pi\hbar}] \in H^2(M,\Zahl )\) ,
that is, \dst[\frac{\Omega}{2\pi\hbar}]\)
is an {\sl integral cohomology class}
\footnote{
$H^2(M,\Zahl )$ is the image of $\check H^2(M,\Zahl )$
by the following composition
$$
0 \to \check H^2(M,\Zahl ) \to \check H^2(M,\Real ) \map{\simeq}
H^2(M,\Real)
$$
}.

In the particular case of $M$ being simply connected,
then the connection $\nabla$ and the compatible hermitian metric
are unique (except equivalence relations).
\label{weil}
\end{teor}

{\bf Remarks}:
\begin{itemize}
\item
Notice that $\Omega$ must be a real form.
\item
The first part of this theorem is theorem \ref{cce}.
The second part is a consequence of section 3.6.
\item
The condition on the first part means (by duality) that the
integral of \dst\frac{\Omega}{2\pi\hbar}\) over every integer
$2$-cocycle in $M$ is integer. This {\it integrability condition}
is called the {\it Bohr-Sommerfeld quantization condition}.
\end{itemize}

Another complementary result is the following:

\begin{teor}
Let $M$ be a manifold and
$(L,\pi ,M)$ a complex line bundle
endowed with a connection $\nabla$
with local connection $1$-form \dst\frac{\theta}{2\pi\hbar}\) .
Then, there exists a $\nabla$-invariant
hermitian metric in $(L,\pi ,M)$
if, and only if, $\theta -\bar \theta$ is an exact $1$-form.
\label{ihm}
\end{teor}
\proof
See \cite{Ko-70}, p. 110.
\qed

A new problem arises now: the integral (\ref{inprod1})
is not necessarily convergent. This means that
$\Gamma (L)$ fails to be a Hilbert space
because the norm $\| \psi \|$ is not defined for every
$\psi \in \Gamma (L)$ and hence
$\Gamma (L)$ cannot be the intrinsic Hilbert space.
In order to avoid this difficulty, we can take the subset of
$\Gamma (L)$ made of sections with compact support.
This subset with the inner product
(\ref{inprod1}) is a {\it pre-Hilbert space}. Then:

\begin{definition}
The completion
$\H_P$ of the set of sections with compact support in
$\Gamma (L)$ is a Hilbert space which is called the
{\rm prequantum Hilbert space}
\footnote{
Remember that this is also the set of
square integrable smooth sections in $\Gamma (L)$.}.
The projective space ${\bf P}\H_P$
is the {\rm space of prequantum states}.
\label{prehs}
\end{definition}

\subsection{Prequantization Operators}

Now, we can prove that the set of operators $O_f$ given by
$$
O_f := -i\hbar\nabla_{X_f} + f =
-i\hbar (X_f + \frac{i}{\hbar}\langle X, \theta \rangle ) + f
$$
satisfies part (b) of definition \ref{fquan}.

\begin{teor}
The assignment $f \mapsto \Op_f$ is a $\Real$-linear map from
the Poisson algebra of the functions ${\mit\Omega}^0(M)$
into the Lie algebra of the self-adjoint operators of $\H_P$, satisfying

the properties
b(i-iv) of definition \ref{fquan}.
\label{prop}
\end{teor}
{\sl (Proof)} \quad
In order to see that the operator $\Op_f$ is self-adjoint we calculate:
\beann
\langle \Op_f(\psi ) | \psi' \rangle
&=&
\int_M \h (\Op_f(\psi ),\psi') \LF
=
\int_M \h ((-i\hbar\nabla_{X_f}+f)\psi ,\psi ') \LF
\\
&=&
\int_M -i\hbar X_f(\h (\psi ,\psi ')) \LF
+ \int_M i\hbar \h (\psi ,\nabla_{X_f}\psi ') \LF
+ \int_M \h (f\psi ,\psi ') \LF
\\
&=&
\int_M \h (\psi ,-i\hbar\nabla_{X_f}\psi ') \LF
+ \int_M \h (\psi ,f\psi ') \LF
+ \int_M -i\hbar X_f( \h (\psi ,\psi ')) \LF
\\
&=&
\int_M \h (\psi ,(-i\hbar\nabla_{X_f}+f)\psi ') \LF
+ \int_M -i\hbar X_f(\h (\psi ,\psi ')) \LF
\\
&=&
\int_M \h (\psi ,\Op_f(\psi ')) \LF
+ \int_M -i\hbar X_f(\h (\psi ,\psi ')) \LF =
\langle \psi | \Op_f(\psi ') \rangle
+ \int_M -i\hbar X_f(\h (\psi ,\psi ')) \LF
\eeann
and it suffices to show that the last integral vanishes.
But observe that  this is an integral of the form
\dst\int_M X_f(g) \LF\) , where
$g \in \Cinfty(M)$ is a function with compact support.
We have that $\d \LF = 0$ and $\d (g\LF ) = 0$,
since they are maximal degree forms.
Moreover, $\Lie(X_f)\Omega=0$, since $X_f$ is a Hamiltonian vector
field, hence
\beann
(X_fg) \LF
&=&
\Lie(X_f)(g\LF - g\Lie(X_f)\LF =\Lie(X_f)(g\LF )
\\ &=&
\d \ \inn(X_f)(g\LF )+ \inn(X_f)\d (g\LF )
=\d \ \inn(X_f)(g\LF )
\eeann
and the integral of
$\d\inn(X_f)(g\LF )$
vanishes because $g$ is a function with compact support.

It is quite evident that the assignment is $\Real$-linear and that it
satisfies the
three first properties.

For the last property we have:
\beann
[\Op_f,\Op_g](\psi )
&=&
(\Op_f\Op_g - \Op_f\Op_g)(\psi )
\\
&=&
(-i\hbar \nabla_{X_f} + f)(-i\hbar\nabla_{X_g}+g)(\psi )
- (-i\hbar \nabla_{X_g} + g)(-i\hbar\nabla_{X_f}+f)(\psi )
\\ &=&
-\hbar^2 (\nabla_{X_f}\nabla_{X_g} - \nabla_{X_g}\nabla_{X_f})(\psi )
-i\hbar (\nabla_{X_f}(g\psi ) + f\nabla_{X_g}(\psi ) + fg\psi
\\ & &
- \nabla_{X_g}(f\psi ) - g\nabla_{X_f}(\psi ) -gf\psi )
\\ &=&
-\hbar^2 [\nabla_{X_f},\nabla_{X_g}](\psi )
-i\hbar (X_f(g)\psi + g\nabla_{X_f}(\psi ) + f\nabla_{X_g}(\psi )
\\ & &
- X_g(f)\psi - f\nabla_{X_g}(\psi ) - g\nabla_{X_f}(\psi ))
\\ &=&
-\hbar^2 [\nabla_{X_f},\nabla_{X_g}](\psi ) + 2i\hbar \{ f,g \}\psi
\\ &=&
 -\hbar^2 (\nabla_{[X_f,X_g]} +
 \frac{i}{\hbar}\Omega (X_f,X_g))(\psi ) +2i\hbar \{ f,g \}\psi
\\ &=&
(\hbar^2 \nabla_{-[X_f,X_g]} - i\hbar \{ f,g \} + 2i\hbar \{ f,g \}
)(\psi)
\\ &=&
(\hbar^2 \nabla_{X_{\{ f,g\} }} + i\hbar \{ f,g \} )(\psi) =
i\hbar(-i\hbar \nabla_{X_{\{ f,g\} }} + \{ f,g \} )(\psi) =
i\hbar\Op_{\{ f,g \}}(\psi )
\eeann
where we have taken into account the property given by proposition
\ref{clacur}.
\qed

Observe that, for every $f\in {\mit\Omega}^0(M)$,
the operator $\Op_f$ is well defined on the set
of compact supported sections of the complex line bundle
$(L,\pi , M)$.

This is the
{\sl prequantization procedure} of Kostant, Souriau and Segal
\cite{Ko-70}, \cite{Se-60}, \cite{So-69}.
Summarizing, given the symplectic manifold
$(M,\Omega )$, it consists in constructing
a complex line bundle $(L,\pi ,M)$
endowed with an hermitian metric $\h$
and a compatible connection $\nabla$
such that \dst{\rm curv}\nabla = \frac{\Omega}{2\pi \hbar}\) .

\subsection{Examples}

\subsubsection{Prequantization of cotangent bundles}

In the particular case of $M$ being a cotangent bundle $\Tan^*Q$,
we have a natural (global) symplectic potential,
which has as local expression
$\theta = -p_j\d q^j$.
The cohomology class of $\frac{\Omega}{2\pi\hbar}$ is zero and we have
simply
$L \simeq M \times \Complex$.
Then, $\H_P \simeq \C (\Tan^*Q)$
and a natural hermitian metric is defined by
$$
\h ((m,z_1),(m,z_2)) := z_1 \bar z_2
$$
which is compatible with the connection defined by $\theta$.

\subsubsection{Prequantization of the harmonic oscillator}

Next, as a typical example, we consider
the $n$-dimensional harmonic oscillator. In this case we have:
$$
M \equiv \{ (q^j,p_j) \in \Real^{2n} \}
\quad ; \quad
\Omega = \d q^j \wedge \d p_j
\quad ; \quad
H = \frac{1}{2}(p_j^2 + q^{j^2})
$$
The conditions assuring the prequantization of the system hold since
$[\frac{\Omega}{2\pi\hbar}]=0$ (so it is integer), and
$L = M \times \Complex$, since $M$ is contractible.

The usual metric is the product.
If $\psi$ is a section of $L$,
the hermitian connection is given by
$$
\nabla_X\psi = X\psi + 2\pi i \omega (X) \psi =
X\psi + \frac{i}{\hbar} \theta (X) \psi
$$
If we take as a (local) symplectic potential
$\theta = \frac{1}{2}(q^j\d p_j -p_j\d q^j)$,
the quantum operators that we obtain are:
$$
\Op_{q^j} = i\hbar \derpar{}{p_j} + \frac{1}{2} q^j
\quad ; \quad
\Op_{p_j} = -i\hbar \derpar{}{q^j} + \frac{1}{2} p_j
$$
and then, with this choice for the symplectic potential,
$$
\Op_H = -i\hbar (p_j\derpar{}{q^j} - q^j\derpar{}{p_j})
$$
therefore the action of the operator associated with the energy is
$$
\Op_H (\psi ) = i\hbar \{ H,\psi \}
$$
Then the classical and prequantum dynamics coincide
and this shows that prequantization is not
the suitable method for quantizing this classical system,
because the spectrum of the energy operator is continuous
and this is not true for the quantized harmonic oscillator.

\subsubsection{An enlightening case}

Finally, the following case reveals another problem
still remaining at the end of this stage.
Consider a cotangent bundle $\Tan^*Q$,
where $Q$ is a compact manifold we have again:
$$
\Omega = \d q^j \wedge \d p_j
\quad ; \quad
\H_P \simeq \C (\Tan^*Q)
$$
and we can take $L = \Tan^*Q \times \Complex$
because $[\frac{\Omega}{2\pi\hbar}] = 0$. Taking as a symplectic
potential
$\theta = q^j \d p_j$, we obtain the following quantum operators:
$$
\Op_{q^j} = -i\hbar \derpar{}{p_j}
\quad ; \quad
\Op_{p_j} = i\hbar \derpar{}{q^j} + p_j
$$
Now, if we take the closed subset of $\C (\Tan^*Q)$
made of the functions which are constant on the basis $Q$,
that is, the set $\C_Q(\Tan^*Q) \equiv \{ f(p_j) \}$,
we have that, for every $f \in \C_Q(\Tan^*Q)$,
$$
\Op_{q^j}(f) = -i\hbar \derpar{f}{p_j} \in \C_Q(\Tan^*Q)
\quad ; \quad
\Op_{p_j}(f) = i\hbar \derpar{f}{q^j} + p_jf  \in \C_Q(\Tan^*Q)
$$
hence, $\C_Q(\Tan^*Q)$
is an invariant subspace under the action
of this set of operators or, what means the same thing,
$\H_P$ is not irreducible by this representation.

Notice that this conclusion can also been reached
observing that the operators
\dst\derpar{}{q^j}\) and \dst\derpar{}{p_j}-\frac{i}{\hbar}q^j\)
commute with $\Op_{q^j}$ and $\Op_{p_j}$.
Hence, the last ones do not form a
complete set of commuting observables and,
as a consequence, there exists some non-empty closed subspace
of $\H_P$ different from $\H_P$ which is invariant with respect to their

action.
So that, the condition (v) of definition \ref{fquan} does not hold for
these systems.

\section{Polarizations}
\protect \label{gqpol}

$\H_P$ is not the suitable choice as the intrinsic Hilbert space
of the system (or, what is the same thing, ${\bf P}\H_P$ is not
the true space of quantum states). In fact, as we have seen in the
last example, in general, in the set ${\cal O}(\H_P)$ (whose
elements are the operators $\Op_f$ defined in the equation
(\ref{oper})) we can find that, from a complete set of classical
observables we get a set of quantum observables which does not
satisfy condition (v) of definition \ref{fquan} (it is not
complete). The origin of this problem is the following: if $M$ is
the phase space of the classical system and $\dim M =2n$, then the
prequantum states in ${\bf P}\H_P$ depend on $2n$ variables, but
according to Quantum Mechanics, the true quantum states depend
just on $n$ variables (the dimension of the configuration space).

In order to solve this problem the idea is to ``restrict'' the Hilbert
space $\H_P$
and the set of quantum operators. Then, a new geometric structure is
defined in
$(M,\Omega )$: {\it polarizations}
\footnote{
Sometimes, this structure is called {\it Planck's foliation}
\cite{AM-78}.
}.

We devote this section to justify, define and develop the main features
related to this concept, which are relevant for geometric quantization.
A more extensive study on (real and complex) polarizations can be found
in \cite{Wo-80}.
Other interesting references are \cite{Sn-80} and \cite{Tu-85}.

\subsection{Previous justifications}

In the case $M=\Tan^*\Real^n$,
$(q^i,p_i)$ is a global set of canonical coordinates.
According to Quantum Mechanics, the wave functions
depend only on $(q^i)$ (or only on $(p_i)$).
Then we must select $n$ coordinates and remove them
in order to obtain the space of wave functions.
Next, we are going to generalize this idea.

The first step of the idea lies in selecting
$n$ directions in $M$ by means of the choice of a
$n$-dimensional {\it distribution} $\P$ in $\Tan M$.
Then we will require that the sections of
the prequantum line bundle $(L,\pi ,M)$ representing the quantum states
are invariant by the transformations
induced in $M$ by this distribution, that is,
\beq
\nabla_\P \psi = 0
\label{inv}
\eeq

Now, let $U \subset M$ be an open set
in which $L$ trivializes and let
$s \colon U \to L$ be a trivializing section.
A section $\psi = f s$ satisfying that
$\nabla_\P \psi = 0$
have to satisfy that
$(\P (f) + \frac{i}{\hbar}\langle \P,\theta \rangle )f)s = 0$.
If we take the distribution to be
{\it adapted to the connection $\nabla$},
that is, such that there exists
a local symplectic potential $\theta$ satisfying that
$\langle \P ,\theta \rangle = 0$,
then the above equations reduce to be
$$
\P (f) = 0
$$
This is a system of
$n$ independent linear partial differential equations and,
in order to assure that it is integrable,
it suffices to demand that the distribution $\P$
is {\it involutive}.

So, $f$ is constant along the fibers of $\P$
and therefore $\psi$ is represented as a constant function
along the fibers of the distribution.
In this way, the states represented by the sections $\psi$
which are solutions of the equations
(\ref{inv})
are represented by functions which
depend just on $n$ variables, in the following sense:
being $\P$ involutive, by the Fr\"obenius theorem,
there exist local coordinate systems
$\{ x_i,y_i \}_{i=1,\ldots ,n}$
such that $\P$ is spanned by
\dst\left\{ \derpar{}{x_i} \right\}\) ; then
$\{ x_i \}$ are local coordinates of the integral manifolds of $\P$,
which are defined by $y_i=ctn.$
Therefore $\P (f) = 0$ is locally equivalent to
 \dst\derpar{f}{x_i} = 0\) , so $f$ depends only
 on the variables $\{ y_i\}$.

But now, an additional consistency condition is required:
equation (\ref{inv}) implies that,
if \dst\frac{\Omega}{2\pi\hbar}\)
is the curvature form of the connection, we have
$$
0 = [\nabla_\P ,\nabla_\P ]\psi =
(\nabla_{[\P ,\P ]} + \frac{i}{\hbar}\Omega (\P ,\P ))\psi
$$
then, if the distribution is involutive,
$\nabla_{[\P ,\P ]}\psi = 0$,
and therefore it must be
$\Omega (\P ,\P )\psi = 0$.
This last condition is assured if we impose that
the distribution is {\it isotropic}
but, since it is $n$-dimensional,
it would be maximal isotropic, that is {\it Lagrangian}.
Thus, at the moment, what we need is an
{\it involutive Lagrangian distribution}.

Suppose $\P$ is spanned by a set of
{\it global Hamiltonian vector fields\/}
$\{ Y_{f_j} \}_{j=1,\ldots ,n}$.
The isotropy condition
$\Omega (Y_{f_l},Y_{f_j}) = 0$
can be written in an equivalent way as
\beq
\{ f_l,f_j \} = 0
\label{conmut}
\eeq
Conversely, if $\{ f_j \}$ is a set of $n$
independent global functions such that equation
(\ref{conmut}) holds, then their Hamiltonian vector fields
span an involutive Lagrangian distribution on $M$.
Nevertheless, for many physical systems, to choose a set of
$n$ independent global functions satisfying the equation
(\ref{conmut}) is not always possible
and, for this reason, our construction must be more general.

Finally, another consideration is needed:
instead of working with real distributions
we will consider
{\it complex distributions},
that is, locally spanned by complex vector fields.
In other words, we are going to work in the
{\it complexified tangent bundle\/}
$\Tan M^{\Complex}$
instead of $\Tan M$
\footnote
{Keep in mind that, for every $m \in M$,
$\Tan_m M^{\Complex} \simeq \Tan_m M \otimes \Complex$}.
There are mainly two reasons for doing so
\footnote{
In addition, in order to make the formalism more coherent,
if we are considering complex Hilbert spaces and
complex functions in $M$,
it seems reasonable working with $\Tan M^{\Complex}$,
and then taking complex distributions.}:
\begin{enumerate}
\item
First, as we will see later, there is a kind of distributions which are
specially interesting for quantization
(those which we will call {\sl K\"ahler polarizations}),
and they are complex distributions.
\item
Second, complex polarizations are necessary to establish the
relationship
between  geometric quantization and
the theory of irreducible unitary representations
of Lie groups of symmetries
\cite{Ki-gq}, \cite{Wn-77}.
\end{enumerate}

\subsection{Definitions and properties}

\subsubsection{Complex distributions and polarizations}

Taking into account the above comments,
we define:

\begin{definition}
Let $(M,\Omega )$ be a symplectic manifold.
A {\rm complex polarization} $\P$
on $(M,\Omega )$ is a distribution in $\Tan M^{\Complex}$
such that:
\begin{description}
\item[{\rm (a)}]
It is {\sl Lagrangian}, that is
\begin{description}
\item[{\rm (i)}]\ \
$\forall m \in M$,
${\rm dim} \P_m = n$ (complex dimension)
\item[{\rm (ii)}]\
$\Omega (\P ,\P ) = 0$ \footnote{ $\Omega$ is extended to $\Tan
M^{\Complex}$ as a $\Complex$-bilinear form.}
\end{description}
\item[{\rm (b)}]
It is {\sl involutive}, that is
\begin{description}
\item[{\rm (iii)}]
$[\P ,\P ] \subset \P$
\end{description}
\item[{\rm (c)}]
It satisfies the additional condition
\begin{description}
\item[{\rm (iv)}]\
$\dim (\P_m \cap \bar \P_m \cap \Tan_m M)$
is constant for every $m \in M$
\end{description}
\end{description}
\label{pol}
\end{definition}

Observe that if $\P$ is a polarization,
so is $\bar \P$
\footnote{
Sometimes we commit an abuse of notation
denoting also by $\P$ the set of (complex) vector fields
taking values on the distribution $\P$.}.

In relation to these four conditions we can remark that:
\begin{itemize}
\item
Condition (i) is stated in order to reduce
the number of independent variables just to $n$.
\item
Every complex polarization $\P$ induces a
real isotropic distribution which is also involutive.
In fact, as $\P \cap \bar \P$
is a complex distribution invariant by conjugation, then
$D := \P \cap \bar \P \cap \Tan M$
is a real distribution satisfying that
$D^{\Complex} = \P \cap \bar \P$;
and it is called the {\sl isotropic distribution}.
Furthermore, the fact of $\P$ being involutive
(as it is required in condition (iii)) implies that $D$
is also involutive.
\item
Hence the {\it Frobenius theorem\/}
assures that $D$ defines a foliation on $M$ and,
as a consequence of the isotropy condition
(ii) of definition \ref{pol},
the tangent bundle $D_m$ of every integrable manifold can be locally
spanned by
vector fields whose Lagrange brackets are equal to zero.
We denote by $\D := M/D$
the space of the integrable manifolds of $D$ and by
$\pi_D : M \to \D$ the projection.
\item
Observe that, due to condition (iv),
the leaves of the foliation defined by $D$
have the same dimension.
A remaining question is if the space ${\cal D}$
is a differentiable manifold.
\item
Sometimes the involution condition (iii) of definition \ref{pol}
is replaced by the following {\it integrability condition}:
there exists a local family of complex functions
$\{ z_j \}_{j=1, \ldots ,n} \in \Cinfty (U) \ ,\ (U \subset M)$,
such that $\P$ is locally spanned by
the set of locally Hamiltonian vector fields $\bar X_{z_j}$
\cite{Wo-80}.
\item
As we are going to see in the following subsection,
for a certain type of polarizations (K\"ahler polarizations),
there exists a local family of complex functions
$\{ z_j \}_{j=1, \ldots ,n} \in \Cinfty (U) \ ,\ (U \subset M)$,
such that $\P$ is locally spanned by
the set of vector fields \dst\derpar{}{\bar z_j}\)
({\sl Nirenberg-Newlander theorem}
\cite{Ho-ica}, \cite{NN-cac} ).
\item
Finally, consider the complex distribution
$\P + \bar \P$.
Since it is invariant by conjugation,
it can be considered as the complexification
of a real distribution $E$; that is,
$E^{\Complex} = \P + \bar \P$;
where $E = (\P + \bar \P) \cap \Tan M$,
which is called the
{\sl coisotropic distribution}.
Notice that $D^{\perp} = E$
\footnote
{\ $D^{\perp}$
denotes the
{\it orthogonal symplectic complement\/}
of $D$; that is,
$D^{\perp} := \{ X \in {\cal X}(M) \mid
\forall Y \in D \ , \ \Omega (X,Y) = 0 \}$.},
as it can be easily proved \cite{Ki-gq}.

Then we define:
\end{itemize}

\begin{definition}
Let $(M,\Omega )$ be a symplectic manifold.
A polarization ${\cal P}$ is called
{\rm strongly integrable} or {\rm reducible} iff:
\begin{description}
\item[{\rm (i)}]\ \
$E$ is involutive.
\item[{\rm (ii)}]\
Both spaces of integral manifolds
$\D$ and $\E := M/E$
are differentiable manifolds.
\item[{\rm (iii)}]
The canonical projection
$\pi_{DE} : \D \to \E$
is a submersion.
\end{description}
\label{sap}
\end{definition}

Given a complex polarization $\P$ on $(M,\Omega )$,
for every $m \in M$, we can define
a (pseudo) hermitian form $\hat \h_m$ in $\P_m$ by
\beq
\hat \h_m(X_m, Y_m) := i\Omega_m(X_m,\bar Y_m)
\label{hermm}
\eeq
for every $X_m,Y_m \in P_m$.
We have that:

\begin{prop}
$\ker \hat \h = \P \cap \bar \P$
\end{prop}
{\sl (Proof)} \quad
In fact $\P \cap \bar\P \subset \ker \hat \h$ because $\P$ is isotropic.

On the other side, for every $X \in \P$, if $X \in \ker \hat \h$, then
$\Omega (X,\bar Y)=0$, for all $Y \in \P$,
that is, $\Omega (X,Y)=0$, for all $Y \in \bar P$,
hence $X \in \bar \P$ because $\P$ is Lagrangian
and so is $\bar \P$.
\qed

Consequently, $\hat \h$ projects onto a non-degenerate form
on the quotient $\P /(\P \cap \bar \P )$.
We denote this form by $\bar \h$. Then:

\begin{definition}
Let $\P$ be a complex polarization on $(M,\Omega )$.
$\P$ is said to be
{\rm of type $(r,s)$} iff
the form $\bar \h$ has signature $(r,s)$.
Then, $\P$ is said to be
{\rm positive} if $s=0$.
\label{typrs}
\end{definition}

Observe that, if $\P$ is of type $(r,s)$,
then $\bar \P$ is of type $(s,r)$.

Since the complex dimension of $\ker \hat \h$ is $n-(r+s)$
and $\P \cap \bar \P$ is the complexification of the real distribution,
$D$, then this number is the dimension of the distribution $D$ in $\Tan
M$.
For this reason $n-(r+s) \equiv n-l$ is called the
{\it number of real directions} in $\P$.
On the other hand, from the definition and properties
of $D$ and $E$, it can be proved that the dimension of $E=D^{\perp}$
is $n+l$ (and hence  $\dim D + \dim E = 2n$).

Furthermore, if $\P$ is strongly integrable then,
in every subset $U \subset M$,
there is a set of local coordinates
$\{ x_1, \ldots ,x_{n-l};y_1,\ldots ,y_{n-l};
u_1,\ldots ,u_l;v_1,\ldots ,v_l \}$
such that $D$ is generated by the vector fields
\dst\left\{ \derpar{}{x_j} \right\}_{j=1,\ldots ,n-l}\)
and $\P$ by  \dst\left\{ \derpar{}{x_j} \right\}_{j=1,\ldots ,n-l}\) ,
\dst\left\{ \derpar{}{\bar z_k} \right\}_{k=1,\ldots ,l}\) ,
with $z_k := u_k + i v_k$ \cite{Ki-76}.

Next we introduce the condition of $\P$ to be {\it adapted} to the
connection:

\begin{definition}
Let $(M,\Omega )$ be a symplectic manifold and $(L,\pi ,M)$ a
complex line bundle endowed with a hermitian connection $\nabla$.
A polarization is called {\rm admissible} for the connection
$\nabla$ (or {\rm adapted} to the connection) iff, in some
neighborhood of each point $m \in M$, there is a symplectic
potential $\theta$ (i.e., $\omega = 2\pi i \theta$ is a connection
form of $\nabla$) such that $\langle \P ,\theta \rangle = 0$.
\end{definition}

In relation to this concept, we remark here the following result:

\begin{prop}
Let $(M,\Omega )$ be a symplectic manifold
and $(L,\pi ,M)$ a complex line bundle
endowed with a hermitian connection $\nabla$.
If $\P$ is a strongly integrable polarization
then it is admissible for the connection $\nabla$.
\end{prop}
{\sl (Proof)} \quad
See \cite{Ra-77}.
\qed

Strongly integrable polarizations
will be the only ones
we are going to consider from now on.

Due to their relevance for quantization,
we devote the following subsections to studying
two particular cases which are specially interesting.

\subsubsection{K\"ahler polarizations}
\protect \label{Kpol}

The first important case of polarization is the following:

\begin{definition}
Let $(M,\Omega )$ be a symplectic manifold.
A polarization $\P$ on $M$ is said to be a
{\rm K\"ahler polarization} if $\P \cap \bar \P = 0$.
\end{definition}

For a K\"ahler polarization $\P$ we have that
$\Tan M^{\Complex} = \P \oplus \bar \P$
and $\pi_{DE}~\colon~\D~=~M~\to~\E$.
Hence, if $\P$ is of type $(r,s)$,
we have that $r+s=n$
(and, thus, a K\"ahler polarization is ``totally'' complex,
in the sense that it does not contain any real direction).

In order to analyze the properties of this kind of polarizations,
we need to introduce several new concepts
(see \cite{GS-77}, \cite{Ib-90}, \cite{Ok-87}, \cite{Wo-80}).

\begin{definition}
Let $M$ be a differentiable manifold.
An {\rm almost-complex structure} in $M$ is a $(1,1)$-tensor field
${\cal J}$ on $M$
such that ${{\cal J}_m}^2=-{\rm Id}_{\Tan_m M}$, for every $m \in M$.
If ${\cal J}$ exists, then $(M,{\cal J})$ is said to be an
{\rm almost-complex manifold}.
\end{definition}

{\bf Comment:}
\begin{itemize}
\item
The existence of an almost-complex structure
implies that the dimension of $M$ is even.
Hereafter, $\{ U; x_1,\ldots ,x_n,y_1,\ldots ,y_n\}$
denote a chart of coordinates on $M$.
\end{itemize}

${\cal J}_m$ can be extended to a $\Complex$-linear endomorphism
in $\Tan_m M^{\Complex}$.
Consider now the set of complex analytical coordinates
$\{ z_j := x_j+iy_j \}$,
then a basis of $\Tan_m M^{\Complex}$ is made of the vectors
$$
\left(\derpar{}{z_j}\right)_m = \frac{1}{2} \left(
\left(\derpar{}{x_j}\right)_m -i\left(\derpar{}{y_j}\right)_m \right)
\quad ; \quad
\left(\derpar{}{\bar z_j}\right)_m = \frac{1}{2} \left(
\left(\derpar{}{x_j}\right)_m +i\left(\derpar{}{y_j}\right)_m \right)
$$

Notice that, if ${\cal J}$ is an almost-complex structure, then
the extension of the endomorphism ${\cal J}_m$ to
$\Tan_mM^{\Complex}$ has eigenvalues $\pm i$. Then, $\Tan_m
M^{\Complex}$ can be decomposed into a direct sum: $$ \Tan_m
M^{\Complex} = \Tan_m^{(1,0)} M \oplus \Tan_m^{(0,1)} M $$ where
\beann \Tan_m^{(1,0)}M&:=&\{X_m \in \Tan_m M^{\Complex}\ \mid \
{\cal J}_m X_m=+iX_m\}
\\
\Tan_m^{(0,1)}M&:=&\{X_m \in \Tan_m M^{\Complex}\ \mid \ {\cal J}_m
X_m=-iX_m\}
\eeann
Vectors belonging to these sets
(and the corresponding vector fields,
denoted ${\cal X}^{(1,0)}(M)$ and
${\cal X}^{(0,1)}(M)$) are called
{\it vectors (vector fields) of $(1,0)$-type}
and
{\it vectors (vector fields) of $(0,1)$-type},
respectively
\footnote{
Observe that, for every $X \in {\cal X}(M)$
we have that
$X^+ := \frac{1}{2}(X-i{\cal J}X) \in {\cal X}^{(1,0)}(M)$
and
$X^- := \frac{1}{2}(X+i{\cal J}X) \in {\cal X}^{(0,1)}(M)$,
and hence $X = X^+ + X^-$. So $\vf (M)\subset \P \cap \bar\P$.}.
Let $\bar \P$ and $\P$ be the complex distributions of
$\Tan M^{\Complex}$ defined by
$\Tan_m^{(1,0)}M$ and $\Tan_m^{(0,1)}M$
respectively for every $m \in M$.

\begin{definition}
An almost complex structure in $M$ is said to be a
{\rm complex structure} if the complex distributions
$\P$ and $\bar \P$ are involutive and $n$-dimensional.

This is equivalent to say {\rm (\cite{Ok-87} p. 379)}
that there are sets of
local complex analytical coordinates $(z_j,\bar z_j)$ such that,
for every $m \in M$,
$$
{\cal J}_m\left(\derpar{}{z_j}\right)_m =
 i\left(\derpar{}{z_j}\right)_m
\quad , \quad
{\cal J}_m\left(\derpar{}{\bar z_j}\right)_m=
-i\left(\derpar{}{\bar z_j}\right)_m
$$
or, what is equivalent, there exist local charts of coordinates $\{ U;
x_j,y_j\}$ such that
$$
{\cal J}_m\left(\derpar{}{x_j}\right)_m = \left(\derpar{}{y_j}\right)_m
\quad , \quad
{\cal J}_m\left(\derpar{}{y_j}\right)_m =
 -\left(\derpar{}{x_j}\right)_m
$$

Then $(M,{\cal J})$ is a {\rm complex manifold}.
\end{definition}

If ${\cal J}$ is a complex structure, then
\dst\left\{ \derpar{}{z_j} \right\}\) and
\dst\left\{ \derpar{}{\bar z_j} \right\}\) are local basis for
${\cal X}^{(1,0)}(M)$ and ${\cal X}^{(0,1)}(M)$),
that is, for $\bar \P$ and $\P$, respectively.

Then we can define:

\begin{definition}
An {\rm almost-K\"ahler manifold}
is a triple $(M,\Omega ,{\cal J})$ where:
\begin{description}
\item[{\rm (i)}]\ \
$(M,\Omega )$ is a symplectic manifold.
\item[{\rm (ii)}]\
$(M,{\cal J})$ is an almost-complex manifold.
\item[{\rm (iii)}]
The almost-complex structure and the symplectic form
are compatible, that is, for every $X,Y \in {\cal X}(M)$,
$$
\Omega ({\cal J}X,{\cal J}Y) =  \Omega (X,Y)
$$
\end{description}
\end{definition}

In this case, the distributions $\P$ and $\bar \P$ are isotropic,
v.g.: if $X;Y \in \Tan_m^{(1,0)}M$ then
$$
\Omega (X,Y) = \Omega ({\cal J}X.{\cal J}Y) = \Omega (iX,iY) = -\Omega
(X,Y)
\quad\Rightarrow\quad \Omega (X,Y)= 0
$$
and, since $\dim \P + \dim \bar \P = 2n$, they are Lagrangian.

\begin{definition}
A {\rm K\"ahler manifold} is a triple $(M,\Omega ,{\cal J})$ where:
\begin{description}
\item[{\rm (i)}]\ \
$(M,\Omega )$ is a symplectic manifold.
\item[{\rm (ii)}]\
$(M,{\cal J})$ is a complex manifold.
\item[{\rm (iii)}]
The complex structure and the symplectic form
are compatible, that is, for every $X,Y \in {\cal X}(M)$,
$$
\Omega ({\cal J}X,{\cal J}Y) =  \Omega (X,Y)
$$
\end{description}

Equivalently, we can say that a K\"ahler manifold
is an almost-K\"ahler manifold for which
the complex distributions $\P$ and $\bar \P$ are integrable.
\end{definition}

There is a close relation between K\"ahler manifolds
and K\"ahler polarizations,
which justifies the name of these last ones.
It is stated in the following proposition:

\begin{prop}
If $(M,\Omega ,{\cal J})$ is a K\"ahler manifold,
then $\P$ and $\bar \P$ are K\"ahler polarizations,
called the {\rm holomorphic polarization} and
the {\rm antiholomorphic polarization} respectively.

Conversely, if $(M,\Omega)$ is a symplectic manifold
which carries a K\"ahler polarization $\P$,
then there is a complex structure ${\cal J}$ defined in $M$
which is compatible with $\Omega$,
and therefore, $(M,\Omega ,{\cal J})$ is a K\"ahler manifold.
\end{prop}
{\sl (Proof)} \quad
The first part of the proposition is immediate.

For the second one we have that
if $\P$ is a K\"ahler polarization, as
$\Tan_mM \subset \Tan_m M^{\Complex} = \P_m \oplus \bar \P_m$,
then we can define an almost-complex structure ${\cal J}$
as follows: for every $X_m \in \Tan_m M$
we can write $X_m = Y_m + \bar Y'_m$ for some
$Y_m \in \P_m$ and $\bar Y'_m \in \bar \P_m$,
and then
$$
{\cal J}_mX_m := iY_m - i\bar Y'_m
$$
The differentiability of ${\cal J}$ follows from
its expression in local coordinates.
On the other hand, ${\cal J}$ is compatible with $\Omega$
because, if $X_m,Z_m \in \Tan_m M$, then
\beann
\Omega_m ({\cal J}_mX_m,{\cal J}_mZ_m)&=&
\Omega_m ({\cal J}_m(Y_m+\bar Y'_m),{\cal J}_m(W_m+\bar W'_m))
= \Omega_m (iY_m-i\bar Y'_m,iW_m-i\bar W'_m)
\\
&=&\Omega_m (Y_m,\bar W'_m) + \Omega_m (\bar Y'_m,W_m)
=\Omega_m (Y_m+\bar Y'_m,W_m+\bar W'_m) = \Omega_m (X_m,Z_m)
\eeann
Hence, $(M,\Omega ,{\cal J})$ is an almost- K\"ahler manifold.
Finally, $\P$ is integrable by definition
and then $(M,\Omega ,{\cal J})$ is a K\"ahler manifold
(for which $\P$ is its holomorphic polarization).
\qed

On the other hand, the following property holds:

\begin{prop}
In every K\"ahler manifold $(M,\Omega ,{\cal J})$
there is a non-degenerate (pseudo) hermitian metric ${\rm k}$
defined on $M$ in the following way:
For every $X,Y \in {\cal X}(M)$,
$$
{\rm k} (X,Y) := {\rm g}(X,Y) - i\Omega (X,Y)
$$
where ${\rm g}$ is a (pseudo) Riemannian metric given by
$$
{\rm g}(X,Y) := \Omega (X,{\cal J}Y)
$$
This hermitian metric is ``compatible with ${\cal J}$'',
that is:
$$
{\rm k} ({\cal J}X,{\cal J}Y) = {\rm k} (X,Y)
$$
\end{prop}
{\sl (Proof)} \quad
To prove that this is a non-degenerate hermitian metric is immediate.
For the compatibility with ${\cal J}$, remember that
$\Omega$ is compatible with ${\cal J}$, hence
\beann
{\rm k} ({\cal J}X,{\cal J}Y)&:=&
{\rm g}({\cal J}X,{\cal J}Y) - i\Omega ({\cal J}X,{\cal J}Y)
=\Omega ({\cal J}X,{\cal J}{\cal J}Y) - i\Omega (X,Y)
\\ &=&
\Omega (X,{\cal J}Y) - i\Omega (X,Y)
={\rm g}(X,Y) - i\Omega (X,Y) := {\rm k} (X,Y)
\eeann
\qed

{\bf Comments:}
\begin{itemize}
\item
It is interesting to point out that if ${\rm k}$ is not positive
definite
then nor is $\rm g$ and therefore the corresponding K\"ahler
polarization $\P$
is of type $(r,s)$, with $s\not=0$.
In the same way ${\rm k}$ is positive if, and only if, so are ${\rm g}$
and $\P$.
Sometimes, in the literature, the term ``K\"ahler polarization''
is applied only when this last condition holds and the
other ones are called {\it pseudo-K\"ahler}.
\item
Observe that this proposition allows to state that,
in every K\"ahler manifold,
the symplectic form can be taken as the imaginary part of a
non-degenerate (pseudo) hermitian metric
which is compatible with the complex structure.
\end{itemize}

In relation to the converse problem we have:

\begin{definition}
Let $(M,{\cal J})$ be an almost-complex manifold endowed
with a (pseudo) hermitian metric ${\rm k}$ defined in $M$,
which is compatible with ${\cal J}$.
Then we can define a (skew-symmetric) non-degenerate
$2$-form $\Omega$ by
$$
\Omega (X,Y) :={\rm Img} ({\rm k} (X,Y))
$$
which is called the
{\rm fundamental form} of $(M,{\cal J})$.

If $\Omega$ is closed (i.e., symplectic),
then it is called a {\rm K\"ahler form}.
In this case, $(M,\Omega ,{\cal J})$ is an almost-K\"ahler manifold
and, in addition, if ${\cal J}$ is a complex structure,
then $(M,\Omega ,{\cal J})$ is a K\"ahler manifold.
\end{definition}

It is possible to find a local smooth function $f(z,\bar z)$
such that the local expression of this K\"ahler form is
$$
\Omega=i \hbar\frac{-\partial^2 f}{\partial \bar z_j \partial z_k}
\d \bar z_j \wedge \d z_k
:= i \hbar \bar \partial \partial f
$$
(where we have introduced the notation
\dst\partial := \derpar{}{z_j}\d z_j\) ,
\dst\bar \partial := \derpar{}{\bar z_k}\d \bar z_k\)
\cite{KN-69}, \cite{MK-71}, \cite{Ok-87};
and their composition is in the sense defined in \cite{Wl-85}).
Then the most natural local symplectic potential is
$$
\Theta = \frac{i\hbar}{2}(\partial f -\bar \partial f)
$$
Taking these local expressions into account,
it is immediate to prove that:

\begin{prop}
Let $(M,\Omega ,{\cal J})$ be a K\"ahler manifold
and $(L,\pi,M,\nabla ,\h )$ the corresponding prequantum line bundle.
Then, there are local symplectic potentials
$\theta$ and $\bar \theta$ adapted to the
holomorphic and antiholomorphic polarizations
$\bar \P$ and $P$, respectively,
(that is, these polarizations are admissible
for the connection $\nabla$).

Their local expressions are:
 $$
 \theta =\ i\hbar \partial f =\ i\hbar\derpar{f}{z_j}\d z_j
 \quad ; \quad
 \bar \theta = -i\hbar \bar \partial f=
 -i\hbar\derpar{f}{\bar z_j}\d \bar z_j
 $$
\label{asp}
\end{prop}

\subsubsection{Real polarizations}

The other interesting case is the opposite one
to the above.

\begin{definition}
Let $(M,\Omega )$ be a symplectic manifold.
A polarization $\P$ on $M$ is said to be a
{\rm real polarization} iff
$\P = \bar \P$.
\end{definition}

As you can observe, every real polarization
is always positive, as $\hat\h=0$; and $D=E$.

Notice that, if ${\cal P}$ is a real polarization,
it holds that
${\cal P}\cap \bar{\cal P} = {\cal P}$.
Consider $D \equiv {\cal P} \cap \Tan M$,
that is, the real elements of ${\cal P}$.
$D$ is a Lagrangian distribution in $\Tan M$.
Conversely, if $D$ is a Lagrangian distribution of $\Tan M$,
then its complexification $D^{\Complex}$
is a real polarization.
Hence, the fact of considering real polarizations in $M$
is equivalent to take Lagrangian distributions in $\Tan M$.
Therefore, $M$ is foliated by Lagrangian submanifolds.

In addition the following result holds:

\begin{prop}
If ${\cal P}$ is a real polarization,
then there exist a local basis of $D$
made up of Hamiltonian vector fields.
\label{blh}
\end{prop}
{\sl (Proof)} \quad
Let $\{ X_1,\ldots ,X_n \}$ be a basis of $D$
in an open set $V \subset M$.
The differential $1$-forms
$\inn(X_j)\Omega$ $(j=1,\ldots ,n)$,
are linearly independent and vanish on $D$ over $V$,
since $D$ is Lagrangian.
Therefore, in the open set $V$, the submodule incident to the one
generated by these forms is $D$.
Let $\{ x_1,\ldots ,x_n,y_1,\ldots ,y_n \}$ be
a local system of coordinates in $U \subseteq V$ such that
$$
\{ X_1,\ldots ,X_n \} \equiv
\left\{ \derpar{}{x_1},\ldots ,\derpar{}{x_n} \right\}
$$
and let $\{ Z_1,\ldots ,Z_n \}$ be
the vector fields defined in $U$ by
$\inn(Z_j)\Omega =\d y_j$.
The existence of these local systems is guaranteed
because $D$ is an involutive distribution.
We have that
$$
\{ \d y_1,\ldots ,\d y_n \}' =
\left\{ \derpar{}{x_1},\ldots ,\derpar{}{x_n} \right\} =
\{ X_1,\ldots ,X_n \}
$$
(where $\{\quad \}'$ denotes the submodule incident to $\{\quad \}$),
but
$$
\{ \d y_1,\ldots ,\d y_n \} =
\left\{ \derpar{}{x_1},\ldots ,\derpar{}{x_n} \right\}' =
\{ \inn(X_1)\Omega ,\ldots ,\inn(X_n)\Omega \}
$$
then, since
$$
\{ \d y_1,\ldots ,\d y_n \} =
\{ \inn(Z_1)\Omega ,\ldots ,\inn(Z_n)\Omega \}
$$
it follows that $\{ X_1,\ldots ,X_n \} = \{ Z_1,\ldots ,Z_n \}$
because $\Omega$ is symplectic.
Hence, the vector fields $\{ Z_l \}$
make a basis satisfying the
conditions of the statement.
\qed

As a consequence of the isotropy condition (ii) of definition
\ref{pol}; and taking into account the above proposition,
you can prove that the Poisson brackets between every two coordinates
of the Lagrangian submanifolds which foliate $M$ vanishes.

\subsection{The polarization condition on the states}

\subsubsection{The general situation}

Once the concept of polarization is established,
we use it in order to correctly define the quantum states.
Thus, reminding definition \ref{fquan}, we claim:

\begin{require}
{\bf (and Definition)}\
Let $(M,\Omega )$ be a symplectic manifold.
Let $(L,\pi ,M;\h ,\nabla )$ be the prequantum line bundle
obtained after the prequantization procedure.
Let $\P$ be a strongly integrable polarization
of $(M,\Omega )$.
The space of quantum states of the system
is constructed starting from the set of smooth sections $\Gamma (L)$
which are {\it covariantly constant} along $\P$;
that is, such that
\beq
\nabla_\P\psi = 0
\label{covcon}
\eeq
We call them
{\rm polarized sections} (related to $\P$)
and denote this set by $\Gamma_\P (L)$.
\label{polsec}
\end{require}

  From this condition it can be observed that,
if we take local symplectic potentials such that
$\langle \P , \theta \rangle = 0$,
then the polarized sections are constant
along the leaves of the foliation induced in $M$ by $D$,
in the sense which we have discussed in the beginning of this section.
In particular, using the local basis of coordinates introduced
in the comments after definition \ref{typrs},
we have that $\P$ is locally spanned by the set of vector fields
\dst\left\{ \derpar{}{x_j} \right\}_{j=1 \ldots n-l}\) ,
\dst\left\{ \derpar{}{z_k} \right\}_{k=1,\ldots l}\)
(with $z_k := u_k + i v_k$) and then,
condition (\ref{covcon}) implies that the polarized sections
are represented by functions which only depend on
the coordinates $\{ y_j \}$, $\{\bar z_k \}$.

The last requirement introduces new complications
in the geometric quantization scheme:
\begin{itemize}
\item
In general, $\Gamma_\P (L) \not\subset \H_P$
since the polarized sections are not necessarily square integrable.
\item
Even more, if
$\psi_1,\psi_2 \in \Gamma_\P (L)$,
then $\h (\psi_1,\psi_2)$ is a function which
is constant along the leaves of the foliation
induced by $D$, but the inner product
$\langle \psi_1 | \psi_2 \rangle$
is not defined, in general,.
In fact, since $\h (\psi_1,\psi_2)$
defines a function in the quotient manifold $\D$,
we would integrate it in $\D$,
but this is not possible because, in the general case, we have not
a measure defined in $\D$.
\end{itemize}
The way to solve these problems will be studied in the following
section.

\subsubsection{Quantization of K\"ahler manifolds}

There is a special situation in which  the last obstructions are almost
overcame:
quantization of symplectic manifolds carrying K\"ahler polarizations
or, what is the same thing, quantization of K\"ahler manifolds.
In fact, K\"ahler manifolds are a distinguished kind of symplectic
manifolds
for quantization, as it has been studied in \cite{GS-gqmgr}.
Next, we show the guidelines of this method
(following \cite{Wo-80}).

Let $(M,\Omega )$ be a prequantizable system
endowed with a K\"ahler polarization $\P$,
and let $(L,\pi ,M)$ be a prequantum line bundle.
We can suppose, without loosing of generality, that
we are considering the complex structure in $(M,\Omega )$,
such that $\P$ is the holomorphic polarization.
Remember that the prequantum Hilbert space $\H_P$
is the set of square integrable smooth sections
in $\Gamma (L)$.

Now, in order to obtain the polarized sections,
we can choose the adapted symplectic potential
$\theta = i\hbar \partial f$ of proposition \ref{asp}.
Then, polarized sections $\psi$  are locally represented by holomorphic
functions
$F(z)$, that is, those such that $\bar \partial F = 0$,
Taking another symplectic potential, for instance
\dst\Theta = \frac{i\hbar}{2}(\partial f -\bar \partial f)\) ,
the polarized sections are locally given by
$$
\Psi (z,\bar z) = F(z) e^{-\frac{f(z,\bar z)}{2}}
$$
Then, you can observe that
$\Theta = \theta + \d \alpha$ with
\dst\alpha = \frac{i\hbar}{2}f(z,\bar z)\)
and the relation between both representations of polarized sections is
$\Psi = \psi e^{\frac{i\alpha}{\hbar}}$,
which is consistent with the complex linear bundle structure.
In any case, since there exists a local trivialization
in which polarized sections are represented by holomorphic functions,
we say that they are
{\it holomorphic sections} of the hermitian line bundle
$(L,\pi ,M)$

However, the inner product of these polarized sections
are not well defined since they are not necessarily
square integrable. Hence, we have to restrict this product to the set
of {\it square integrable holomorphic sections}.
We will denote this set by
$\Gamma_{hol}(L)$.

Now we are going to prove that
$\Gamma_{hol}(L)$ is a closed subspace of $\H_P$.
In order to achieve this result,
we need the following lemma due to Weil \cite{We-58}
(although the proof we present here is adapted from
\cite{Tu-gbkgq}).

\begin{lem}
Let $\psi$ a polarized section,
that is, $\psi_\alpha = F_\alpha s^\alpha$,
with $F_\alpha$ an holomorphic function;
and let $\{U_\alpha\}$ be an open covering of $M$
satisfying the following conditions:
\begin{enumerate}
\item
$\{ U_\alpha ,\phi_\alpha \}$ is a differentiable atlas on $M$.
\item
$\{ U_\alpha ,s^\alpha \}$ is a trivializing covering of $L$ and
$\h (s^\alpha ,s^\alpha ) = 1$, for every $\alpha$.
\item
For each $U_\alpha$ there exists a differentiable function
$f_\alpha$ such that
\dst  \Omega \mid _{U_\alpha} =
i \hbar\frac{-\partial^2 f_\alpha}{\partial \bar z_j \partial z_k}
\d \bar z_j \wedge \d z_k\) .
\item
For each $U_\alpha$, the polarization $\P$ is generated by the
vector fields \dst\left\{ \derpar{}{\bar z_j^\alpha}\right\}\)
and $\theta = i\hbar \partial f_\alpha$
is an adapted symplectic potential.
\end{enumerate}
Then, for each compact $K \subset U_\alpha$,
there exists a constant $C(K)>0$ such that,
if $m \in K$, we have
$$
\h_m (\psi_\alpha (m) , \psi_\alpha (m))
= F_\alpha (m) \bar F_\alpha (m)
\leq C(K) \langle \psi \mid \psi \rangle
$$
\end{lem}
{\sl (Proof)}  \quad
Given $K \subset U_\alpha$, consider $\delta >0$ such that,
for every $p \in K$,
$B'(p,2\delta ) := \phi_\alpha^{-1}(B(\phi_\alpha (p),2\delta ))$
where
$B(\phi_\alpha (p),2\delta ) \subset \phi_\alpha(U_\alpha )
\subset \Real^{2n}$.
Then
\beann
\langle \psi | \psi \rangle &=&
\int_M \h (\psi ,\psi ) \LF
\geq
\int_{U_\alpha} \h (\psi ,\psi ) \LF =
\int_{U_\alpha} F_\alpha \bar F_\alpha \LF =
\int_{\phi_\alpha (U_\alpha )}\phi_\alpha^{{-1}^*}(F_\alpha \bar
F_\alpha \LF)
\\ &\equiv&
\int_{\phi_\alpha (U_\alpha )}F'_\alpha \bar F'_\alpha \LF '
\geq
\int_{B(\phi\alpha (p),\delta )}F'_\alpha \bar F'_\alpha \LF '
\geq
C(K') \int_{B(\phi_\alpha (p),\delta )}F'_\alpha \bar F'_\alpha \d
V^{2n}
= (*)
\eeann
where $\d V^{2n}$ denotes the Lebesgue measure in $\Real^{2n}$,
 $K':=\overline{\cup_{p\in K}B(\phi_\alpha (p),\delta)}
 \subset\phi_\alpha(U_\alpha)$
and $C(K') > 0$ is a constant such that $\LF ' \geq C(K') \d V^{2n}$.
Observe that the existence of such a constant is justified since
$$
\d V^{2n} = g \LF ' \leq {\rm sup}_{K'} g \LF '
\ \Rightarrow \
\LF ' \geq C(K') \d V^{2n}
$$
where \dst C(K') :=  \frac{1}{{\rm sup}_{K'}g}\) .
\beann
(*) &=&
C(K')F'_\alpha (\xi )\bar F'_\alpha (\xi ){\cal V}(B(\phi_\alpha
(p),\delta ))
\\ &\geq&
C(K')F_\alpha (\phi_\alpha^{-1}(\xi ))\bar F_\alpha
(\phi_\alpha^{-1}(\xi ))
{\cal V}(B(\phi_\alpha (p),\delta ))
\frac{F_\alpha (p)\bar F_\alpha (p)}
{{\rm sup}({\rm sup}_{K'}(F_\alpha \bar F_\alpha ),1)}
\\ &\equiv&
C(K')C(\delta )F_\alpha (p)\bar F_\alpha (p)
\eeann
${\cal V}(B(\phi_\alpha (p),\delta )$ being the volume of the sphere
$B$.
The product $C(K')C(\delta ) > 0$ depends only on $K$, taking into
account the
initial choice of $\delta$ and the construction of $K'$.
Therefore
\dst F_\alpha (p)\bar F_\alpha (p) \leq C(K) \langle \psi | \psi
\rangle\) ;
where \dst C(K) \equiv \frac{1}{C(K')C(\delta )}\) .
\qed

Then, we have:

\begin{prop}
If $\P$ is a K\"ahler polarization,
then $\Gamma_{hol}(L)$ is a closed subspace of $\H_P$,
and hence $\Gamma_{hol} (L)$ is also a Hilbert space.
\end{prop}
{\sl (Proof)}  \quad
Let $\{ \psi_k \}$ be a Cauchy sequence in $\Gamma_{hol}(L)$,
that is,
$$
\langle \psi_k-\psi_{k'} \mid \psi_k - \psi_{k'} \rangle
\mapping{k,k' \rightarrow \infty} 0
$$
  From the above lemma, $\{ \psi_k \}$
is pointwise convergent to a section $\psi \in \Gamma (L)$. Then,
since the convergence is uniform in a neighborhood of every point,
and the uniform limit of a sequence of holomorphic functions is
also holomorphic, it follows that $\psi \in \Gamma_{hol}(L)$.
Furthermore, the sequence $\{ \psi_k \}$ converges to an element
$\psi' \in \H_P$, as $\H_P$ is a Hilbert space. Moreover, if $K
\subset M$ is compact then, using the triangle inequality, it
follows that, for each $k$, $$ \left[ \int_K \h (\psi -\psi ',\psi
- \psi ') \LF \right]^{\frac{1}{2}} \leq \left[ \int_K \h (\psi
-\psi_k,\psi - \psi_k) \LF \right]^{\frac{1}{2}} +\left[ (2\pi
\hbar )^n\langle \psi ' - \psi_k \mid \psi ' - \psi_k \rangle
\right]^{\frac{1}{2}} $$ Taking into account the above mentioned
convergences, we have that $\psi$ and $\psi '$ differ at most on a
zero-measure set, so $\Gamma_{hol}(L)$ is a closed Hilbert
subspace for the uniform convergence. Really we have to consider
equivalent sections differing just on a zero-measure set, and the
induced Hilbert space structure on the corresponding quotient set.
\qed

Hence, the quantization procedure would end at this stage.
Nevertheless certain problems
concerning the spectrum of the quantum operators
remains unsolved, as it will see in the examples
(see the quantization of the harmonic oscillator).

Using the antiholomorphic polarization,
polarized sections would be the antiholomorphic sections,
but the procedure is the same as above.

\subsection{The polarization condition on the operators}

To take the polarized sections as the
standpoint set for constructing the true
quantum states obliges to restrict
the set of admissible quantum operators or,
what means the same thing, the set of quantizable classical observables.

In fact, if we want that condition (v) of definition
\ref{fquan} holds, we have previously to assure that
$\Gamma_\P (L)$ is invariant under the action of the quantum operators.
This leads to set the following:

\begin{definition}
The set of {\rm polarized quantum operators}, ${\cal O} (\Gamma_\P
(L))$,
is made of the operators $\Op_f$
corresponding to functions $f \in \Cinfty(M)$ such that,
for every $\psi \in \Gamma_\P (L)$,
\beq
\Op_f \sta \in \Gamma_\P (L)
\label{polop}
\eeq
\end{definition}

There are different ways of characterizing the functions
$f$ representing the quantizable classical observables.
In order to obtain them and establish their equivalence
with the above definition we need to prove that:

\begin{lem}
Let $(M,\Omega )$ be a prequantizable system
and $(L,\pi ,M,\h ,\nabla )$ the corresponding prequantum line bundle.
Let $\P$ be a polarization in $(M,\Omega )$ and
$X \in {\cal X}^{\Complex}(M)$ such that
$\nabla_X(\Gamma_\P(L))=0$.
Then $X \in \P$.
\end{lem}
{\sl (Proof)} \quad
It suffices to prove it locally.

Thus, consider the local set of coordinates
$\{ x_1, \ldots ,x_{n-l}; y_1,\ldots ,y_{n-l};
u_1,\ldots ,u_l; v_1,\ldots ,v_l \}$
introduced in the comments after definition \ref{typrs},
and a local symplectic potential $\theta$
adapted to the polarization $\P$.
Since $\P$ is locally spanned by the vector fields
\dst\left\{\derpar{}{x_j}\right\}_{j=1,\ldots ,n-l}\) ,
\dst\left\{ \derpar{}{\bar z_k} \right\}_{k=1,\ldots l}\) ,
(with $z_k := u_k + i v_k$), then
a local symplectic potential adapted to $\P$ is given by
$\theta = \alpha^j \d y_j + \beta^k \d z_k$.

Let $\psi$ be a complex function representing a polarized section.
Then $\psi \equiv \psi (y_j,z_k)$. Now, if
\dst X \equiv A^j \derpar{}{x_j} + B^j \derpar{}{y_j}
+ C^k \derpar{}{z_k} + D^k \derpar{}{\bar z_k}\)
is a vector field such that  $\nabla_X\psi = 0$,
we are going to see that this implies that
$B^j=0$ and $C^k=0$.
In fact, we have that
\beq
0 = \nabla_X\psi = X(\psi ) + \langle X,\theta \rangle =
B^j \derpar{\psi}{y_j} + C^k \derpar{\psi}{z_k}
+ B^j\alpha_j + C^k\beta_k
\label{uno}
\eeq
Now, the set $\{ y_j \}$,  $\{ z_k \}$
is a basis for the functions representing the polarized sections.
The above condition for this set splits into
a system of $n$ linear equations
(with the coefficients $B^j, C^k,$
as unknowns):
$$
\left[\matrix{
1+\alpha_1y_1 & \alpha_2y_1 & \ldots & \alpha_{n-l}y_1 &
\beta_1y_1 & \ldots & \beta_ly_1 \cr
\alpha_1y_2 & 1+ \alpha_2y_2 & \ldots & \alpha_{n-l}y_2 &
\beta_1y_2 & \ldots & \beta_ly_2 \cr
\vdots & \vdots & & \vdots & \vdots & & \vdots \cr
\alpha_1y_{n-l} & \alpha_2y_{n-l} & \ldots &1+\alpha_{n-l}y_{n-l} &
\beta_1y_{n-l} & \ldots & \beta_ly_{n-l} \cr
\alpha_1z_1 & \alpha_2z_l & \ldots &\alpha_{n-l}z_l &
1+\beta_1z_l & \ldots & \beta_lz_l \cr
\vdots & \vdots & & \vdots & \vdots & & \vdots \cr
\alpha_1z_l & \alpha_2z_l & \ldots &\alpha_{n-l}z_l &
\beta_1z_l & \ldots & 1+\beta_lz_l \cr
}\right]
\left[\matrix{
B^1 \cr B^2 \cr \vdots \cr B^{n-l} \cr C^1 \cr \vdots \cr C^l}\right]
= \left[\matrix{0 \cr \vdots \cr 0 \cr}\right]
$$
and we are going to prove that it has only
the trivial solution. In fact, this system
has the form $TX=-X$, where
$$
T = \left[\matrix{
\alpha_1y_1 & \alpha_2y_1 & \ldots & \alpha_{n-l}y_1 &
\beta_1y_1 & \ldots & \beta_ly_1 \cr
\vdots & \vdots & & \vdots & \vdots & & \vdots \cr
\alpha_1y_{n-l} & \alpha_2y_{n-l} & \ldots &\alpha_{n-l}y_{n-l} &
\beta_1y_{n-l} & \ldots & \beta_ly_{n-l} \cr
\alpha_1z_1 & \alpha_2z_l & \ldots &\alpha_{n-l}z_l &
\beta_1z_l & \ldots & \beta_lz_l \cr
\vdots & \vdots & & \vdots & \vdots & & \vdots \cr
\alpha_1z_l & \alpha_2z_l & \ldots &\alpha_{n-l}z_l &
\beta_1z_l & \ldots & \beta_lz_l \cr
}\right]
$$
and $X$ is the matrix of unknowns.
Then, observe that $T$ has rank equal to $1$
(each row is a linear combination of other ones),
except if $\theta = 0$.
Therefore its only eigenvalue is the trace of $T$,
that is, $\sum_j\alpha_jy_j+\sum_k\beta_kz_k$,
which is different from $-1$
(because, if it were constant, then it would be equal to zero:
it suffices to calculate it at the point such that
the image under the homomorphism of the system is the
point zero in $\Real^{2n}$).
So, $-1$ is not the eigenvalue of $T$,
therefore the determinant of the system is different from zero
and then the solution is $X=0$,
that is, $B^j = 0 = C^k$.
Therefore, we have
$$
X = A^j \derpar{}{x_j} + D^k \derpar{}{\bar z_k} \in \P
$$
Notice that if $\theta = 0$ this result follows
in a straightforward way from
(\ref{uno}).
\qed

Now, we can prove that:

\begin{prop}
The following statements are equivalent:
\begin{enumerate}
\item
$\Op_f \in {\cal O} (\Gamma_\P (L))$
\item
$\Op_f$ preserves the polarization $\P$; in the sense that,
$[\Op_f,\nabla_{\P}] \psi = 0$;
for every $\psi \in \Gamma_\P (L)$,
\item
The function $f$, from which the operator $\Op_f$ is constructed, is
such that
$[X_f,\P] \subset \P$.
\end{enumerate}
\label{cnspo}
\end{prop}
{\sl (Proof)}
\begin{enumerate}
\item
The condition
$\Op_f \in {\cal O} (\Gamma_\P (L)) \quad \Leftrightarrow \quad
[\Op_f,\nabla_{\P}] \psi = 0$
is an immediate consequence of the equation (\ref{covcon}).
\item
Now, we are going to prove that
$[\Op_f,\nabla_{\P}] \psi = 0
\quad \Leftrightarrow \quad
[X_f,\P] \subset \P$.
for every $\psi \in \Gamma_\P (L)$. We have that
\beann
[\Op_f,\nabla_{\P}] \psi
&=&
\Op_f\nabla_\P \psi - \nabla_\P \Op_f \psi =
-\nabla_\P \Op_f \psi =
-\nabla_\P (-i\hbar\nabla_{X_f}+f)\psi
\\ &=&
i\hbar\nabla_\P \nabla_{X_f}\psi - \nabla_\P (f\psi ) =
i\hbar\nabla_\P \nabla_{X_f}\psi - \P (f)\psi - f\nabla_\P \psi =
i\hbar\nabla_\P \nabla_{X_f}\psi - \P (f)\psi
\\ &=&
i\hbar(\nabla_\P \nabla_{X_f}-\nabla_{X_f}\nabla_{\P})\psi - \P (f)\psi
=
i\hbar(\nabla_{[\P ,X_f]}+\frac{i}{\hbar}\Omega (\P ,X_f))\psi - \P
(f)\psi
\\ &=&
i\hbar\nabla_{[\P ,X_f]}\psi - (\Omega (\P ,X_f) + \Omega (X_f,\P
))\psi  =
i\hbar\nabla_{[\P ,X_f]}\psi
\eeann
Therefore, if $[X_f,\P ] \subset \P$, then
$[\Op_f,\nabla_{\P}] \psi = 0$.
Conversely, taking into account the above lemma,
we have immediately that
$[\Op_f,\nabla_{\P}] \psi = 0
\quad \Rightarrow \quad
[X_f,\P] \subset \P$.
\end{enumerate}
\qed

So we have:

\begin{require}
The set of quantum operators
${\cal O} (\Gamma_\P (L))$
is made of the operators $\Op_f$ satisfying condition
(\ref{polop}) or, what is equivalent,
the conditions $(2)$ or $(3)$ of proposition \ref{cnspo}.
\label{quanop}
\end{require}

It is evident that, once the complex line bundle is known,
the quantization of a system
(i.e., the Hilbert space and the set of quantum operators)
depend on the choice of the polarization
(but it does not on the choice of the symplectic potential).
To select a polarization determines the
{\it representation} of the quantum system
(see the examples).

In order to compare different representations, that is,
quantizations obtained from different choices of polarizations,
there is a procedure known as the method of the {\it
Blattner-Kostant-Sternberg kernels},
\cite{Bl-73}, \cite{Bl-75}, \cite{Bl-77}, \cite{GS-77}.
In particular, since the condition given in proposition
\ref{cnspo} does not allow to quantize all the classical observables,
this method is applied in order to quantize observables
which do not satisfy that condition.

\subsection{Examples}

\subsubsection{Cotangent bundles: the Schr\"odinger and
the momentum representations}

In the particular case of the example 1 in section 4.4,
where  $M = \Tan^* Q$,
there is a very special real polarization:
the {\it vertical polarization},
which is spanned by the vector fields
\dst\left\{ \derpar{}{p_j} \right\}\) .
Then, taking the adapted symplectic potential
$\theta = -p_j \d q^j$,
the polarized sections are the functions
$\psi \in \C (\Tan^*Q)$
such that \dst\derpar{\psi}{p_j} = 0\) ;
i.e., those being constant along the fibers of $\Tan^*Q$,
that is, $\psi = \psi (q^j)$.
The quantum operators corresponding to the observables
{\it positions} and {\it momenta} are
$$
\Op_{q^j} = q^j \  ; \
\Op_{p_j} = -i\hbar \derpar{}{q^j}
$$
but the energy is not quantizable.
When we quantize in this manner, the pair
$(\Gamma_\P (L),{\cal O}(\Gamma_\P (L)))$
is known as the {\it Schr\"{o}dinger representation}
of $(\Tan^*Q,\Omega )$.

Using the polarization spanned by the vector fields
\dst\left\{ \derpar{}{q^j} \right\}\)
(which is also real), and taking as adapted symplectic potential
$\theta ' = q^j \d p_j$,
we would obtain the functions
$\psi = \psi (p_j)$ as the polarized sections and
the quantum operators corresponding to the observables
position and momenta are now
$$
\Op_{q^j} = i\hbar \derpar{}{p_j}
\ ; \ \Op_{p_j} = p_j
$$
This is the so-called
{\it momentum representation} of $(\Tan^*Q,\Omega )$.

Observe that the relation between these representations
is the Fourier transform.
(See \cite{Sn-80} for more information on these topics).

\subsubsection{Cotangent bundles: the Bargman-Fock representation}

For the case $M = \Tan^* Q$,
we have just seen that the Schr\"odinger and the momentum
representations
are obtained by choosing the real polarizations spanned by
the Hamiltonian vector fields corresponding to the
momenta and positions respectively.
Next we are going to analyze another typical representation
obtained when we use a particular K\"ahler polarization.

Instead of the canonical coordinates,  we introduce the complex ones
$\{ z_j, \bar z_j \}$ where $z_j := p_j+iq^j$.
The expression of the symplectic form in these coordinates is
$$
\Omega=\frac{i}{2} \d \bar z_j \wedge \d z_j
$$
Now, $(\Tan^*Q,\Omega ,{\cal J})$ is a K\"ahler manifold,
where the complex structure is given by
$$
{\cal J}\left(\derpar{}{p_j} \right) = \left(\derpar{}{q^j}\right)
\quad , \quad
{\cal J}\left(\derpar{}{q^j}\right) = -\left(\derpar{}{p_j}\right)
$$
Now, we take the polarization $\P$,
spanned by \dst\left\{ \derpar{}{\bar z_j}\right\}\) .
As symplectic potentials we can chose
$$
\Theta = \frac{i}{4}(\bar z_j \d z_j - z_j \d \bar z_j)
$$
or the adapted one
$$
\theta = \frac{i}{2}\bar z_j\d z_j
$$
In any case, the polarized sections are
the holomorphic sections of the complex line bundle
$\Tan^*Q \times \Complex$,
and, using the symplectic potential $\Theta$, the quantum operators are
\beq
\Op_{z_j}=-2\hbar\derpar{}{\bar z_j}+\frac{z_j}{2}
\quad ; \quad
\Op_{\bar z_j} \equiv \Op_{z_j}^+=2\hbar\derpar{}{z_j}+\frac{\bar
z_j}{2}
\label{opers}
\eeq
which, in the physical terminology, are called the
{\it creation} and {\it annihilation} operators, respectively.
In this representation, a relevant role is played by the
so-called {\it number operator}:
$$
\Op_{\bar z_j z_j} =
2\hbar \left( z_j\derpar{}{z_j}-\bar z_j \derpar{}{\bar z_j} \right)
$$

This way of quantizing is known as the {\it holomorphic} or {\it
Bargman-Fock representation}
of $(\Tan^* Q, \Omega )$. On the contrary, if the polarization used is
the holomorphic one
$\bar \P$, the polarized sections would be the antiholomorphic sections
and
the representation so obtained is called the
{\it antiholomorphic representation} of $(\Tan^* Q, \Omega )$.

\subsubsection{The harmonic oscillator: Bargmann representation}

The next example that we consider
is the quantization of the $n$-dimensional harmonic oscillator,
using K\"ahler polarizations.
The prescriptions we are going to follow
will lead us to obtain the
{\it holomorphic} or {\it Bargmann representation}
of the harmonic oscillator.

We begin  identifying the phase space
$M \equiv \Real^{2n}$
(which is an Euclidean vector space endowed with the usual metric)
with $\Complex^n$,
and introducing the complex analytical coordinates
$\{ z_j, \bar z_j \}$
defined in the above subsection.
Hence, the Hamiltonian function
and the symplectic form are written
$$
H(z_j,\bar z_j)=\frac{1}{2}\bar z_j z_j
\quad ; \quad
\Omega=\frac{i}{2} \d \bar z_j \wedge \d z_j
$$
$(\Complex^n ,\Omega ,{\cal J})$ is a K\"ahler manifold,
where the complex structure ${\cal J}$ is given in the usual way,
and the hermitian metric is given also by the Hamiltonian.
Then, we can take the K\"ahler polarization $\P$ spanned by
\dst\left\{ \derpar{}{\bar z_j} \right\}\) , and as symplectic
potentials
(which are global, in this case)
we can chose
$$
\Theta = \frac{i}{4}(\bar z_j \d z_j - z_j \d \bar z_j)
$$
or the adapted one
$$
\theta = \frac{i}{2}\bar z_j\d z_j
$$

Using the first one, the polarized sections
(eigenstates of the system) are
$$
\psi (z_j,\bar z_j) = F(z) e^{-\frac{z_j \bar z_j}{4\hbar}}
$$
which are holomorphic sections on $\Complex$,
and the inner product is given by
$$
\langle \psi_1 \mid \psi_2\rangle  =\left( \frac{1}{2\pi \hbar}
\right)^n
\int_M F_1(z)\bar F_2(z) e^{-\frac{z_j\bar z_j}{2\hbar}}\LF
$$

For the observables, we obtain
$\Op_{z_j}$ and $\Op_{\bar z_j}$ given in
(\ref{opers}) and
$$
\Op_H = \frac{1}{2}\Op_{\bar z_j z_j} =
\hbar \left( z_j\derpar{}{z_j}-\bar z_j \derpar{}{\bar z_j} \right)
$$
It can be checked out that these operators
satisfy the condition given in proposition
\ref{cnspo}, and, so, the Hamiltonian function is quantizable.
It is also immediate to observe that
\beann
\Op_{z_j} \sta &=& z_j \psi = z_j F(z)e^{-\frac{z_j\bar z_j}{4\hbar}}
\\
\Op_{z_j}^+\sta &=& 2\hbar\derpar{\psi}{z_j}+\frac{\bar z_j}{2}\psi
= 2\hbar\derpar{F}{z_j}e^{-\frac{z_j\bar z_j}{4\hbar}}
\\
\Op_H\sta &=& \hbar\left( z_j\derpar{\psi}{z_j} -\bar
z_j\derpar{\psi}{\bar z_j}\right) = \hbar z_j \derpar{F}{z_j}
\eeann

Now, the problem of quantization of the harmonic oscillator
is almost solved. It remains to obtain the
energy spectrum. In order to make it,
we have to solve the eigenvalues equation
for the Hamiltonian operator
$$
\Op_H \sta = {\rm E} \sta
$$
but, taking into account the expression of $\Op_H$,
this equation is
$$
\hbar z_j \derpar{F(z)}{z_j} = {\rm E} F(z)
$$
Therefore, we obtain that
the eigenfunctions $F(z)$ are homogeneous polynomials
of degree $N$, $F(z_j) \equiv a z_1^{N_1}\ldots z_n^{N_n}$
(where $N_j \in \Zahl$ and $N=N_1+\ldots +N_n$), and
the eigenvalues of the Hamiltonian
(i.e., the energy) are $N\hbar$.

As you can observe, if use is made of the adapted symplectic potential
$\theta$, the eigenstates are
$\psi (z_j,\bar z_j) = F(z)$
and the Hamiltonian operator acting on these functions is just
\dst\Op_H = \hbar z_j \derpar{}{z_j}\) .
Hence the results are the same as above.

Nevertheless these results are not physically correct,
since the true eigenvalues equation is really
$$
\hbar\left(z_j \derpar{}{z_j} +\frac{n}{2}\right)F(z) = {\rm E} F(z)
$$
and the eigenvalues are $(N+\frac{n}{2})\hbar$.
This shows that, even in the case of using
K\"ahler polarizations, quantization does not end
at this stage and a final step must be done:
the so-called {\sl metaplectic correction}
\cite{Sn-80}, \cite{Tu-85}, \cite{Wo-80}.

\section{Metalinear bundles. Bundles of densities and half-forms.
\\ Quantization}

As we have already pointed out in the above section,
the set $\Gamma_\P (L)$ of polarized sections are not necessarily square

integrable;
and, in addition, if $\psi_1,\psi_2 \in \Gamma_\P (L)$,
the inner product $\langle \psi_1 | \psi_2 \rangle$
is not defined, in general, because we have not
a measure defined in $\D$.
Although, in general, the use of K\"ahler polarizations
allows us to overcome these problems, the spectrum of the quantum
operators
which is obtained is incorrect in certain cases.

The way to solve these problems consists in introducing
the {\sl bundles of densities and half-forms}, as new geometrical
structures
for quantization.

Some basic references on this topic are \cite{Ga-83} and \cite{Wo-80}
(In addition, some interesting and important examples of application of
these techniques
can be examined in \cite{Sn-80}, \cite{Tu-85}, \cite{Wo-80},
and other references quoted on them).
Finally, in the appendix we give some explanations about bundles
associated to
group actions, which can be of interest for the understanding of this
section.

\subsection{Complex metalinear group}

Consider the group $\GL$. Its subgroup $\SL \subset \GL$
is defined as the group of matrices $A \in \SL$  with $det\, A = 1$.
Consider now the following exact sequence of groups:
\beq
0 \map{} \Zahl \map{j} \Complex \times \SL \map{p} \GL \map{} 0
\label{es1}
\eeq
where
$$
j(k) := \left(\frac{2\pi i k}{n},e^{-\frac{2\pi i k}{n}}I\right)
\quad ; \quad
p(u,A) := e^{u}A
$$
On the other hand, if $p(z,A)=I$, then $e^{z}A=I$, hence $A=e^{-z}I$ and

therefore
$(z,A)=(z,e^{-z}I)$. Since $1=det\, (e^{-z}I=e^{-zn}$, then \dst
z=\frac{2\pi i k}{n}\) ,
with $k \in \Zahl$; hence $(z,A)=(\frac{2\pi i k}{n},e^{-\frac{2\pi i
k}{n}}I)= j(k)$.
Thus, we have proved that $\ker p = {\rm Im}\, j$.
We can also construct the following exact sequence
\beq
0 \map{} 2\Zahl \map{j} \Complex \times \SL \map{\pi}
\frac{\Complex \times \SL}{2\Zahl} \map{} 0
\label{es2}
\eeq
where now $j$ is restricted to $2\Zahl$; that is,
\dst j(2k) =\left(\frac{4\pi i k}{n},e^{-\frac{4\pi i k}{n}}I\right) \)
.

\begin{definition}
The quotient group \dst \ML :=\frac{\Complex \times \SL}{2\Zahl}\)
is called the {\rm complex metalinear group} of dimension $n$.
\end{definition}

Observe that $(z_1,A_1),(z_2,A_2)\in\Complex\times\SL$
 give equivalent elements in $\ML$ if there exists $k \in \Zahl$ such
that
\dst z_1 = z_2 + \frac{4\pi i k}{n}\) and $A_1=e^{-\frac{4\pi i
k}{n}}A_2$;
in other words, the coset of $(z,A)$ is
 $$
 \overline{(z,A)} =\left\{ \left( z + \frac{4\pi i k}{n} ,
 e^{-\frac{4\pi i k}{n}}A\right) \ ;
 \ k \in \Zahl\right\}
 $$
  From the above definition we have a surjective natural morphism
\beann
\rho : &\ML& \to \GL
\\
&\overline{(z,A)}& \mapsto \ e^{z}A
\eeann
which makes commutative the following diagram
$$
\begin{array}{ccccc}
\rho & \colon & \ML &
\begin{picture}(30,10)(0,0)
\put(13,8){\mbox{$\rho$}}
\put(0,4){\vector(1,0){30}}
\end{picture}
&\GL
\\
& & &
\begin{picture}(30,30)(0,0)
\put(5,5){\mbox{$\pi$}}
\put(30,0){\vector(-1,1){30}}
\end{picture}
&
\begin{picture}(10,30)(0,0)
\put(5,10){\mbox{$p$}}
\put(0,0){\vector(0,1){30}}
\end{picture}
\\
& & & &\Complex \times \SL
\end{array}
$$ If $\rho\overline{(z,A)} = I$ then $e^{z}A = I$, therefore
$e^{nz} = 1$ and $A = e^{-z}I$, hence \dst z = \frac{2\pi i
h}{n}\) , for some $h \in \Zahl$. But $z$ is determined up to an
integer multiple of \dst \frac{4\pi i k}{n}\) , therefore $$ \ker
\rho \equiv\left\{ \left(\frac{2\pi i h}{n} + \frac{4\pi i k}{n},
e^{-\frac{2\pi i h}{n}} e^{-\frac{4\pi i k}{n}}I \right) \ ; \ h,k
\in \Zahl \right\} $$ Observe that, if $h_1,h_2$ have the same
parity, they give the same element of $\ML$. Then we have that $$
\ker \rho= \left\{ \left(\frac{2\pi i h}{n} + \frac{4\pi i k}{n},
e^{-\frac{2\pi i h}{n}} e^{-\frac{4\pi i k}{n}}I \right) \ ; \
h=0,1 \ ; \ k \in \Zahl \right\} = \left\{ \overline{(0,I)},
\overline{\left(\frac{2\pi i}{n},e^{-\frac{2\pi i}{n}}I\right)}
\right\} \simeq \Zahl_2 $$ Therefore, we have the exact sequence
\beq 0 \map{} \Zahl_2 \map{\iota} \ML \map{\rho} \GL \map{} 0
\label{es3} \eeq where \dst \iota (0) = \overline{(0,I)}\) and
\dst\iota (1)=\overline{\left(\frac{2\pi i}{n},e^{-\frac{2\pi
i}{n}}I\right)}\) . Taking into account the natural structures of
complex Lie group in $\ML$ and $\GL$, this exact sequence
characterizes $\ML$ as a double covering of $\GL$. Observe that
the construction of this double covering arises from the two exact
sequences (\ref{es1}) and (\ref{es2}) since $$ \frac{\Complex
\times \SL}{j(\Zahl)} \simeq \GL \quad ; \quad \frac{\Complex
\times \SL}{j(2\Zahl)} \simeq \ML $$ and taking them together
$$\begin{array}{ccccccc} 0 \map{} &\Zahl& \map{j} &\Complex \times
\SL& \map{p} &\GL& \map{} 0
\\
&\big\uparrow& &\big\uparrow\ id & &\big\uparrow\ \rho &
\\
0 \map{} &2\Zahl& \map{j} &\Complex \times \SL & \map{\pi} &\ML& \map{}
0
\end{array}$$
Now, $\rho$ is defined in a natural way using these sequences.
On the other hand, in order to evaluate $\ker \rho$,
it suffices to see that, if $\overline{(z,A)} \in \ML$
has the same image (by $\rho$) than $\overline{(0,I)}$, then
$(z,A)$ and $(0,I)$ have the same image by $p$, that is,
$(z,A) \in j(\Zahl)$. Therefore, either $(z,A) \in j(2\Zahl)$,
and then $\overline{(z,A)} = \overline{(0,I)}$; or
$(z,A) \not\in j(2\Zahl)$, and hence we have that
$$
\ker \rho = \frac{j(\Zahl)}{j(2\Zahl)} \simeq \Zahl_2
$$
as we have already pointed out.

Notice that the sequence (\ref{es1}) allows us to write the elements of
$\GL$ as
 $$
 \overline{(z,A)} =
 \left\{ \left(z + \frac{2\pi i h}{n},e^{-\frac{2\pi i h}{n}}A \right)
 \ ; \ h \in \Zahl \right\}
 $$
with $z \in \Zahl$, $A \in \SL$.

For every $B \in \GL$ we can write $B = e^{z}B_0$,
with $det\, B_0=1$ and $e^{nz}=det\, B$.
Then, it is clear that $z$ is determined up to a factor
\dst\frac{2\pi i h}{n}\) , with $h\in\Zahl$. Then $B = p(z,B_0)$.
In constructing $\ML$, we split the elements $\overline{(z,B_0)}$
into two classes: the one having $h$ even and the other having $h$ odd.

Observe that,
in $\GL$, the square root of the determinant is not defined as a
holomorphic function.
However, so is it in $\ML$. In fact, we have
$$
\begin{array}{ccccc}
\ML &\mapping{\rho}& \GL &\longrightarrow& \Complex^*
\\
\overline{(z,A)} &\mapsto& e^{z}A &\mapsto&
det\, (e^{z}A) = e^{nz} det\, A = e^{nz}
\end{array}
$$
and we can construct the function
$$
\begin{array}{ccccc}
\chi &\colon &\ML& \mapping{det} & \Complex^*
\\
& & \overline{(z,A)} & \mapsto & e^{\frac{nz}{2}}
\end{array}
$$
which is well defined, since it does not depend on the representative of

$\overline{(z,A)}$.  In fact, if we consider $\overline{(z,A)} \in \ML$,

then $\chi(\overline{(z,A)}) = e^{nz/2}$ is well defined because
\dst z=z_0+\frac{4\pi i h}{n}\) , and therefore
\dst e^{nz/2}=e^{\frac{nz_0}{2}+2\pi i h}=e^{\frac{nz_0}{2}}\)
which does not depend on $h$. Moreover, it satisfies that
$$
\chi^2(\overline{(z,A)}) = (e^{\frac{nz}{2}})^2 = e^{nz} = (det \circ
\rho )(\overline{(z,A)})
$$
and then $\chi = \sqrt{(det \circ \rho )}$.
That is, we can lift the map $det$ to $\ML$ by means of $\rho$ and
obtain a square root of this lifting. Observe that this function cannot
be defined
in $\GL$ in this way. In fact, if $B \in \GL$, according to the above
considerations we have
$B=e^{z} B_0$, with $e^{nz}=det\, B$. Now, $B=p(z,B_0)$ and writing
\beann
\GL &\mapping{det}& \Complex^*
\\
B=p(z,B_0) &\mapsto& e^{nz}=det\, B
\eeann
the function $B \to e^{nz/2}$ is not well defined since
\dst z=z_0+\frac{2\pi i h}{n}\) , with $h\in \Zahl$, and hence
\dst e^{nz/2} =e^{\frac{n}{2}(z_0+\frac{2\pi i h}{n})} =
e^{\frac{nz_0}{2}+\pi i h}\) ,
whose value depends on $h$.

\subsection{Principal metalinear bundles. Classification}

\begin{definition}
Let $p : P \to M$ be a principal fiber bundle with structural group
$\GL$.
A {\rm metalinear bundle associated with} $(P,p,M)$
is a principal fiber bundle $\bar p : \bar P \to M$
 with structural group $\ML$ and a differentiable map
 $\bar \rho : \bar P\to P$
between bundles over $M$ giving the identity on $M$,
such that the following diagram commutes
$$
\begin{array}{ccc}
\bar P \times \ML & \mapping{\bar \iota} & \bar P
\\
\bar \rho \times \rho \big\downarrow & & \big\downarrow \rho
\\
P \times \GL & \mapping{\iota} & P
\end{array}
$$
where $\iota , \bar \iota$ are the actions of the groups on the bundles,

and such that $\bar \rho$ is a double covering.
\label{amfb}
\end{definition}

Now, given a principal fiber bundle $(P,p,M)$ with structural group
$\GL$,
we can ask when a metalinear associated bundle exists and,
in this case, if it is unique. We will answer these questions
afterwards.

Let $(U_i,s_i)$ be a trivial system of $P$ with transition functions
$g_{ij} : U_{ij} \to \GL$. We can construct transition functions of
$\bar P$,
$\bar g_{ij} : U_{ij} \to \ML$, making commutative the following diagram

$$
\begin{array}{ccccc}
0 \to \Zahl_2 \to & \ML & \map{\rho} & \GL \to 0
\\
& & \qquad \bar g_{ij}
&
\begin{picture}(15,15)(0,0)
\put(15,0){\vector(-1,1){15}}
\end{picture}
& \big\uparrow g_{ij}
\\
& & & & U_{ij}
\end{array}
$$
In order to construct $\bar g_{ij}$, it suffices to take $U_{ij}$
in such a manner that $g_{ij}(U_{ij})$
is contained in a fundamental open set of the covering.

Let us suppose that $U_{ij}$ are contractible. Then the function
$$
c_{ijk} := \bar g_{ij} \bar g_{jk} \bar g_{ki} \colon U_{ijk} \to \ML
$$
satisfies that $\rho \circ c_{ijk} = 1$, therefore
$c_{ijk} \in {\rm Im}\, \Zahl_2$, hence $c_{ijk}$ is constant.
In this way we have defined a $2$- cochain
$$
c(P) : (U_i,U_j,U_k) \to c_{ijk}
$$
with coefficients in $\Zahl_2$,
which is associated with every trivializing covering of $P$.
It can be proved that
\begin{enumerate}
\item
$c(P)$ is closed.
\item
$[c(P)] \in H^2(M,\Zahl_2)$ does not depend neither on the local system
$(U_i,s_i)$ used, nor on the liftings $\bar g_{ij}$ constructed.
It is a cohomology class associated with $p\colon P \to M$.
\item
A $\GL$-principal fiber bundle $(P,p,M)$ admits an associated metalinear

bundle
if, and only if, $c(P)=0$. Then:

\begin{definition}
Let $(P,p,M)$ satisfying that $c(P)=0$; and
$\bar p_1:\bar P_1 \to M$ and $\bar p_2:\bar P_2 \to M$
two metalinear fiber bundles associated with $P$,
with covering maps $\bar \rho_1\colon\bar P_1 \to P$ and
$\bar \rho_2\colon\bar P_2 \to P$. then $P_1$ and $\bar P_2$ are
{\rm equivalent} if there exists a diffeomorphism
$\tau : \bar P_1 \to \bar P_2$ such that
\begin{description}
\item[{\rm (a)}]
It is an isomorphism between $\Zahl_2$-principal fiber bundles over $P$,

and
\item[{\rm (b)}]
It is an isomorphism between $\ML$-principal fiber bundles over $M$.
\end{description}
\label{amfbeq}
\end{definition}

\item
With the above definition of equivalence, the group $H^1(M,\Zahl_2)$
acts freely and transitively on the set of equivalence classes of
metalinear fiber bundles associated with $P$.
\end{enumerate}

A last question is to find the condition for a function $\bar f\colon
\bar P \to \Complex$
to be $\bar \rho$-projectable.
 The necessary and sufficient condition for this is that, if
 $\bar a,\bar b \in \bar P$ and
 $\bar \rho (\bar a) = \bar \rho (\bar b)$,
 then $\bar f(\bar a) = \bar f(\bar b)$.
Then, you can observe that, if $\bar \rho (\bar a) = \bar \rho (\bar
b)$,
then $\bar p (\bar a) = \bar p (\bar b)$; that is, there exists
$\bar g \in \ML$ such that $\bar b = \bar a \bar g$,
hence $\rho(\bar g) = id$, that is, $\bar g \in \ker \rho$.
Therefore, the desired condition is that $\bar a \in \bar P$ and
$\bar g \in \ker \rho$, then $\bar f(\bar a) = \bar f(\bar a \bar g)$

\subsection{Bundles of densities}

In order to introduce integration in non-integrable manifolds, we need
new
structures: the so-called {\it densities}.

\subsubsection{Bundle of orientations}

\begin{definition}
Let $M$ be a differentiable manifold with $\dim M = n$,
and let $(U_{\alpha},\psi_{\alpha})$ be an atlas of $M$.
Consider the line bundle $L$ over $M$ defined by taking
$\{ U_{\alpha} \times \Real \}$ as open sets and
$$
f_{\alpha \beta} :=
sign\, \det J(\psi_{\alpha} \circ \psi_{\beta}^{-1})=
sign\, \det J(g_{\alpha \beta})
$$
as transition functions; where $g_{\alpha \beta} = \psi_{\alpha} \circ
\psi_{\beta}^{-1}$
are the functions of change of variables in $M$ for the atlas
$(U_{\alpha},\psi_{\alpha})$; and $J$ denotes the Jacobian matrix.
$L$ is called the {\rm bundle of orientations} of the manifold $M$.
\label{fo}
\end {definition}

Note that, from the condition
$g_{\alpha \beta} \circ g_{\beta \gamma} = g_{\alpha \gamma}$,
we deduce that $f_{\alpha \beta} \circ f_{\beta \gamma} = f_{\alpha
\gamma}$,
and then the family $\{ f_{\alpha \beta} \}$ satisfies the cocycle
condition and hence
the bundle structure is unique.

\begin{prop}
$M$ is orientable if, and only if, $L$ is trivial.
\label{jusfo}
\end{prop}
{\sl (Proof)} \quad
$(\Longrightarrow)$ \quad
If $M$ is orientable then there is an atlas with changes of coordinates
with positive jacobians,
therefore $f_{\alpha \beta}$ are the identity and, then $L$ trivializes.

$(\Longleftarrow)$ \quad
If $L$ trivializes, there exists a trivializing covering with transition

functions
$g_{\alpha \beta}$ which are the identity, that is, $M$ is orientable
since the sign of the jacobians are positive.
\qed

\subsubsection{Bundle of densities}

\begin{definition}
Consider the real line bundle $Q(M) := \Lambda^n(\Tan^* M)
\otimes_{\Real} L$.
Its transition functions
$$
h_{\alpha \beta} :=
\det [^tJ(g_{\alpha \beta})]^{-1} sign\, \det J(g_{\alpha \beta})=
\det J(g_{\alpha \beta})^{-1}
$$
$(U_{\alpha},\psi_{\alpha})$ being an atlas of $M$ and
$g_{\alpha \beta} = \psi_{\alpha} \circ \psi_{\beta}^{-1}$.
$Q(M)$ is called the {\rm bundle of densities} of $M$.
Sections of $Q(M)$ are called {\rm densities} on $M$.
\label{dens}
\end{definition}

Now we are going to see how the densities change
when the coordinates change. Let $\sigma \colon M \to Q(M)$ be a density

on $M$
and let $(U_{\alpha},\psi_{\alpha})$ be an atlas of $M$.
Put $\psi_{\alpha} \equiv \{ \psi_{\alpha}^1 \ldots \psi_{\alpha}^n \}$.

A basis of the sections of $Q(M)$ in the open set $U_{\alpha}$ is given
by
a basis of ${\mit\Omega}^n(\Tan^* M) \mid_{U_{\alpha}}$
which is $\{ \d \psi_{\alpha}^1 \wedge \ldots \wedge \d \psi_{\alpha}^n
\}$;
and a basis of $L \mid_{U_{\alpha}}$ which is $\{ u_{\alpha} \} \in
\Real$.
So then $\{ \d \psi_{\alpha}^1 \wedge \ldots \wedge \d \psi_{\alpha}^n
\}\otimes u_{\alpha} \}$
is a basis of the sections of $Q(M)$ in $U_{\alpha}$.

Thus, the density $\sigma$ induces the map
$\sigma_{\alpha}\colon U_{\alpha} \to U_{\alpha} \times L$
given by
$$
\sigma_{\alpha}(x) :=(x,s_{\alpha}(x)
(\d \psi_{\alpha}^1 \wedge \ldots \wedge \d \psi_{\alpha}^n
\otimes u_{\alpha})_x)
$$
that is, in $U_{\alpha}$ the section is given by
$s_{\alpha}(x) (\d
\psi_{\alpha}^1\wedge\ldots\wedge\d\psi_{\alpha}^n\otimes
u_{\alpha})_x$,
and in $U_{\beta}$ by
 $s_{\beta}(x)(\d \psi_{\beta}^1\wedge
 \ldots\wedge\d\psi_{\beta}^n\otimes u_{\beta})_x$.
If we take $x \in U_{\alpha \beta}$ we have
$s_{\alpha}(x) =s_{\beta}(x) \det (J(g_{\alpha \beta}(x)))$.
Hence, the integration of $\sigma$ in $M$ has sense,
since we have only to restrict to trivializing open sets $\{ U_{\alpha}
\}$;
and taking a partition of the unity $\{ \eta_{\alpha} \}$,we define
$$
\int_M \sigma :=
\sum_{\alpha}\int_{U_{\alpha}} \eta_{\alpha}  \sigma_{\alpha}
$$
which has the same properties than the
usual integration of forms in orientable manifolds.

As final remark, it is usual to consider {\it complex densities},
that is, to take sections of the fiber bundle $Q(M) \otimes_{\Real}
\Complex$,
which is a complex line bundle over $M$. You can observe that a complex
density
$\sigma$ is nothing but a sum $\sigma = \sigma_1 + i\sigma_2$;
$\sigma_1$ and $\sigma_2$ being real densities.

\subsection{Bundles of half-forms}

Let $p\colon P \to M$ be a principal fiber bundle wit group $\GL$,
which admits an associated metalinear bundle $\bar p : \bar P \to M$.

\begin{definition}
\ben
\item
Consider the action of $\ML$ on $\Complex$
$$
\begin{array}{ccc}
\ML \times \Complex & \to & \Complex
\\
(\bar g,z) & \mapsto & \chi (\bar g)z
\end{array}
$$
(remember that, if $\bar g = \overline{(z,A)}$
with $z \in \Complex$ and $A \in \SL$, then $\chi (\overline{(z,A)}) =
e^{nz/2}$).
The line bundle associated with this action is denoted $N^{1/2}$; that
is,
$$
N^{1/2} := \bar P \times_{\ML} \Complex \mapping{\pi_{\ML}} M
$$
where we remind that $\bar P \times_{\ML} \Complex$
is defined through the equivalence relation in $\bar P \times \Complex$
$$
(\bar a,z) \simeq (\bar b,w)
\Leftrightarrow
\exists \bar g \in \ML \ \mid \ (\bar b,w)=(\bar a \bar g,g^{-1}z)
$$
\item
Consider now the action of $\GL$ on $\Complex$
$$
\begin{array}{ccc}
\GL \times \Complex & \to & \Complex
\\
(g,z) & \mapsto & \mid det\, g \mid z
\end{array}
$$
The line bundle associated with this action is denoted $| N |$; that is,

$$
| N | := P \times_{\GL} \Complex \mapping{\pi_{\GL}} M
$$
where we remind that $P \times_{\GL} \Complex$
is defined through the equivalence relation in $P \times \Complex$
$$
(a,z) \simeq (b,w) \Leftrightarrow \exists g \in \GL \ \mid \
(b,w)=(ag,g^{-1}z)
$$
\een
\label{absn}
\end{definition}

Next we study the sections of $N^{1/2}$ and $| N |$.

If $\sigma : M \to N^{1/2}$ is a section of $\pi_{\ML}$, we can
interpret it as a function $\sigma : \bar P \to \Complex$
satisfying the following condition: if $\bar a, \bar b \in \bar P$
with $\bar p (\bar a) = \bar p (\bar b)$ (and then there exists
$\bar g \in \ML$ with $\bar b = \bar a \bar g$), then $\sigma
(\bar a) = \chi (\bar g) \sigma (\bar b)$. In a similar way, If
$\mu : M \to | N |$ is a section of $\pi_{\GL}$, we interpret it
as a function $\mu : P \to \Complex$ satisfying the following
condition: if $a,b \in P$ with $p(a) = p(b)$ (that is, there
exists $g \in \GL$ with $b=ga$), then $\mu (a) = \mid det\, g \mid
\mu (b)$. Bearing this interpretation in mind we have the
following result:

\begin{prop}
Let $\sigma ,\sigma '\colon M \to N^{1/2}$ be sections of
$\pi_{\ML}$, which we interpret as associated functions $\sigma
,\sigma ' \colon\bar P \to \Complex$. Then the function $\sigma
\sigma '^* : \bar P \to \Complex$ is projectable by $\bar \rho :
\bar P \to P$ and its projected function $\bar \rho (\sigma \sigma
'^*) : \bar P \to \Complex$ arises from a section of $| N |$. This
allows to construct a map $$
\begin{array}{ccc}
\Gamma (N^{1/2}) \times \Gamma (N^{1/2})
& \mapping{\langle , \rangle} &
\Gamma (| N |)
\\
(\sigma ,\sigma ') & \mapsto & \bar \rho (\sigma \sigma '^*)
\end{array}
$$
which is sesquilinear with respect to the module structures of the
sections of
$N^{1/2}$ and $| N |$ on $\Cinfty(M)\otimes\Complex$.
\label{propi1}
\end{prop}
{\sl (Proof)} \quad
{\sl (a)} \quad
$\sigma \sigma '^*$ is $\bar \rho$-projectable.

We have to prove that, if $\bar a \in \bar P$ and $\bar g \in \ker
\rho$, then
$\sigma \sigma '^*(\bar a \bar g) = \sigma \sigma '^*(\bar a)$.
In fact, as the action of $\bar g = \overline{(z,A)}$ on $\Complex$
consists in multiplying by
$e^{nz/2}$, we obtain that
\beann
\sigma \sigma '^*(\bar a \bar g)
& = &
\sigma (\bar a \bar g) \sigma '^*(\bar a \bar g) =
\chi (\bar g)^{-1} \sigma (\bar a)
(\chi (\bar g)^{-1})^* \sigma '(\bar a)^*
\\
& = &
\chi (\bar g)^{-1} (\chi (\bar g)^{-1})^*
\sigma (\bar a) \sigma '(\bar a)^* =
\mid \chi (\bar g) \mid ^{-2} (\sigma \sigma '^*)(\bar a) =
\mid \chi (\bar g)^2 \mid ^{-1} (\sigma \sigma '^*)(\bar a)=
(\sigma{\sigma '}^*)(\bar a)
\eeann
(Notice that $\chi (\bar g)^2 = det\, \rho (g) = det\, I = 1$).
So, the result is proved.

\quad \quad
{\sl (b)} \quad
$\rho (\sigma \sigma '^*)$ arises from a section of $| N |$.

We have to prove that, if $a \in P$ and $g \in \GL$, then
$$
\bar \rho (\sigma \sigma '^*)(ag) =
\mid det\, g \mid^{-1} \bar \rho (\sigma \sigma '^*)(a)
$$
In order to see that, consider $\bar a \in \bar P$ and $\bar g \in \ML$
with $\bar \rho (\bar a) = a$ and $\rho (\bar g) = g$. Then we have that

\beann
\bar \rho (\sigma \sigma '^*)(ag)
& = &
(\sigma \sigma '^*)(\bar a \bar g) =
\chi (\bar g)^{-1} (\chi (\bar g)^{-1})^*
(\sigma \sigma '^*)(\bar a) =
\mid \chi (\bar g) \mid ^{-2} (\sigma \sigma '^*)(\bar a)
\\
& = &
\mid det\, \rho (\bar g) \mid ^{-1} (\sigma \sigma '^*)(\bar a) =
\mid det\, g \mid ^{-1} (\sigma \sigma '^*)(\bar a)
\eeann
as we wanted to prove.
\qed

Now we introduce the Lie derivative of sections of $N^{1/2}$ and $| N
|$.

Let $X \in {\cal X}(P)$ be a $\GL$-invariant vector field. Since
$\bar \rho : \bar P \to P$ is a local diffeomorphism, we can lift
$X$ to a vector field $\bar X \in {\cal X}(\bar P)$ which is
$\ML$-invariant (because $X$ is $\GL$-invariant and it has
coverings). Let $\sigma : M \to N^{1/2}$ be a section. If we
consider it as a function from $\bar P$ to $\Complex$, it makes
sense to calculate $\Lie(\bar X) \sigma$. In the same way, if $\mu
: M \to | N |$ is a section, it makes sense to calculate $\Lie(X)
\mu$, when we interpret it as a function from $P$ to $\Complex$.
Then the following property holds:

\begin{prop}
Let $X \in {\cal X}(P)$ be a $\GL$-invariant vector field
and $\bar X \in {\cal X}(\bar P)$ its $\ML$-invariant lifting.
Consider $\sigma , \sigma ' : M \to N^{1/2}$. Then
\begin{description}
\item[{\rm (a)}]
$\Lie(\bar X) \sigma$, $\Lie(\bar X) \sigma '$
are sections of $N^{1/2}$.
\item[{\rm (b)}]
$\Lie(X) \langle \sigma ,\sigma ' \rangle$
is a section of $| N |$.
\item[{\rm (c)}]
$\Lie(X) \langle \sigma ,\sigma ' \rangle =
\bar \rho ((\Lie(\bar X)  \sigma )\sigma '^*) +
\bar \rho (\sigma \Lie(\bar X) \sigma '^*)$
\end{description}
\label{propi2}
\end{prop}
{\sl (Proof)} \quad
{\sl (a)} \quad
We have to prove that, if $\bar a \in \bar P$ and $\bar g \in \ML$, then

$$
(\Lie(\bar X) \sigma )(\bar a \bar g) =
\chi (\bar g)^{-1} (\Lie(\bar X) \sigma )(a)
$$
In fact, if $\bar X$ is $\ML$-invariant, this means that
$\bar X_{\bar a \bar g} = \Tan_{\bar a} \mu (\bar g) \bar X_{\bar a}$,
where
$$
\begin{array}{ccccc}
\mu (\bar g) &:& \bar P & \to & \bar P
\\
& & \bar x & \mapsto & \bar x \bar g
\end{array}
$$
therefore
\beann
(\Lie(\bar X) \sigma )(\bar a \bar g)
&=&
\bar X_{\bar a \bar g} \sigma =
(\Tan_{\bar a} \mu (\bar g) \bar X_{\bar a}) \sigma =
\bar X_{\bar a} (\sigma \circ \mu (\bar g)) =
\bar X_{\bar a} (\chi (\bar g)^{-1}\sigma )
\\ &=&
\chi (\bar g)^{-1} \bar X_{\bar a} \sigma =
\chi (\bar g)^{-1} (\Lie(\bar X_{\bar a})\sigma)(\bar a)
\eeann
since $X$ is $\Complex$-linear and
$(\sigma \circ \mu (\bar g))(\bar x) =
\sigma (\bar x \bar g) = \chi (\bar g)^{-1}\sigma (\bar x)$.

\quad \quad {\sl (b)} \quad
We have to see that, if $a \in P$ and $g \in \GL$, then
$$
(\Lie(X) \langle \sigma ,\sigma ' \rangle)(ag) =
\mid det\, g \mid^{-1} (\Lie(\bar X) \langle \sigma ,\sigma '
\rangle)(ag)
$$
In general, this condition holds for every section
$\alpha : M \to | N |$; in fact:
\beann
(\Lie(X)\alpha)(ag) & = & X_{ag} \alpha = (\Tan_a \mu (g) X_a) \alpha =
X_a (\alpha \circ \mu (g)) =X_a (\mid det\, g \mid^{-1} \alpha )
\\ & = &
\mid det\, g \mid^{-1} X_a \alpha = \mid det\, g \mid^{-1} \Lie(X)
\alpha \eeann by the same reasons than in the above paragraph.

\quad \quad {\sl (c)} \quad
Consider $a \in P$ and $\bar a \in \bar P$ such that
$\bar \rho (\bar a) = a$. We have that
\beann
(\Lie(X) \langle \sigma ,\sigma ' \rangle )(a)
& = &
X_a \langle \sigma ,\sigma ' \rangle =
X_a \bar \rho (\sigma \sigma '^*) =
 (\Tan_{\bar a} \bar \rho \bar X_{\bar a})
 (\bar \rho (\sigma \sigma '^*))=
\bar X_{\bar a}(\bar \rho (\sigma \sigma '^*) \circ \tilde\rho)
\\ & = &
\bar X_{\bar a}(\sigma \sigma '^*) =
(\bar X_{\bar a}\sigma ) \sigma '^*(\bar a) +
\sigma (\bar a) \bar X_{\bar a}\sigma '^* =
(\Lie(\bar X)\sigma )(\bar a)\sigma '^*(\bar a) +
\sigma (\bar a)(\Lie(\bar X)\sigma '^*)(\bar a)
\\ & = &
((\Lie(\bar X)\sigma )\sigma ')(\bar a) + (\sigma \Lie(\bar X)\sigma
'^*)(\bar a) =
[\bar \rho (\Lie(\bar X)  \sigma \sigma '^*) +
\bar \rho (\sigma \Lie(\bar X) \sigma '^*)](a)
\eeann
\qed
Observe that, in general, the equality
$$
\Lie(X) \langle \sigma ,\sigma ' \rangle =
\langle \Lie(\bar X)  \sigma , \sigma ' \rangle +
\langle \sigma , \Lie(\bar X) \sigma ' \rangle
$$
does not hold since $\Lie(\bar X) \sigma '^*\not= (\Lie(\bar X) \sigma
')^*$,
except if $X$ is real. In this case the right relation is
$$
\Lie(X) \langle \sigma ,\sigma ' \rangle =
\langle \Lie(\bar X)  \sigma , \sigma ' \rangle +
\bar \rho (\sigma \Lie(\bar X) \sigma '^*)
$$
and it is easy to prove that the function $\sigma \Lie(\bar X) \sigma
'^*$
is $\bar \rho$-projectable and, hence, the expression makes sense.
It can also be written as
$$
\Lie(X) \langle \sigma ,\sigma ' \rangle =
\langle \Lie(\bar X)  \sigma , \sigma ' \rangle +
\langle \sigma , \Lie(\bar X^*) \sigma ' \rangle
$$
where $\bar X^*$ is the conjugate vector field of $\bar X$.

In general, we are going to be interested in deriving
with respect to vector fields which are tangent to $M$.
In order to do it we must have a procedure for lifting them
to vector fields tangent to $P$ being $\GL$-invariant.
In addition, if we restrict to real vector fields, then
$$
\Lie(X) \langle \sigma ,\sigma ' \rangle =
\langle \Lie(\bar X)  \sigma , \sigma '^* \rangle +
\langle \sigma , \Lie(\bar X) \sigma ' \rangle
$$

\subsection{Some particular cases}

\subsubsection{Frame bundle of a differentiable manifold}

Let $M$ be a differentiable manifold with $dim\, M = n$ and
$p_* \colon P_* \to M$ the bundle of covariant complex references of
$M$.
$P_*$ is a principal fiber bundle with structural group $\GL$.
Suppose that $P_*$ admits an associated metalinear bundle
$\bar p_* \colon \bar P_* \to M$ and let $| N |$ be the associated
bundle.

\begin{prop}
$| N |$ is diffeomorphic to the bundle of complex densities of $M$.
\end{prop}
{\sl (Proof)} \quad
The bundle of complex densities  of $M$ is $Q(M) \otimes_{\Real}
\Complex$.
In order to prove that they are diffeomorphic, we must see that they
have the
same transition functions.

Observe that $Q(M)$ and $Q(M) \otimes_{\Real} \Complex$
have the same transition functions, since $\Complex$ is the trivial
bundle;
then its transition functions are the identity.
On the other hand, the transition functions of $Q(M)$
are the absolute value of those of $\Lambda^*(\Tan^*M)$,
therefore, the transition functions of $Q(M)$ are $| det ^t[\Tan
g_{\alpha \beta }]^{-1} |$.
The transition functions of $| N |$ are the same than the ones of $P_*$,

which are $(\Tan g_{\alpha \beta })^*$,
and act on $\Complex$ in order to obtain $| N |$ as
$\mid det\, (\Tan g_{\alpha \beta }^{-1})^*\mid$. Hence, they have the
same transition functions.
\qed

\begin{definition}
With the above considerations, the sections of the bundle $N^{1/2}$
are called {\rm uniformly complex sections}.
\end{definition}

According to the above results, if $\sigma , \sigma '\in \Gamma
(N^{1/2})$,
then $\langle \sigma ,\sigma ' \rangle$ is a complex density on $M$.

As an example, consider the bundle $p^* \colon P^* \to M$
of contravariant complex references of $M$,
and suppose that it admits an associated metalinear bundle
$\bar p^* \colon \bar P^* \to M$. In this case and,
in the same way, the sections of the associated bundle
$| N |$ can be identified with the complex multiples
of the absolute values of the skew symmetric contravariant tensors of
higher degree.

Now, let $X$ be a real vector field in $M$. It can be lifted to a vector

field in $\bar P^*$
in the following way:
Let $\tau_t$ be a local uniparametric group associated with $X$.
This group acts on the references of $X$ by means of the differential,
and so we obtain a vector field $\tilde X \in {\cal X}(P^*)$.
In addition, $\tilde X$ is $\GL$-invariant, that is, if
$\{ u_1,\ldots ,u_n \}$ is a complex reference of $\Tan_x M$ and $g \in
\GL$, then
$$
\Tan \tau_t ((u_1,\ldots ,u_n)g) = (\Tan \tau_t (u_1,\ldots ,u_n))g
$$
Since $\tilde X$ is $\GL$-invariant and $\bar P^*$ is a covering of
$P^*$,
the vector field $\tilde X$ lifts to a vector field $\bar X$ in $\bar
P^*$.
In order to construct $\bar X$ it suffices to do it in basic open sets
of a covering.
Observe that the so-obtained vector field $\bar X$ is $\ML$-invariant by

construction.

If $X$ is a complex vector field, it suffices to take its real and
imaginary parts
 in order to obtain the associated vector field
 $\bar X \in {\cal X}(\bar P^*)$
which is $\ML$-invariant.

\subsubsection{Frame bundle of a real polarization in a symplectic
manifold}

(See section 5.2 for notation).

Let $(M,\Omega )$ be a symplectic manifold with $dim\, M = 2n$,
and let ${\cal P}$ be a real polarization.
Let $p:P \to M$ be the frame bundle of ${\cal P}$
which is a principal bundle with structural group
$\GL$, and we suppose that it admits a
metalinear bundle $\bar P$ associated with $P$.
 Let $| N |$ and $N^{1/2}$ the line bundles associated with
 $P$ and $\bar P$.
The elements of $\Gamma (M,| N | )$ and $\Gamma (M,N^{1/2})$
are called {\it complex densities} and {\it complex half-forms}
on $M$ {\it associated with the polarization ${\cal P}$}.

Consider $X \in {\cal X}^{\Complex}(M)$ such that lets the polarization
${\cal P}$ invariant;
that is, $[X,{\cal P}] \subset {\cal P}$.
Denote by ${\cal X}^{\Complex}({\cal P})$ the set of these vector
fields.
If $\tau_t$ is a local uniparametric group associated with $X$ and
$\Tan \tau_t$ its lifting to $\Tan M^{\Complex}$, we can obtain a vector

field
$\tilde X \in {\cal X}(\Tan M^{\Complex})$ associated with the group
$\Tan \tau_t$.
Since $X$ lets invariant the polarization, the transformations $\Tan
\tau_t$
change references of ${\cal P}$ into references of ${\cal P}$,
then $\tilde X$ can be interpreted as a vector field tangent to $P$.
This is equivalent to saying that the trajectories of $\Tan \tau_t$ in
$\Tan M^{\Complex}$
are contained in $P$ or that $\tilde X$ is tangent to $P$.

 In addition, $\tilde X$ is $\GL$-invariant since
 if $p \in P$ and $A \in\GL$ we have
$\Tan_x \tau (pA) = (\Tan_x \tau (p))A$;
as a consequence of the associativity of the product of matrices.
Therefore, $\tilde X$ lifts to a vector field $\bar X$ which is
$\ML$-invariant,
using the local diffeomorphism $\bar\rho:\bar P \to P$
and the action of $\Zahl_2$ on $\ML$.

We can define the action of ${\cal X}^{\Complex}({\cal P})$ on the
densities and half-forms
associated with the polarization ${\cal P}$.
Let $X \in {\cal X}^{\Complex}({\cal P})$; and
$\alpha \in \Gamma (M,| N | )$ and $\beta \in \Gamma (M,N^{1/2})$.
Since $\alpha$ can be considered as a function $\alpha : P \to
\Complex$, we can write
$$
\Lie(X)\alpha:=\tilde X(\alpha )
$$
and $\Lie(X)\alpha \in \Gamma (M,| N | )$,
since $\tilde X$ is a $\GL$-invariant vector field.
In the same way, since $\beta$ can be interpreted as a function
$\beta :\bar P \to \Complex$, we can write
$$
\Lie(X)\beta :=\bar X(\beta )
$$
and $\Lie(X)\beta \in \Gamma (M,N^{1/2})$,
since $\bar X$ is a $\ML$-invariant vector field.

Now, consider $X \in {\cal X}^{\Complex}(M)$, its extension $\tilde X$
to $P$ and
a section $\alpha\in\Gamma (M,| N | )$.
Suppose that it is a Hamiltonian vector field (according to the
proposition \ref{blh}),
and denote it by $X_f$.
If $X_g$ is another Hamiltonian vector field in $D$
\footnote{
Remember that, for a real polarization, $D=E=\P\cap\Tan M$,
},
we have that
$\inn([X_f,X_g])\Omega =0$;
and taking into account that $\Omega$ in non-degenerate
we conclude that $[X_f,X_g]=0$.
Therefore, $X_g$ is invariant under the action
of the uniparametric group $\tau_t$ generated by $X_f$;
that is, $\Tan\tau_t(X_{g_x})=X_{f_{\tau_t}}(x)$.
Then, let $\{ X_{f_1},\ldots ,X_{f_n}\}$ be
a basis of $\vf ({\cal P})$ in an open set $U \subset M$
made of Hamiltonian vector fields.
We can consider the following functions in $U$
associated with the section $\alpha$
$$
F_{\alpha}(x) := (\Lie(X)\alpha )(X_{f_1}(x),\ldots ,X_{f_n}(x))
\quad , \quad
G_{\alpha}(x) := [X(\alpha(X_{f_1},\ldots ,X_{f_n}))](x)
$$
for a given $X$ in $D$. Therefore:

\begin{prop}
If $X \equiv X_f$ is a Hamiltonian vector field then $F_{\alpha} =
G_{\alpha}$
\label{fg}
\end{prop}
{\sl (Proof)} \quad
If $\tilde X$ is the canonical lift of $X$ to $P$, we have that
\beann
F_{\alpha}(x) &:=&
(\Lie(X)\alpha )(X_{f_1}(x),\ldots ,X_{f_n}(x)) =
\\ &=&
 \lim_{t \to 0}(1/t)
 (\alpha (\Tan_x\tau_t X_{f_1}(x),\ldots
,\Tan_x\tau_t X_{f_n}(x)) -\alpha(X_{f_1}(x),\ldots ,X_{f_n}(x)))
\\ &=&
\lim_{t \to 0}(1/t)(\alpha (X_{f_1}(\tau_t(x)),\ldots
,X_{f_n}(\tau_t(x)))
-\alpha(X_{f_1}(x),\ldots ,X_{f_n}(x)))
\eeann
since the vector fields are invariant under the action of $\tau_t$. On
the other hand,
\beann
G_{\alpha}(x) &:=&
[X(\alpha(X_{f_1},\ldots ,X_{f_n}))](x)
\\ &=&
\lim_{t \to 0}(1/t)(\alpha (X_{f_1},\ldots ,X_{f_n}))(\tau_t(x))
-\alpha (X_{f_1},\ldots ,X_{f_n})(x)
\\ &=&
\lim_{t \to 0}(1/t)(\alpha (X_{f_1}(\tau_t(x)),\ldots
,X_{f_n}(\tau_t(x)))
-\alpha(X_{f_1}(x),\ldots ,X_{f_n}(x)))
\eeann
\qed

Taking into account the above considerations we have:

\begin{prop}
Consider $x \in U\subset M$. Let $V$ be the integral manifold of $D$
passing through~$x$; and the section
$$
\begin{array}{ccccc}
\sigma & \colon & U\cap V & \longrightarrow & P
\\
& & x & \mapsto & (X_{f_1},\ldots ,X_{f_n})
\end{array}
$$
where $\{ X_{f_1},\ldots ,X_{f_n}\}$ is a basis of $D$ made of
Hamiltonian vector fields.
Then, $\tilde X_f$ is tangent to the image of~$\sigma$, for every
$f\in\Cinfty (M)$.
\end{prop}
{\sl (Proof)} \quad
If $\tau$ is the local uniparametric group associated with $X_f$,
then $\Tan\tau$ lets invariant the vector fields $X_{f_i}$,
hence the integral curves of $\tilde X$
are contained in $\sigma(U\cap V)$
and therefore it is tangent to the image of $\sigma$.
\qed

Let $\pi \colon M \to {\cal D}$ be the projection defined by the
integral manifolds
of the distribution $D$. Let $\{ \bar \alpha \}$ be
the set of complex densities of ${\cal D}$. We are going to define a map

$\varphi$
from $\{ \bar \alpha \}$ to $\Gamma (M, | N |)$
which allows us to integrate sections of $| N |$.
Consider $x \in M$ and $X_x \in D_x$, then $\inn(X_x)\Omega \colon
\Tan_xM\to \Complex$
is $\Real$-linear and vanishes on $D_x$, hence it passes to the quotient

$\Tan_xM/D_x$ which is canonically isomorphic to $\Tan_{\pi (x)}{\cal
D}$.
Let $\Phi(X_x)$ the map induced in $\Tan_{\pi (x)}{\cal D}$.
Then, we have a $\Real$-linear map
$$
\begin{array}{ccccc}
\Phi & \colon & D_x & \to & Hom_{\Real}(\Tan_{\pi (x)}{\cal D},\Complex
)
\\
& & X_x & \mapsto & \Phi (X_x)
\end{array}
$$
which is an isomorphism because $\Omega$ is a symplectic form.
If $\{ X_x^1,\ldots ,X_x^n \}$ is a $\Complex$-basis of $D_x$, then
$\{ \Phi (X_x^1),\ldots ,\Phi (X_x^n) \}$ is a basis of
$(\Tan^*_{\pi (x)}{\cal D})^{\Complex}$.
 Let $\{ \bar X_x^1,\ldots ,\bar X_x^n \}$
be the dual $\Complex$-basis. Then:

\begin{prop}
For every complex density $\bar \alpha\in{\cal D}$,
let $\{ X_x^1,\ldots ,X_x^n \}$ be a reference of $D_x$ (that is a point

of $P$).
Define a map $\varphi\colon \{\bar\alpha\} \to \Gamma(M,| N |)$,
such that $\alpha :=\varphi (\bar \alpha)$, as follows
$$
\alpha (X_x^1,\ldots ,X_x^n) :=\bar \alpha (\bar X_x^1,\ldots ,\bar
X_x^n)
$$
Therefore
\begin{description}
\item[{\rm (i)}] \ \
$\varphi$ is well defined.
\item[{\rm (ii)}] \
$\varphi$ is injective.
\item[{\rm (iii)}]
$Im \varphi =\{ \alpha \colon P \to \Complex \ \mid \ \alpha \in
\Gamma(M,| N |) \ , \ \Lie(X)\alpha = 0 \ , \ \mbox{for every X in
D (loc. Hamiltonian)} \}$
\end{description}
\end{prop}
{\sl (Proof)} \quad
{\sl (i)} \ \
We have to see that the so-defined function satisfies that
$$
\alpha (Rg) = g^{-1} \alpha (R) =| det\, g |^{-1} \alpha (R)
$$
where $R \in P$ and $g \in \GL$. Let $R=(v^1,\ldots ,v^n)$ be
a reference of $D_x$ and $g=(g_{ij})$ a matrix of $\GL$; then
$$
Rg = (\sum g_{i1}v^i,\ldots ,\sum g_{in}v^i) = (u^1,\ldots ,u^n)
$$
therefore
$$
\Phi (u^j) = \Phi ( \sum g_{ij}vi) = \sum g_{ij} \Phi (v^i)
$$
and hence $\Phi (Rg) = \Phi (R) g$.
Now, we are going to see the relation between the dual basis of $\Phi
(R)$ and
$\Phi (R) g$. Denoting by $\overline {\Phi (R)}$
the dual basis of $\Phi(R)$, we have that $^t\overline {\Phi (R)} \cdot
\Phi (R) = I$, and hence
$^t\overline {\Phi (R) g} \cdot \Phi (R) g = I$.
But $\overline {\Phi (R) g} = \overline {\Phi (R)} h$,
for some $h \in \GL$, then $^th ^t\overline {\Phi (R)} \cdot \Phi (R) g
= I$;
and thus $h = ^tg^{-1}$, therefore we obtain
$\overline {\Phi (R) g} = \overline {\Phi (R)} ^tg^{-1}$, so we have
that
$$
\alpha (Rg)= \bar \alpha (\overline {\Phi (Rg)}) =
\bar \alpha (\overline {\Phi (R) g}) =
\bar \alpha (\overline {\Phi (R)}  ^tg^{-1}) =
| det \ ^tg^{-1} | \bar \alpha (\overline {\Phi (R)}) =
\mid {\rm det}\, g\mid^{-1} \alpha (R)
$$

\qquad {\sl (ii)} \
If $\varphi (\bar \alpha )=0$ then $\bar \alpha$ vanishes
in a reference of $(\Tan M)^{\Complex}$, thus it is null.

\qquad {\sl (iii)} \
Let $\alpha \in Im \varphi$ with $\alpha = \varphi (\bar \alpha )$,
we have to see that $\Lie (X)\alpha = 0$
for every locally Hamiltonian vector field $X$ in $D$
(it suffices to prove it for the vector fields in $D$).
Observe that it suffices to see that, for all $x \in M$, it holds
$(\Lie(X)\alpha) (X_x^1, \ldots ,X_x^n)=0$ for a reference of $D_x$.
Let $\{X_x^1,\ldots ,X_x^n \}$ be a reference of $D$ and
$\{X_{f_1},\ldots ,X_{f_n} \}$
a basis made of Hamiltonian vector fields in an open set $U \subset M$
with $x \in U$,
which prolongs the above reference. Let $\tilde X$ be the extension of
$X$ to $P$.
If $V_x$ is the integral manifold of $D$
passing through $x$, $\tilde X$ is tangent to $\sigma (U\cap V_x)$,
being
$\sigma (x) = (X_{f_1}(x),\ldots ,X_{f_n}(x))$, therefore, in order to
see that
$\Lie(X)(X_x^1,\ldots ,X_x^n)=0$, it suffices to see that $\alpha$
is constant on $\sigma (U\cap V_x)$. In fact, consider
$\bar x = \pi (x) \in {\cal D}$ and let $\bar U \subset {\cal D}$ be an
open set
with $\pi (U) \subset \bar U$. Since $X_{f_j}$ is in $D$,
the functions $f_j$ are constant  along the integral submanifolds of
$D$.
In fact, $\inn(X_{f_j})\Omega = \d f_j$ and $\d f_j(X_{f_i})=0 \ , \
\forall i$;
then $\d f_j = 0$, when restricted to the integral submanifolds of $D$.
Therefore, there exist functions $\bar f_j \colon \bar U \to \Complex$
such that $\pi^* \bar f_j = f_j$. Now, consider $y \in U \cap V_x$,
hence $\pi (y) = \bar x$ and we have that
$$
\Phi (X_{f_j}(y)) = \d f_j(x)\mid_{\Tan_{\bar x}{\cal D}} =
\d \bar f_j(\bar x)
$$
therefore the dual basis of $\{ \Phi (X_{f_1}(y)), \ldots ,\Phi
(X_{f_n}(y)) \}$
is \dst\left\{ \derpar{}{\bar f_1}\Big\vert_{\bar x},\ldots
,\derpar{}{\bar f_n}\Big\vert_{\bar x}\right\} \)
and thus
$$
\alpha (X_{f_1}(y), \ldots ,X_{f_n}(y))=
\bar \alpha (\bar X_{f_1}(\bar x), \ldots ,\bar X_{f_n}(\bar x)) =
\bar \alpha \left(\derpar{}{\bar f_1}\Big\vert_{\bar x},\ldots ,
\derpar{}{\bar f_n}\Big\vert_{\bar x}\right)
$$
which is constant when $y$ varies in $U \cap V_x$, as we wanted to
prove.

\qquad \qquad \
Conversely, consider $\alpha \in \Gamma(M,| N |)$, $\alpha \colon P \to
\Complex$,
satisfying that $\Lie(X)\alpha = 0$
for every locally Hamiltonian vector field $X$ in $D$.
We are going to construct a density $\bar \alpha$  on ${\cal D}$ such
that
$\alpha = \varphi (\bar \alpha)$. In order to do this,
it suffices to give its action on a local basis in ${\cal D}$.
Thus, consider $\bar x \in {\cal D}$ and let $\{ \bar f_1,\ldots , \bar
f_n \}$
be local coordinates in ${\cal D}$. Put $f_j = \pi^*\bar f_j$
which are defined in $U = \pi^{-1}(\bar U)$ and let
$X_{f_1}, \ldots ,X_{f_n}$ be their associated Hamiltonian vector
fields;
then they are a local basis of $D$ made of Hamiltonian vector fields
and,
for every $y \in V = \pi^{-1} (\bar x)$ ($V$ being an integral
submanifold of $D$),
$\alpha (X_{f_1}(y), \ldots ,X_{f_n}(y))$ is constant. In fact,
since $V$ is connected, it suffices to see that, if $X_x \in \Tan_xV$
then $X_x(\alpha (X_{f_1}, \ldots ,X_{f_n}))=0$.
Let $X \in {\cal X}(M)$ be a Hamiltonian vector field
which prolongs $X_x$ locally; according to proposition \ref{fg}
we have
$$
0 = (\Lie(X)\alpha)_x(X_{f_1}, \ldots ,X_{f_n}) =
X_x(\alpha(X_{f_1}, \ldots ,X_{f_n}))
$$
hence $\alpha (X_{f_1}, \ldots ,X_{f_n})$ is constant on $V$.

Now we  can define the density $\bar \alpha$ in ${\cal D}$ given
by $$ \bar \alpha \left(\derpar{}{\bar f_1}\Big\vert_{\bar
x},\ldots , \derpar{}{\bar f_n}\Big\vert_{\bar x}\right) := \alpha
(X_{f_1}, \ldots ,X_{f_n}) \Big\vert_V $$ (since we can take any
point of $V$). In this way we have defined locally $\alpha$. Now
we can define it globally. Let $\{\bar g_1, \ldots ,\bar g_n \}$
be another local coordinate system in a neighborhood $\bar V$ of
$\bar x$. Then $$ \bar \alpha\left(\derpar{}{\bar
g_1}\Big\vert_{\bar x},\ldots , \derpar{}{\bar g_n}\Big\vert_{\bar
x}\right) = \alpha (X_{g_1}, \ldots ,X_{g_n}) \mid_V $$ If $\bar g
= \psi (\bar f)$ and $(J\psi )$ denotes the Jacobian matrix of
$\psi$ (the change of coordinates), then, if $\alpha$ must be a
density, it must verify that $$ \left(\bar
\alpha\left(\derpar{}{\bar f_1}\Big\vert_{\bar x},\ldots ,
\derpar{}{\bar f_n}\Big\vert_{\bar x}\right)\right)_x = |det\,
J\psi| \left(\bar \alpha\left(\derpar{}{\bar g_1}\Big\vert_{\bar
x},\ldots , \derpar{}{\bar g_n}\Big\vert_{\bar x}\right)\right)_x
$$ since \dst\derpar{}{\bar f} = \derpar{}{\bar g} (J\psi )\) . In
fact, we have that $$ \d \bar g = \d \bar f (J\psi ) \Rightarrow
X_{g_i} = X_{f_i} (J\psi) $$ since the map $\bar f\mapsto X_f$ is
$\Complex$-linear. Therefore \beann \bar
\alpha\left(\derpar{}{\bar g_1},\ldots ,\derpar{}{\bar
g_n}\right)\Big\vert_x &=& \alpha (X_{g_1}, \ldots ,X_{g_n})
\mid_V= \alpha ((X_{f_1}, \ldots ,X_{f_n})(J\psi )) \mid_V
\\ &=&
|detJ\psi |^{-1}\alpha (X_{f_1}, \ldots ,X_{f_n}) \mid_V=
|detJ\psi |^{-1} \bar \alpha\left(\derpar{}{f_1}, \ldots
,\derpar{}{f_n}\right)\Big\vert_{\bar x}
\eeann
as we wanted to prove. Hence $\bar \alpha \in \{\bar \alpha \}$.
\qed

In relation with densities over ${\cal D}$ and $N^{1/2}$ we have:

\begin{prop}
Let $\sigma$, $\sigma '$ be two sections of $N^{1/2}$
such that $\Lie(X)\sigma = 0$, $\Lie(X)\sigma ' = 0$,
for every locally Hamiltonian vector field $X$ in $D$.
Then $\langle \sigma ,\sigma ' \rangle$ is a complex density on ${\cal
D}$
(that is, $\langle \sigma ,\sigma ' \rangle$ belongs to $Im \phi$).
\end{prop}
{\sl (Proof)} \quad
If $X$ is locally Hamiltonian then
$$
\Lie(X)\langle \sigma ,\sigma ' \rangle =
\langle \Lie(X)\sigma ,\sigma ' \rangle +
\langle \sigma ,\Lie(X)\sigma ' \rangle = 0
$$
and the result follows as a consequence of the above proposition.
\qed

\subsection{Quantization: Space of quantum states. Operators}

Let $(M,\Omega )$ be a symplectic manifold,
$[\frac{\Omega}{2\pi\hbar}] \in H^2(M,\Real )$ being an integer class.
Let $\pi \colon L \to M$ be a complex line bundle endowed with a
hermitian connection $\nabla$ with curvature $2$-form
$\frac{{\bf \Omega}}{2\pi\hbar}$.
Suppose that $M$ is also endowed with a polarization ${\cal P}$ and that

the
frame bundle of ${\cal P}$, $p \colon P \to M$,
admits a metalinear bundle $\bar p \colon \bar P \to M$.
Let $N^{1/2}$ and $|N|$ be the line bundles of half-forms and densities
associated with $\bar P$.

We are going to define the space of quantum states in the following way:

We consider the line bundle $p \colon L \otimes N^{1/2} \to M$
and, in the set of sections $\Gamma (M,L \otimes N^{1/2})$, we
denote by $H^{\cal P}$ the $\Complex$-vector space generated by
the sections $s \otimes \sigma$; where $s \in \Gamma (M,L)$ and
$\sigma \in \Gamma (M,N^{1/2})$ with compact support and such that
$\nabla_Xs = 0$ and $\Lie(X)\sigma = 0$, for all locally
Hamiltonian vector field $X \in {\cal P} \cup \bar {\cal P}$; that
is, the sections $s$ and $\sigma$ are invariant by the vector
fields of ${\cal P} \cup \bar {\cal P}$. Observe that, if $s
\otimes \sigma$ and $s' \otimes \sigma '$ are elements of $H^{\cal
P}$, then $\langle s,s' \rangle$ is a projectable function on
${\cal D} = M/D$. On the other hand, $\langle \sigma ,\sigma^*
\rangle$ is identified with a complex density on ${\cal D}$.
Therefore, $\langle s,s' \rangle \langle \sigma ,\sigma^* \rangle
\in \{ \bar \alpha \}_{\Complex}$, (where  $\{ \bar \alpha
\}_{\Complex}$ denotes the set of complex densities on ${\cal D}$
with compact support). This allows us to define a pre-Hilbert
product in the following way: \beq (s \otimes \sigma ,s' \otimes
\sigma ') := \int_{{\cal D}} \langle s,s' \rangle \langle \sigma
,\sigma ' \rangle \label{herprod} \eeq which can be extended
linearly to $H^{\cal P}$. Then:

\begin{require}
In the geometric quantization programme, the completion ${\cal H}^{\cal
P}$
of the set $H^{\cal P}$, endowed with the hermitian product
(\ref{herprod}), is the intrinsic Hilbert space ${\cal H}_Q$
and the projective space ${\bf P}{\cal H}^{\cal P}$
is the space of quantum states ${\bf P}{\cal H}_Q$ of the definition
\ref{fquan}.
\end{require}

Now, as it was stated in Section 2, we must represent the Poisson
algebra of
$(M,\Omega )$ in ${\cal H}^{\cal P}$.
Nevertheless, according to the proposition \ref{cnspo},
it is not possible to represent
all its elements but only those belonging to
$A^{\cal P} \equiv\{ f \in \Cinfty (M) \ \mid \ [X_f,\cal P] \subset
\cal P \}$
(the set of observables whose Hamiltonian vector fields preserve the
polarization).
Therefore:

\begin{require}
{\bf (and Definition)}
In the geometric quantization programme,
the quantum operator associated with the classical observable
$f \in A^{\cal P}$, is the operator $^{\cal P}O_f$ defined in ${\cal
H}^{\cal P}$,
with values in $\Gamma (M,L \otimes N^{1/2})$, which is defined by
$$
^{\cal P}O_f (s \otimes \sigma) :=
(O_f s)\otimes \sigma + i s \otimes \Lie(X_f)\sigma
$$
where $O_f$ is the prequantization operator defined by (\ref{oper}).
\end{require}

And then we can prove:

\begin{teor}
The map
$$
\begin{array}{ccccc}
{\cal Q} & \colon & A^{\cal P} & \to &
\mbox{{\rm self-adjoint operators in} ${\cal H}^{\cal P}$}
\\
& & f & \mapsto & ^{\cal P}O_f
\end{array}
$$
is well defined and satisfies the conditions (b(iii)) and (b(iv)) of the

definition \ref{fquan}.
\end{teor}
{\sl (Proof} \quad
In order to see that ${\cal Q}$ is well defined we have to prove that
\begin{enumerate}
\item
${\cal Q}(f)(s \otimes \sigma) \in {\cal H}^{\cal P}$,
$\forall s \otimes \sigma\in H^{\cal P}$,
\item
${\cal Q}(f)$ is self-adjoint.
\end{enumerate}
Therefore, let $U$ be an open set of $M$ and $X_g$ in ${\cal P}
\cap \bar {\cal P}$ a locally Hamiltonian vector field for $g \in
\Cinfty (U)$;
 since ${\cal Q}(f)(s \otimes \sigma)
 =O_f s \otimes\sigma + i s \otimes
\Lie(X_f)\sigma$, we can easily prove that \beann \nabla_{X_g}(O_f
s) = \nabla_{X_g}(\nabla_{Xf}s-2 \pi i f s) &=& 0
\\
\Lie(X_g)\Lie(X_f)\sigma &=& 0
\eeann
and the part (1) follows. The proof of the part (2) is a simple matter
of calculation.

Finally the verification of the conditions (b(iii)) and (b(iv)) of the
definition \ref{fquan}
is immediate.
\qed

\subsection{Example: Simms quantization of the harmonic oscillator}

The physical system we are considering is specified by the following
features:
\begin{itemize}
\item
Phase space:
$M = \{ (q,p) \in \Real^2-\{ 0 \} \} \cong \Complex^*$.
\item
Symplectic form:
$\Omega = \d p \wedge \d q$,

Symplectic potential:
$\theta = 1/2 (p \d q - q \d p)$.
\item
Hamiltonian function:
$H = \frac{1}{2} (p^2 + q^2)$.
\end{itemize}
We take as line bundle the trivial bundle $L=M \times \Complex$,
with the usual metric $\h ((x,z),(x,z'))=zz'$
and the hermitian connection given by
$$
\nabla_Xf = Xf+2\pi i \theta(X)f
$$
where $f \colon M \to \Complex$ is a section of $L$.
Consider in $M$ the real polarization ${\cal P}$ determined by the
circumferences
with center at the origin of $\Real^2$. In this case, ${\cal P}$
admits a global basis given by the vector field
$$
X_H = -p\derpar{}{q} +q\derpar{}{p} \equiv \derpar{}{\Theta}
$$
which is just the Hamiltonian vector field associated with $H$.
Therefore, the ${\bf GL}(1,\Complex )$-principal fiber bundle, $P$,
of references of ${\cal P}$ is a trivial bundle and a global
trivialization is given by
$$
\begin{array}{ccc}
M \times {\bf GL}(1,\Complex ) & \rightarrow & P
\\
(x,z) & \mapsto & z X_H(x)
\end{array}
$$
which is a diffeomorphism. (Observe that $(x,1) \mapsto X_H(x)$).
On the other hand, according to the definitions we have that
$H^2(M,\Zahl_2)=0$ and $H^1(M,\Zahl_2)=\Zahl_2$,
(both because $S^1$ is a deformation retract of $M$),
and hence $P$ admits two equivalence classes of
associated metalinear bundles.
We are going to describe and quantify them.

{\sl First case \/}: \underline{Trivial metalinear bundle}.

In this case we have $\bar P_1 = M \times {\bf ML}(1,\Complex )$,
and the covering map $\bar \rho \colon P_1 \to P$
is given by $\bar \rho (x,\bar g) := ((x,\rho (\bar g))$,
where $\rho \colon {\bf ML}(1,\Complex ) \to {\bf GL}(1,\Complex )$
is the natural covering.

In order to identify the sections of the bundle
$L \otimes N^{1/2} \to M$ you can observe that, if
$s_0 \colon M \to M \times L$ is the unit section and
$\sigma_0 \in \Gamma (M,N^{1/2})$ satisfies that
$\sigma_0(x,1)=1$, (we consider it as a function
$\sigma_0 \colon \bar P_1 \to \Complex$),
then the sections of $L \otimes N^{1/2}$ have the form
$fs_0 \otimes \sigma_0$, where $f$ is a function of $M$ in $\Complex$.

For constructing $H^{\cal P}$ we have to consider the sections
$fs_0 \otimes \sigma_0$ such that
$\Lie(X_H)\sigma_0=0$, $\nabla_{X_H}(fs_0)=0$,
since the polarization ${\cal P}$ is real and $X_H$ generates ${\cal P}$

globally.
The first condition holds trivially and, for the second one, we have
that
$$
0=\nabla_{X_H}fs_0 =(X_Hf)s_0 + 2\pi i f \theta (X_H)s_0=
\derpar{f}{\Theta}s_0 + 2\pi i f(-\frac{1}{2}r^2)s_0 =
(\derpar{f}{\Theta} - \pi i r^2 f)s_0
$$
and the condition for $f$ is
\beq
\derpar{f}{\Theta} - \pi i r^2 f = 0
\label{eqcla}
\eeq
besides of having compact support. This equation has no solution.
In fact: its general solution should be $f(r,\Theta) = C(r)e^{\pi i r^2
\Theta}$
but, it must satisfy that $f(r,\Theta) = f(r,\Theta+2\pi n)$, $n \in
\Zahl$,
and hence $e^{\pi i r^2 2 \pi n} = 1$,
that is $\pi r^2 \in \Zahl$, which is absurd if $f$ depends
differentiabily on $r$.
Therefore $H^{\cal P}=0$ and hence this system cannot be quantified.

Nevertheless, instead of ordinary functions we can take distribution
sections of $L$.
This consists of taking sections of the form $fs_0$ where $f$ is a
distribution in $M$.
In this way, we will arrive now to the same equation (\ref{eqcla}) which

has solution.
In fact, we can prove a solution of the form
$f(r,\Theta)=g(r)e^{ik\Theta}$, with $k \in \Real$ and $g(r)$
being a distribution in $\Real^+$. We have that
$$
ikg(r)e^{ik\Theta}-\pi ir^2g(r)e^{ik\Theta}=0
$$
therefore $(k-\pi r^2)g(r)=0$.
Thus $g(r)$ has to be a multiple of the distribution $\delta
(r-\sqrt{\frac{k}{\pi}})$
and, in addition, $k$ has to be a non-negative integer since
$f(r,\Theta ) = f(r,\Theta +2\pi h)$, $h \in \Zahl$.
Therefore we obtain that a set of non-vanishing vectors of
${\cal H}^{\cal P}$ is given by
$e^{ik\Theta}\delta (r-\sqrt{\frac{k}{\pi}})s_0 \times \sigma_0$,
with $k\in\{ 0\}\cup\Zahl$.
In this space we can apply the quantization procedure previously
described
for the case of functions. The result is that these vectors are
eigenvectors of the operator $O_H$ corresponding to the Hamiltonian
function.
Moreover, since $X_H$ belongs to the polarization,
as a consequence of the corollary of the quantization theorem,
we have that $O_H$ consists in multiplying by the function $2\pi H =
2\pi r^2$.

{\sl Second case \/}: \underline{ Non trivial metalinear bundle}.

In this case, in order to construct the bundle
$\bar p \colon \bar P_2 \to P$, we consider the set
$\Real^+ \times \Real \times {\bf ML}(1,\Complex )$.
We take on it the equivalence relation defined by
$$
(r,\Theta ,\lambda ) \sim (r,\Theta +2\pi h,\varepsilon^h\lambda )
\ , \ h \in \Zahl
$$
where $\varepsilon$ is the non trivial element
of the ker of the morphism
$\rho \colon {\bf ML}(1,\Complex ) \to {\bf GL}(1,\Complex )$.
Let $[r,\Theta ,\lambda ]$ be the equivalence class of
$(r,\Theta ,\lambda )$ and let $\bar P_2$ the quotient set.
We have the natural map
$$
\begin{array}{ccccc}
\bar \rho & \colon & \bar P_2 & \rightarrow & P
\\
& & [r,\Theta ,\lambda ] & \mapsto & (r,e^{i\Theta},p(\lambda ))
\end{array}
$$
It is clear that $\bar P_2$ is not
the trivial bundle $\bar P_1$,
since the points we have identified when we construct
$\bar P_2$ are not the same as in $\bar P_1$.

In order to construct a trivial system in $\bar P_2$,
observe that $M$ is the quotient of $\Real^+ \times \Real$,
by the same equivalence relation,
but restricted to this set, that is,
$$
(r,\Theta ) \sim (r,\Theta +2\pi h) \ , \ h \in \Zahl.
$$
Now, we take the following open sets in $M$:
$$
U_1 \equiv \{re^{i\Theta}, \ r \in \Real^+, \ \Theta \in (0,2\pi ) \}
\quad ; \quad
U_2 \equiv \{re^{i\Theta}, \ r \in \Real^+, \ \Theta \in (-\pi,\pi ) \}
$$
and, from the natural projection
$\bar p \colon \bar P_2 \to M$, the following sections:
$$
\begin{array}{ccccccccccc}
\bar s_1 & \colon & U_1 & \to & \bar P_2
& \qquad ; \quad &
\bar s_2 & \colon & U_2 & \to & \bar P_2
\\
& & re^{i\Theta} & \mapsto & [r, \Theta ,1]
& \qquad &
& & re^{i\Theta} & \mapsto & [r, \Theta ,1]
\end{array}
$$
Then, $\{U_i,\bar s_i \}$ is a trivial system of the bundle
$\bar p \colon \bar P_2 \to M$. The transition function
$c_{12} = \bar s_2 \circ \bar s_1^{-1}$ is
\beann
c_{12}([r,\Theta ,1]) =
(\bar s_2 \circ \bar s_1^{-1})[r,\Theta ,1] &=&
\left\{ \begin{array}{cc}
\bar s_2(re^{i\Theta}), & \mbox{$\Theta \in (0,\pi )$}
\\
\bar s_2(re^{i(\Theta -2\pi )}), & \mbox{$\Theta \in (\pi ,2\pi )$}
\end{array}
\right.
\\ &=&
\left\{ \begin{array}{cc}
[r,\Theta ,1], & \mbox{$\Theta \in (0,\pi )$}
\\
{}[r,\Theta -2\pi ,1]=[r,\Theta ,\varepsilon ],
& \mbox{$\Theta \in (\pi ,2\pi )$}
\end{array}
\right.
\eeann
(remember that $U_{12}=(0,\pi )\cup (\pi ,2\pi)$).
Let $\sigma \colon M \to \bar P_2$ be a section of $\bar p$,
then $\bar p \circ \sigma \colon M \to P$ is a section of $p$.
For $\bar s_1$ and $\bar s_2$ we have
\beann
(p \circ \bar s_1)(re^{i\Theta}) =\bar p([r,\Theta ,1])
=(re^{i\Theta},1)
\quad ,\quad {\rm in}\ U_1
\\
(p \circ \bar s_2)(re^{i\Theta}) = \bar p([r,\Theta ,1])
=(re^{i\Theta},1)
\quad ,\quad {\rm in}\ U_2
\eeann
Observe that, according to the global trivialization defined in $P$,
$X_H(re^{i\Theta})$ corresponds to $(re^{i\Theta},1)$,
and then both sections $p \circ \bar s_1$ and $p \circ \bar s_2$
give $X_H$ in their domains.

In order to quantize the non trivial bundle, we start describing the
bundle
$N^{1/2} \to M$. In $U_1$ and $U_2$ we take the sections
$\sigma_i \colon U_i \to N^{1/2}$ as follows:
we define $\sigma_i$ as a function
 $\sigma_i \colon \bar p^{-1}(U_i) \to\Complex$
and we work at a point (taking into account that the invariance
condition
must be verified). If $x \in U_1$, we define $\sigma_1(x) \in N_x^{1/2}$

such that $\sigma_1(x)([\bar s_1(x)])=1$, and if $x \in U_2$,
we define $\sigma_2(x)$ as $\sigma_2(x)([\bar s_2(x)])=1$.
The relation between both sections is as follows:
if $\Theta \in (0,2\pi )$, then $\bar s_1(re^{i\Theta}) = \bar
s_2(re^{i\Theta})$, therefore
$\sigma_1(re^{i\Theta}) = \sigma_2(re^{i\Theta})$.
If $\Theta \in (\pi ,2\pi )$, then
$\bar s_1(re^{i\Theta}) = \bar s_2(re^{i\Theta})\varepsilon$, and hence
$$
\sigma_1((\bar s_1(re^{i\Theta})) = \sigma_1(\bar s_2(re^{i\Theta}))
=-\sigma_2(\bar s_2(re^{i\Theta}))
$$
and, in this case, $\sigma_1 = -\sigma_2$.

The sections of $L \otimes N^{1/2}$ are studied in the following way:
let $s_0 \colon M \to L$ be the unit section. The elements of
$H^{\cal P}$ are of the form $f_1 s_0 \otimes \sigma_1$ in $U_1$,
and $f_2 s_0 \otimes \sigma_2$ in $U_2$;
and since $X_H$ is a global generator for ${\cal P}$, we have that
$$
\begin{array}{ccccc}
\Lie (X_H)\sigma_1 = 0 , & {\rm in}\ U_1 &\quad ; \quad &
\Lie (X_H)\sigma_2 = 0, & {\rm in}\ U_2
\\
\nabla_{X_H}(f_1s_0) = 0, & {\rm in}\ U_1 &\quad ; \quad &
\nabla_{X_H}(f_2s_0) = 0, & {\rm in}\ U_2
\end{array}
$$
Now we can see that the conditions on $\sigma_i$ are automatically
satisfied.
In fact, consider $\bar x \in \bar P_2$ with $\bar \rho(\bar x) = x$,
then there exists $\lambda \in {\bf ML}(1,\Complex )$
with $\bar x = \bar s_1(x)\lambda$, therefore
$$
(\Lie(X_H)\sigma_1)(\bar x) =
(\Lie(X_H)\sigma_1)(\bar s_1(x)\lambda ) =
\lambda^{-1}(\Lie(X_H)\sigma )(\bar s_1(x))
$$
where the invariance condition of $\bar X_H$ (the lifting of $X_H$ to
$\bar P_2$)
is taken into account. Now, if $\tau_t$ is the uniparametric group of
$\bar X_H$, we have that
$$
(\Lie(X_H)\sigma )(\bar s_1(x)) =
\lim_{t \to 0}\frac{1}{t}
(\sigma(\bar \tau_t(\bar s_1(x)))-
\sigma(\bar s_1(x)))
$$
and using the corresponding local diffeomorphisms we have that
\beann
\bar \tau_t(\bar s_1(x))
&=&
\bar \rho^{-1}(\Tan \tau_t \bar p(\bar s_1(x))=\bar \rho^{-1}(\Tan
\tau_t X_H(x)))
\\ &=&
\bar \rho^{-1}(X_H(\tau_t(x)))= \bar \rho^{-1}(\bar p \circ \bar
s_1(\tau_t(x))) = \bar s_1(\tau_t(x)) \eeann where we have taken
into account that $\bar p \circ s_1 = X_H$; and hence $$
(\Lie(X_H)\sigma_1)(\bar s_1(x))= \lim_{t \to 0}\frac{1}{t}
((\sigma_1 \circ \bar s_1)(\tau_t(x))-(\sigma_1 \circ \bar
s_1)(x))= (X_H(\sigma_1 \circ \bar s_1))(x) = 0 $$ since $\sigma_1
\circ \bar s_1$ is the unit constant function. In the same way we
obtain that $\Lie(X_H)\sigma_2 = 0$ in $U_2$.

The conditions on $f_is_0$ give again the equation (\ref{eqcla}),
therefore $H^{\cal P}=0$ unless we work with distribution sections as in

the above case.

For other examples see \cite{Sn-80}, \cite{Tu-85}, \cite{Wo-80}.

\section{Some ideas on geometric quantization of constrained systems}

The geometric quantization programme  which we have explained along the
above sections
deals with {\it regular} dynamical systems; that is,
systems whose classical phase space is represented
(totally or partially) by a symplectic manifold.
Nevertheless, Dirac and Bergmann started early the study of
{\it singular} (or {\it constrained}) dynamical systems
and their quantization \cite{BG-tps}, \cite{Di-rmp}, \cite{Di-lqm}.
Hence we expect that the methods of geometric quantization can
be applied for quantizing these systems (perhaps after some minor
modifications).
Geometrically this means to quantize presymplectic manifolds.
Thus, in this brief section we make an introduction on how
the geometric quantization programme is applied
to singular systems and the problems arising in this procedure.
Representative references on this subject are
\cite{AS-86}, \cite{Blau-88a}, \cite{Blau-88b}, \cite{Go-86},
\cite{GS-81},\cite{Lo-90}, \cite{Ml-89}, \cite{Sn-83}, \cite{Tu-91};
and we refer to them for a more detailed expositions on the topics of
this section.
Survey expository works on the geometrical description of classical
singular systems are, for instance,
\cite{BK-ymtcs}, \cite{Ca-tsl}, \cite{GNH-pca}, \cite{GP-ggf},
\cite{MR-gs}.

\subsection{General setting and arising problems}

As it is known,  the phase space of a constrained system is a
presymplectic manifold $(C,\omega_C)$. But, in order to apply the
geometric quantization procedure we need a symplectic manifold.
Then, there are three kinds of symplectic manifolds associated
with $(C,\omega_C)$, namely: some extended phase space $(M,\Omega
)$, the reduced phase space $\rps$, and some ``gauge fixed
manifold'' $(P,\Omega_P)$. The first one can be constructed, in
general, by applying the {\sl coisotropic imbedding theorem}
\cite{Go-cit}, \cite{Ma-src}. Then $M$ is a tubular neighborhood
of $C$, considered as the zero section of the dual bundle
$(\ker\,\omega_C)^*$ (where $\ker\,\omega_C$ denotes the
characteristic bundle of $C$); and $M$ is unique up to local
symplectomorphisms between tubular neighborhoods. In the second
one, $\tilde C$ is the quotient $C/\ker\,\omega_C$ (which is
assumed to be a differentiable manifold). Finally, a gauge fixed
manifold is a global section of the canonical projection
$\rho\colon C\to\tilde C$. Thus, the quantization of a constrained
systems is based on the quantization of some of these symplectic
manifolds.

Geometric quantization via gauge fixing conditions
is not the most suitable way
The most usual forms of quantizing
a constrained system consist in using the reduced or some extended phase

space,
and these are the only cases that we discuss here
(comments and results on this subject can be found in \cite{Sn-83}).

As it is known the true space of physical states of a constrained
systems
is the reduced phase space $\rps$.
Hence, the most reasonable thing seems to quantize it directly.
Nevertheless, in general, this is a difficult task because:
\begin{itemize}
\item
The topology of $\tilde C$ is very complicated.
\item
Covariance is lost in the reduction procedure.
\item
Additional assumptions on $C$ are required in order to
$\tilde C$ have a suitable structure for quantization.
\item
$\tilde C$ has singularities, etc.
\end{itemize}
In spite of these problems, the geometric quantization
of the reduced phase space can be successfully carried out
in a significant number of cases \cite{AS-86}, \cite{Go-86},
\cite{Va-83};
and then we will denote $(\Hr ,{\cal O}(\Hr ))$
the Hilbert space  and the set of operators so obtained.
In these cases, the advantage of the method is that,
both constraints and gauge symmetries are incorporated
and divided out at the classical level and they
have not to be considered at the quantum level:
$\Hr$ is the intrinsic Hilbert space and ${\cal O}(\Hr )$ is the set of
quantum operators
corresponding to the constrained system.

Although it is not the true physical phase space of a constrained
systems,
quantization of the extended phase space is more natural by several
reasons:
\begin{itemize}
\item
The topology and geometry of $M$ are simpler than those of $\tilde C$.
\item
No additional requirements must be assumed for $C$.
\item
In many cases, the extended phase space $M$ is
an initial datum of the problem and it has not to be constructed;
although, in these cases, $C$ must be a coisotropic submanifold of $M$.
If $C$ is not coisotropically imbedded in $M$, then
second class constraints must be previously removed in order to achieve
a consistent quantization; and then the Poisson bracket must be replaced

by
a new operation called {\it Dirac bracket}.
In geometrical terms, this correspond to take
a symplectic manifold where the submanifold $C$
is coisotropically imbedded, and then use the Poisson bracket
operation defined on it by the corresponding symplectic structure
\cite{Sn-74}.
\end{itemize}
This method involves two steps:
starting from $(M,\Omega )$, we first obtain the corresponding pair
$(\He ,{\cal O}(\He ))$.
But $\He$ is not the true intrinsic Hilbert space of the constrained
systems, since constraints
have not been taken into account. Then, they have to be enforced at the
quantum level.
The way to implement this is known as the
{\it Dirac's method} of quantization of constrained systems,
whose guidelines are the following:
\begin{itemize}
\item
It is assumed that the final constraint submanifold $C$
is defined in $M$ as the zero set of a family of constraint functions
$\{ \zeta \}$,
and that these constraints are quantizable, that is,
$\{ \Op_\zeta \} \subset {\cal O}(\H )$.
\item
Then, constraints are enforced at the quantum level,
demanding that the set of admissible quantum states is
$\Hc := \{ \sta \in \H \ \mid \ \Op_\zeta \sta = 0 \}$.
\end{itemize}

The translation of this procedure in terms of geometric quantization
was carried out mainly by Gotay {\it et al}  \cite{Go-cit},
\cite{GS-81}.
This method leads to consistent results only if
it allows us to obtain a representation of the Lie algebra
of the gauge group on $\He$ and, in order to achieve it,
some previous requirements are needed (see \cite{GS-81}).
When the Dirac's method goes on,
a subset $\Hc \subset \He$ is obtained and it contains
the physical quantum wave functions of the constrained system.

Then, a subsequent problem to be taken into account is that $\Hc$
is not always a Hilbert space, and the way to make it
into a Hilbert space ${\cal H}_0$ is not clear in general.
When this is possible, ${\cal H}_0$ and ${\cal O}({\cal H}_0))$
are the intrinsic Hilbert space and the quantum operators of the
constrained system.

\subsection{Comparison between methods}

In this way, several new problems arise in relation to the geometric
quantization
of a constrained system. In fact:
\begin{enumerate}
\item
In some cases, to quantize both the reduced and the extended phase space

simultaneously is not always possible, since:
\ben
\item
Sometimes the reduced phase space $\rps$ cannot be quantized.
\item
Some (or none) extended phase space $(M,\Omega )$ is not quantizable,
because the Dirac's method is not applicable.
\item
The set $\Hc$ (if it exists) is no longer a Hilbert space.
\een
\item
If both methods of quantization go on, it is expected to be equivalent
\footnote{
In fact, at the classical level it is equivalent to construct the
reduced phase space $\rps$
or, starting from the extended phase space $(M,\Omega )$,
to carry out the constraints and divide the symmetries out.
Hence we can expect the same equivalence to be true at the quantum
level.
},
that is, the Hilbert spaces $\Hr$ and ${\cal H}_0$ would be unitarily
isomorphic;
and the same for the corresponding sets of quantum operators.
Unfortunately, as it was analyzed first in \cite{AH-82}, this is not the

case
because there are two kind of obstructions:
\begin{enumerate}
\item
The geometrical structures needed for the quantization of $(M,\Omega )$
(hermitian line bundles, polarizations, metalinear bundles)
could not be invariant under the action of the gauge group and
then they cannot be $\rho$-projectable in order to obtain
compatible geometrical structures for the quantization of $\rps$.
Thus, both quantization procedures are incompatible.
\item
Both quantization procedures can be compatible in the above sense,
but a second obstruction can appear when we introduce
the inner product on $\He$ and $\Hr$.
In fact,  ${\cal H}_0$ could not inherit an inner product from $\He$ or,

if does it,
the equivalence between both quantizations
could not be extended to an unitary isomorphism between ${\cal H}_0$ and

$\Hr$.
\end{enumerate}
\end{enumerate}

In the literature, these questions are referred as the
noncommutativity of the procedures of
{\sl reducing} (or {\sl constraining}) and {\sl quantizing};
and they constitute one of the main problems in the quantization of
constrained systems.

A lot of works have been devoted to discuss these topics:

For example, geometric quantization of constrained systems is studied in

\cite{Go-86}
for the particular case when the extended symplectic manifold
is a cotangent bundle;
showing that, under some general hypothesis, reduction and quantization
commute.
In relation to this last case, in \cite{DET-90}, some examples for which

these
hypothesis do not hold are discussed (and then, the problem
of quantization is solved in the ambient of the BRST theory).

Another more generic example of noncommutativity between reduction and
quantization
is given in \cite{Lo-90}, where the quantization of the extended phase
space
leads to non-equivalent different possibilities, in general.

\cite{GS-81} is another classical reference, where geometric
quantization via
coisotropic imbedding is studied. After giving the conditions for which
quantization is independent on the choice of the ambient symplectic
manifold;
other problems concerning this way of quantizing are commented;
namely: how the quantization procedure depends on the choice of a basis
of constraints
defining $C$ in $M$, or what happens if some of these constraint
functions is not quantizable or, even, how to quantize when there are
not
constraints defining $C$ in $M$.

Finally, \cite{AS-86} is mainly devoted to the study and comparison of
polarizations in both ways of quantization.

As a final remark, it is important to point out that many of the
problems concerning geometric quantization of singular systems
are solved using the more recent techniques of the
BRST theory. The explanation of this method goes far from the aim
of this work. Some references on this topic are
\cite{ALN-90}, \cite{ALN-91}, \cite{DEGST-91}, \cite{Ib-90},
\cite{Ko-77}, \cite{Lo-92},
\cite{Tu-92a}, \cite{Tu-92b} (as it is obvious, this list is far to be
complete).

\appendix

\section{Bundles associated with group actions}

Let $p : P \to M$ be
a principal fiber bundle with structural group $G$
and let $Q$ be a differentiable manifold
with a differentiable left action of $G$ in $Q$.
Consider the set $P \times Q$
and the action of $G$ defined by
$$\begin{array}{ccc}
G \times (P \times Q) & \to & P \times Q
\\
g,(a,q) & \mapsto & (ag, g^{-1}q)
\end{array}$$
Let $P \times_G Q$ be the orbit space of this action.
This is the quotient of $P \times Q$ by the equivalence relation
$(a,q) \sim (r,s) \Leftrightarrow \exists g \in G \
\mid \ (ag,g_{-1}q) = (r,s)$
We have the natural projection
$$
\begin{array}{ccccc}
\pi & : & P \times_G Q & \to & M
\\
& & \overline{(a,q)} & \mapsto & p(a)
\end{array}
$$

Let $\{ U_{\alpha},\psi_{\alpha} \}$ be
a trivializing system of $P$. Then
$\psi_{\alpha} : \pi^{-1}(U_{\alpha}) \to U_{\alpha} \times G$
is a diffeomorphism and
$\psi_{\alpha} \circ \psi_{\beta}^{-1}$
means to multiply by an element of $G$ in the fiber.
We can construct the diffeomorphisms
$$
p^{-1}(U_{\alpha}) \times Q \mapping{\psi_{\alpha} \times id}
U_{\alpha} \times G \times Q
$$
which allow us to define the bijections
$$\begin{array}{ccc}
\pi^{-1}(U_{\alpha}) & \mapping{\varphi_{\alpha}} & U_{\alpha} \times Q
\\
\overline{(a,q)} & \mapsto & (p(a),\bar\varphi_{\alpha}(a,q))
\end{array}$$
where $\bar\varphi_{\alpha}(a,q)$
is the unique element of $Q$ such that
$(p(a),e,\bar\varphi_{\alpha}(a,q)) \sim (\psi_{\alpha}(a),q)$.
Notice that $\varphi_{\alpha}$
is well defined; in fact,
consider $(a,q) \sim (a_1,q_1)$,
that is, $\exists g \in G \ \mid \ (a_1g,g_{-1}q_1) = (a,q)$
then
\beann
(\psi_{\alpha} \times id)(a,q) &=&
(\psi_{\alpha} \times id)(a_1g,g^{-1}q_1) =
(\psi_{\alpha}(a_1g),g^{-1}q_1)
\\ &=&
(\psi_{\alpha}(a_1g)g^{-1},q_1) =
(\psi_{\alpha}(a_1),q_1) =
(\psi_{\alpha} \times id)(a_1,q_1)
\eeann

Imposing the condition of $\varphi_{\alpha}$
to be diffeomorphisms,
we obtain on $\pi^{-1}(U_{\alpha})$
a structure of differentiable manifold which,
on its turn, gives a fiber bundle structure on
$P \times_G Q$ with the projection $\pi : P \times_G Q \to M$,
with fiber $Q$. So we have that $\pi : P \times_G Q \to M$
is a fiber bundle over $M$ with fiber $Q$ which is say to be
{\it associated} with $P$ by the action of $G$ in $Q$.
If $\{ U_{\alpha},\psi_{\alpha} \}$ is a trivializing system of $P$,
then we can construct another one of $P \times_G Q$
by means of
$$
\begin{array}{ccc}
\pi^{-1}(U_{\alpha}) & \mapping{\psi_{\alpha}} & U_{\alpha} \times Q
\\
\overline{(a,q)} & \mapsto & (p(a),\bar\varphi_{\alpha}(a,q))
\end{array}
$$
where $(\psi_{\alpha}(a),q) \sim (p(a),1,\bar\varphi_{\alpha}(a,q))$.

Now, let $\sigma : M \to P \times_G Q$ be a section of $\pi$. We
can interpret $\sigma$ as a map $\tilde \sigma : P \to Q$ in the
following way: let $a \in P$ and $x=p(a)$, we take $\sigma (x) =
\overline{(a,z)}$ and then we define $\tilde \sigma (a) := z$,
that is, we take $\tilde \sigma (a)$ in such a way that $\sigma
(x) = \overline{(a,\tilde \sigma (a))}$. You can observe that
there is only one element of the class of $\sigma (x)$ whose
representative has $a \in P$ as the first element, that is, such
that $\sigma (p(a)) = \overline{(a,\tilde \sigma (a))}$. It is
easy to prove that $\tilde \sigma$ is differentiable.

Thus we have a map
$$
\begin{array}{ccccc}
\eta & : & \Gamma (P \times_G Q) & \to & \Cinfty(P,Q)
\\
& & \sigma & \mapsto & \tilde \sigma
\end{array}
$$
Now we can ask for the image of $\eta$,
that is, given a map $\phi$,
which condition satisfies $\phi$
in order to be the map associated with a section of $\pi$?.
Let $\sigma$ be a section of $\pi$ and
$a,b \in P$ such that $p(a)=p(b)$,
then,
if $b=ag$,
we have
$$
\sigma(p(a))=
\sigma(p(b)) = \overline{(a,\tilde \sigma(a))} =
\overline{(b,\tilde \sigma(b))} =
\overline{(ag,\tilde \sigma(b))} =
\overline{(a,g \sigma(b))}
$$
That is, if $b=ag$, then
$$
\tilde \sigma(a) = g \tilde \sigma(b)
\Rightarrow
\tilde \sigma(ag) = g^{-1} \tilde \sigma(a)
$$
This condition characterizes the image of $\eta$.
In fact, let $\phi : P \to Q$ be a map satisfying that
$\phi (a) = g \phi (b)$ if $b = ag$. We construct a section
$\tilde \phi : M \to P \times_G Q$ in the following way:
$\tilde \phi (x) = \overline{(a,\phi (a))}$, $a$ being any element of
$P$ such that
$\pi (a) = x$. Obviously $\eta (\tilde \phi ) = \phi$.

\subsection*{Acknowledgments}

We are highly indebted with Prof.\ P.L. Garc\'{\i}a (Univ. Salamanca)
for making the reference \cite{Ga-83} available to us and allowing to
use it
as a guideline for some parts of this report.
We are grateful to Prof.\ L.A. Ibort (Univ. Carlos III of Madrid)
because he motived our interest and introduced us
in the subject of geometric quantization and many
other related areas. Moreover he has clarified some questions
which are not clearly explained in the usual literature.
We thank specially Prof.\ X.~Gr\`acia (Univ.\ Polit\`ecnica de
Catalunya)
for his carefully reading and, in general, for his helping in the
elaboration of this report.
It is also a pleasure to thank
Profs.\ J.F. Cari\~nena, M. Asorey, M.F. Ra\~nada, C. L\'opez, E.
Mart\'{\i}nez
and other members of the Dept. F\'{\i}sica Te\'orica (Univ.\ Zaragoza)
for the discussions and comments made in relation
to many aspects of geometric quantization;
and for inviting us to give a course on these topics.
Likewise, we thank
Profs.\ J. Girbau, A. Revent\'os (Univ.\ Aut\`onoma Barcelona)
and C. Curr\'as (Univ.\ Barcelona),
as well as other members of their departments,
for their suggestions and patience in attending our explanations.

We are also grateful for the financial support
of the CICYT TAP97-0969-C03-01.

\end{document}